\documentclass[twocolumn,times,
               ]{aastex631}

\usepackage{amsmath}	        
\usepackage{textcomp,gensymb}	

\newcommand{\thisstar}{HD~137010}
\newcommand{\thisstarb}{HD~137010~b}

\newcommand\ms{\ensuremath{\text{m}\,\text{s}^{-1}}}
\newcommand\kms{\ensuremath{\text{km}\,\text{s}^{-1}}}
\newcommand\masyr{\ensuremath{\text{mas}\,\text{yr}^{-1}}}

\definecolor{my_color}{HTML}{CF0000}

\begin{document}

\title{A Cool Earth-sized Planet Candidate Transiting a Tenth Magnitude K-dwarf From K2}

\correspondingauthor{Alexander Venner}
\email{alexandervenner@gmail.com}

\author[0000-0002-8400-1646]{Alexander Venner}
\altaffiliation{Planet Hunters K2 contributors}
\affiliation{Centre for Astrophysics, University of Southern Queensland, Toowoomba, QLD 4350, Australia}

\author[0000-0001-7246-5438]{Andrew Vanderburg}
\affiliation{Center for Astrophysics, Harvard \& Smithsonian, Cambridge, MA 02138, USA}

\author[0000-0003-0918-7484]{Chelsea X. Huang}
\affiliation{Centre for Astrophysics, University of Southern Queensland, Toowoomba, QLD 4350, Australia}

\author[0000-0001-6263-4437]{Shishir Dholakia}
\affiliation{Centre for Astrophysics, University of Southern Queensland, Toowoomba, QLD 4350, Australia}

\author[0000-0002-1637-2189]{Hans Martin Schwengeler}
\altaffiliation{Planet Hunters K2 contributors}
\affiliation{Citizen scientist, c/o Zooniverse, Department of Physics, University of Oxford, Denys Wilkinson Building, Keble Road, Oxford, OX1 3RH, UK}

\author[0000-0002-2532-2853]{Steve B. Howell}
\affiliation{NASA Ames Research Center, Moffett Field, CA 94035, USA}

\author[0000-0001-9957-9304]{Robert A. Wittenmyer}
\affiliation{Centre for Astrophysics, University of Southern Queensland, Toowoomba, QLD 4350, Australia}

\author[0000-0002-2607-138X]{Martti H. Kristiansen}
\altaffiliation{Planet Hunters K2 contributors}
\affiliation{Brorfelde Observatory, Observator Gyldenkernes Vej 7, DK-4340 Tølløse, Denmark}

\author{Mark Omohundro}
\altaffiliation{Planet Hunters K2 contributors}
\affiliation{Citizen scientist, c/o Zooniverse, Department of Physics, University of Oxford, Denys Wilkinson Building, Keble Road, Oxford, OX1 3RH, UK}

\author[0000-0002-0654-4442]{Ivan A. Terentev}
\altaffiliation{Planet Hunters K2 contributors}
\affiliation{Citizen scientist, c/o Zooniverse, Department of Physics, University of Oxford, Denys Wilkinson Building, Keble Road, Oxford, OX1 3RH, UK}



\begin{abstract}

The transit method is currently one of our best means for the detection of potentially habitable ``Earth-like" exoplanets. In principle, given sufficiently high photometric precision, cool Earth-sized exoplanets orbiting Sun-like stars could be discovered via single transit detections; however, this has not previously been achieved. In this work, we report a 10-hour long single transit event which occurred on the $V=10.1$ K-dwarf HD~137010 during K2 Campaign~15 in 2017. The transit is comparatively shallow ($225\pm10$~ppm), but is detected at high signal-to-noise thanks to the exceptionally high photometric precision achieved for the target. Our analysis of the K2 photometry, historical and new imaging observations, and archival radial velocities and astrometry strongly indicate that the event was astrophysical, occurred on-target, and can be best explained by a transiting planet candidate, which we designate HD~137010~b. The single observed transit implies a radius of $1.06^{+0.06}_{-0.05}~R_\oplus$, and assuming negligible orbital eccentricity we estimate an orbital period of $355^{+200}_{-59}$~days ($a=0.88^{+0.32}_{-0.10}$~AU), properties comparable to Earth. We project an incident flux of $0.29^{+0.11}_{-0.13}~I_\oplus$, which would place HD~137010~b near the outer edge of the habitable zone. This is the first planet candidate with Earth-like radius and orbital properties that transits a Sun-like star bright enough for substantial follow-up observations.


\end{abstract}



\section{Introduction} \label{sec:intro}

The detection of potentially habitable ``Earth-like" planets is a major and enduring goal of exoplanet research. At present, planets of Earth-like mass and/or radius orbiting in the circumstellar habitable zone (HZ), where liquid water could potentially exist, lie near to or beyond current sensitivity limits for Sun-like FGK stars. However, out of all exoplanet detection techniques, the transit method has demonstrated particularly high efficacy for the detection of these exoplanets. The search for Earth-sized exoplanets in the HZs of Sun-like stars was the core scientific aim of the \textit{Kepler} mission \citep{Kepler, Koch2010}, which observed a 116 square degree field in the northern sky quasi-continuously for 4~years. Of its many exoplanet discoveries, \textit{Kepler} detected the first known small transiting planets orbiting in the HZ (Kepler-22, \citealt{Borucki2012}; Kepler-62, \citealt{Borucki2013}; Kepler-452, \citealt{Jenkins2015}), including the first ``Earth-sized" HZ planet (Kepler-186, \citealt{Quintana2014}).\footnote{For the purposes of this work we arbitrarily define ``Earth-sized" as being within 25\% of the radius of Earth, i.e. $0.8<R_\text{p}<1.25~R_\oplus$.} However, as the majority \textit{Kepler} target stars are faint ($V>13$), the transits of these planets often lie close to the detection limits for Sun-like stars and the validity of some of these planets has been questioned \citep{Mullally2018, Burke2019}. This was exacerbated by the premature end of the \textit{Kepler} mission in 2013 due to instrument failure.

In recent times much of the focus of transiting exoplanet research has moved to the \textit{Transiting Exoplanet Survey Satellite} \citep[TESS;][]{TESS}. Like \textit{Kepler} before it, TESS has had an immense impact on the study of transiting planets. However, the shorter observing baselines and lower photometric precision of TESS means that it is not optimised for detection of temperate terrestrial planets orbiting Sun-like stars. Earth-sized planets discovered by TESS orbiting FGK-type tend to orbit bright stars with short orbital periods \citep[e.g.][]{Dragomir2019, Kunimoto2025}. In contrast, M-dwarfs present an advantage in that the transits of terrestrial planets are proportionally deeper and temperate planets can be found at shorter orbital periods. As a result, temperate terrestrial planets discovered by TESS have exclusively been found in orbit of M-dwarfs (e.g. TOI-700, \citealt{Gilbert2020, Gilbert2023, Rodriguez2020}; TOI-715, \citealt{Dransfield2024}; Gliese~12, \citealt{Dholakia2024, Kuzuhara2024}); indeed, the detection capabilities of TESS for M-dwarf planets even extend to planets significantly cooler than Earth \citep[e.g. $T_{\text{eq}}\approx$~200~K,][]{Scott2025}. However, equivalent planets orbiting Sun-like stars lie largely beyond the reach of TESS. In the near future, missions like PLATO \citep[][expected 2026]{PLATO2014, PLATO2024} and ET \citep[][expected 2028]{Ge2022} aim to detect Earth-sized planets orbiting in the habitable zones of Sun-like stars through a  \textit{Kepler}-like ``stare" observing strategy. If past \textit{Kepler} results provide a comparable benchmark, detection of planets with Earth-like orbital periods will require several years of observations.

Acknowledging this historical context, the K2 mission \citep{K2} occupies a unique position in this paradigm. After instrument failures prematurely ended the \textit{Kepler} prime mission, K2 repurposed the telescope to conduct a transit survey of the ecliptic plane focusing on brighter stars. After implementing corrections to the challenging photometric systematics caused by spacecraft motion, K2 observations provided high photometric precision that match \textit{Kepler} performance, especially for bright stars \citep[e.g.][section 6.3]{K2SFF_joint}. In the best of cases, the precision of K2 photometry allowed for the detection of planets with Earth-like transit depths ($\approx$100~ppm), making it potentially sensitive to temperate terrestrial exoplanets.

Since K2 depended on radiation pressure from the Sun for its pointing, the duration of each K2 pointing was restricted to $\lesssim$80~days per campaign. This meant that unlike \textit{Kepler}, transiting planets with longer orbital periods could not be detected conventionally through repeat transit detections. However, discoveries through observations of single transits does remain a possibility. The study of exoplanets from single transits began in earnest with \textit{Kepler} \citep[e.g.][]{Wang2015, ForemanMackey2016, Uehara2016}, developed further during K2 \citep[e.g.][]{Osborn2016}, and has become especially opportune in the TESS era \citep[e.g.][]{Eisner2021, Dalba2022, Mann2023, Sgro2024}. Some notable K2 discoveries based on single transits include the first K2 exoplanet discovery \citep[K2-2;][]{Vanderburg2015, Thygesen2024}, the unique planetary system of K2-93 \citep[HIP~41378;][]{Vanderburg2016}, a long-period Super-Earth in the Hyades cluster \citep[HD~283869;][]{Vanderburg2018}, and what may be the longest-period K2 planet from the extraordinary 54~hr-long transit observed on K2-311 \citep[EPIC~248847494;][]{Giles2018}.

Given the high photometric precision attained by K2, there is meaningful potential to detect single transit events from small long-period exoplanets. For stars smaller than the Sun, including the $\approx$41\% plurality of K/M-dwarf K2 targets \citep{Huber2016}, this could even extend to Earth-sized planets. Though the book has closed on K2 following the 2018 \textit{Kepler} end-of-life \citep{Howell2020}, ongoing study of K2 data continues to bring new discoveries to light \citep[e.g.][]{Incha2023}.

In this work, we report the detection of a shallow single transit in K2 observations of the relatively bright K-dwarf \thisstar{} ($V=10.1$). We find that the transit event, which was detected through visual inspection, is best explained by a planet of size ($1.06^{+0.06}_{-0.05}~R_\oplus$) and orbital period ($355^{+200}_{-59}$~d) comparable to Earth. We estimate that the planet candidate receives a low incident flux which may place it near the outer edge of the habitable zone, or potentially even beyond. This is a significant addition to the small sample of cool Earth-sized exoplanets, and presents a small milestone in the search for Earth-like exoplanets around nearby Sun-like stars.

\vspace{-2mm}
\section{Data} \label{sec:data}

\subsection{K2 Photometry} \label{subsec:photometry}

\begin{figure*}[t]
    \centering
    \includegraphics[width=\textwidth]{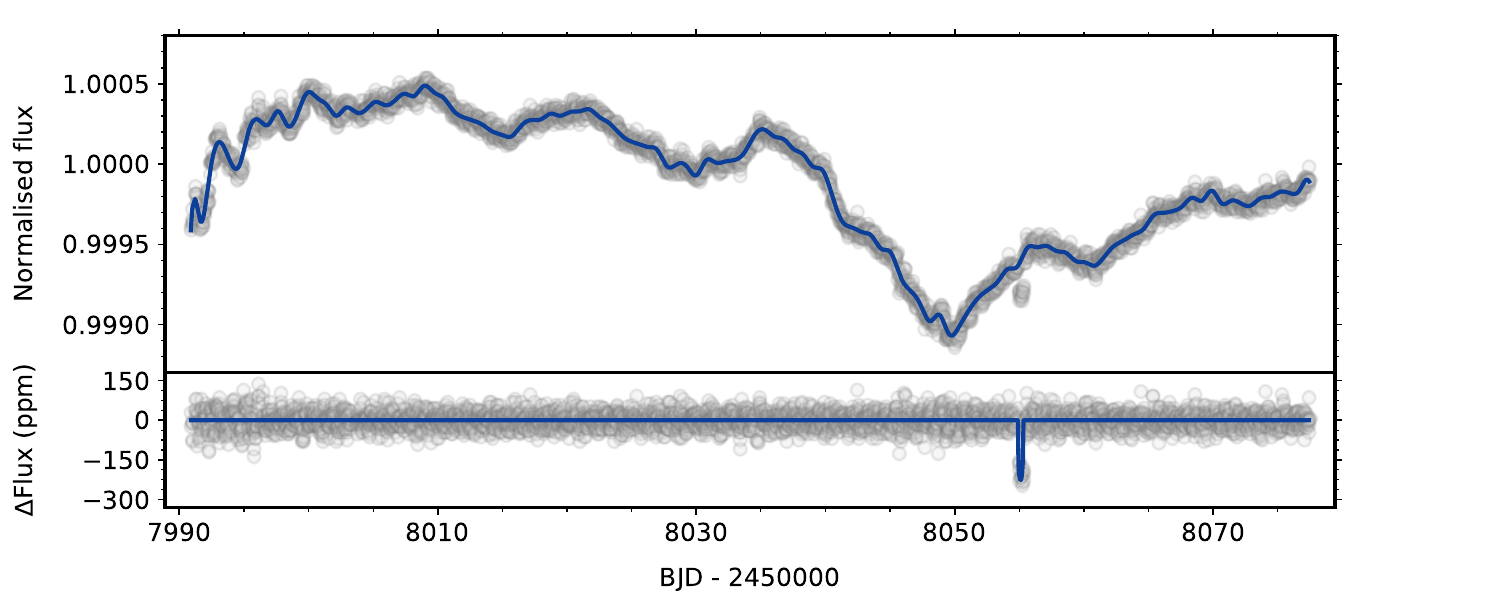}
    \includegraphics[width=\textwidth]{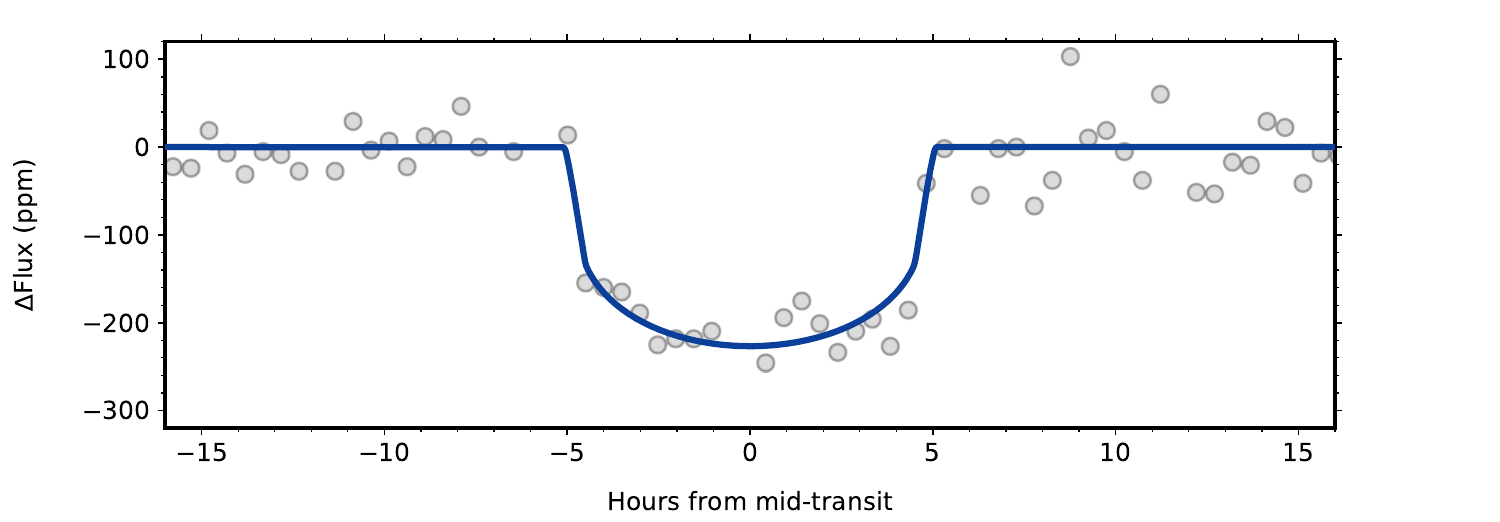}
    \caption{(\textit{Top}) Systematics-corrected K2 Campaign~15 light curve of \thisstar{} (EPIC 249661074). The individual K2 long-cadence flux measurements are marked by grey points, while the blue curve represents a spline model to detrend the long-term photometric variability. A noticeable drop in flux, consistent with a planetary transit, is visible around BJD $\approx$ 2458055.1. (\textit{Middle}) K2 light curve with long-term variability removed. A best-fit transit model has been applied to the transit event. (\textit{Bottom}) Inset focusing on the transit event. The event has a depth of $\sim$225~ppm and $\sim$10~hr duration, and we estimate it is detected at a high signal-to-noise of $\approx$30. The transit event is consistent with an Earth-sized exoplanet with a $\approx$1~yr orbital period.}
    \label{fig:photometry}
\end{figure*}

\thisstar{} (EPIC~249661074) was observed by \textit{Kepler} for 88 days between 2017 August 23 and 2017 November 19 during Campaign~15 of the K2 mission. Following the method of \citet{K2SFF}, a systematics-corrected light curve was generated and made available online.\footnote{\url{https://lweb.cfa.harvard.edu/~avanderb/k2c15/ep249661074.html}} Subsequently, one of us (H.M.S.) identified a single transit-like feature in the light curve through visual inspection using the \texttt{LcTools} software (\citealt{Schmitt2019}; see \citealt{VSG}). The transit-like feature occurred at BJD$\approx$2458055.1, with a shallow depth of around 225~ppm and a duration of about 10~hr. For this work, we produce a new systematics-corrected light curve from the target pixel files jointly fitting for a transit model (following \citealt{K2SFF}, \citealt{K2SFF_joint}), which is shown in Figure~\ref{fig:photometry} and tabulated in Appendix~\ref{appendix:data}. This event is easily noticed visually thanks to the unusually high precision of the K2 photometry; the 6.5~hr combined differential photometric precision (CDPP) is $\sim$8.5~ppm, close to the limiting dispersion achieved in either \textit{Kepler} or K2 \citep[ex.][]{Gilliland2011, Luger2016}. We also observe some low-level variability in the light curve acting at a timescale of weeks, though it is difficult to determine whether this reflects stellar activity or K2 systematics.

Next, we attempt to quantify the statistical significance of the event. We first estimate the signal-to-noise ratio (SNR) of the event assuming classical white Gaussian noise,

\begin{equation}
    \text{SNR}=\delta \frac{\sqrt{N}}{\sigma}
    \:,
\end{equation}

where $\delta$ is the transit depth, $N$ is the number of in-transit photometric data points, and $\sigma$ is the standard deviation of the out-of-transit light curve \citep[][section~5.2]{Rowe2014}. In our best-fit transit model (Figure~\ref{fig:photometry}), $\delta=225$~ppm, $N=18$, and $\sigma=32$~ppm, which yields $\text{SNR}=30$. Alternatively, as in \citet{Kunimoto2020}, we may estimate $\sigma$ as $1.48\times$ the median absolute deviation (MAD) of the out-of-transit light curve; here the MAD is 20~ppm, yielding $\text{SNR}=32$. This implies high significance for the signal.

A more realistic estimate of the true SNR of the event, taking into account red noise in the light curve, can be obtained by performing a matched filter analysis. Following \citet{Gilbert2023}, we convolved the K2 light curve with the best-fit transit model from our simultaneous fit to the transit and spacecraft systematics. We compared the peak of the matched filter response with the standard deviation of the filter response far from the transit to measure a signal to noise ratio of 11.2. If we ignore the first 20 days of the light curve which show slightly higher red noise levels, the SNR of the transit rises above 13. While lower than the white-noise SNR, this nonetheless indicates high statistical significance.

We then attempted to evaluate the veracity of the event in the K2 data. To confirm that the event is not simply caused by instrument effects, we inspected the K2 photometry of other stars near to our target. For all stars with K2 observations that lie within 5~arcminutes of \thisstar{}, we do not identify any convincing analogues of the transit-like feature.\footnote{The nearby $V=12$ star EPIC~249661654 is an eclipsing binary which has an eclipse coincident with the event on \thisstar{}. However, this eclipse has a shorter duration and a ``V-shaped" profile inconsistent with the \thisstar{} event. We therefore rule out contamination from this star.} We also confirmed that the profile and duration of the event is independent of our choice of aperture, consistent with a transit event on the target star.

We next inspected the K2 pixel images to test whether there is any evidence that the event occurred off-target. \thisstar{} is sufficiently bright ($K_p=9.79$) to significantly saturate the \textit{Kepler} detector, which makes it challenging to measure the in-transit centroid drift. Nonetheless, we do not observe any coherent changes in the photocentre position during the event at the pixel level. We also do not detect any moving objects in the \thisstar{} field around the time of transit, which serves to rule out the possibility that the signal could be caused by a passing solar system object.

We therefore conclude that the event observed in the K2 photometry of \thisstar{} is statistically significant and is consistent with having occurred on-target. Furthermore, owing to the high detection significance of the event, we can determine that it possesses a curved flux minimum and short ingress/egress, both of which are typically characteristic properties of a planetary transit.

We return to verify the transit hypothesis in more detail in Section~\ref{subsec:interpretation}. If the event is indeed a planetary transit on \thisstar{}, the 225~ppm depth implies a radius of $\sim$1~$R_\oplus$, which would be remarkably small for an object detected from a single transit. In addition, due to the absence of repeat events in the K2 light curve, its orbital period must be longer than $>$64~days. Estimating the orbital period of a planet from a single transit is challenging, but a first-order estimate can be made by comparison to the orbit of Earth. The idealised duration of the transit of Earth across the Sun is $\approx$13~hr \citep{Heller2016}, resulting in a transverse velocity across the idealised transit chord of $\sim$3.7~$R_\odot/\text{d}$. While the observed 10-hr duration of the transit is shorter and implies a transit velocity of $\sim$4.8~$R_*/\text{d}$, as \thisstar{} is a K-dwarf only $\sim$70\% the size of the Sun this entails a transverse velocity similar to Earth ($\sim$3.4~$R_\odot/\text{d}$) if the impact parameter is low. Hence, given the idealised assumptions of low eccentricity and impact parameter, the transit of \thisstar{} is consistent with a planet with an Earth-like orbital period ($P\approx1$~yr). We explore this more rigourously in Section~\ref{subsec:transit_model}.

\subsection{HARPS Spectroscopy} \label{subsec:spectroscopy}

Independent of the K2 observations, \thisstar{} was observed with the HARPS spectrograph \citep{HARPS} as part of the volume-limited planet search \citep{LoCurto2010, Sousa2011}. This survey was designed to include RV-amenable solar-type stars in the southern hemisphere within a distance of $<$57.5~pc. Though these observations are not sensitive to the expected RV signal of the transiting planet candidate, they allow us to check for the presence of other companions orbiting the star.

We downloaded the HARPS observations from the ESO Science Archive.\footnote{\url{https://archive.eso.org/wdb/wdb/adp/phase3_spectral/form?phase3_collection=HARPS}} \thisstar{} was observed with HARPS 8 times between 2006-04-09 and 2010-04-21, with a median RV uncertainty of 1.4~\ms{}. The observing cadence is discontinuous, as the first four observations date to 2006 whereas the latter four belong to 2010. The HARPS RVs do not show any evidence of short-term RV variability, which might occur in false positive scenarios. However, there is a $\approx$7~\ms{} offset between the 2006 and 2010 RVs, which is too large and has too long of a timescale to plausibly be attributed to the transiting planet candidate. We return to explore this variability in Section~\ref{subsec:companions}. Normalising for the offset between the 2006 and 2010 RVs, we find that the RMS of the RVs is 1.3~\ms{}, which is on par with the RV uncertainties. We therefore conclude that the available RV data do not contradict the planetary hypothesis. We reproduce the HARPS RVs in Appendix~\ref{appendix:data}.

\subsection{Hipparcos-Gaia Astrometry} \label{subsec:astrometry}

\thisstar{} has been observed by both \textit{Hipparcos} \citep[HIP~75398;][]{Hipparcos} and \textit{Gaia} \citep{Gaia}. This allows us to use \textit{Hipparcos-Gaia} astrometry to constrain the tangential motion of the star over the 25-year inter-mission baseline. We extract the astrometric data from the \textit{Gaia}~EDR3 version of the \textit{Hipparcos-Gaia} Catalog of Accelerations \citep[HGCA;][]{Brandt2018, Brandt2021}. We reproduce this data in Appendix~\ref{appendix:data}. There is no significant evidence for an astrometric acceleration in the \textit{Hipparcos-Gaia} astrometry; the offset between the \textit{Gaia} proper motion and the \textit{Hipparcos-Gaia} mean proper motion is
$\Delta\mu=$~($-0.003\pm0.074$, $+0.025\pm0.053$~mas~yr$^{-1}$) in right ascension and declination respectively, equivalent to $\Delta v=$~($-1\pm16$, $+5\pm11$~m~s$^{-1}$) in SI units \citep[][equations 9, 10]{Venner2021}. We therefore set a 3$\sigma$ upper limit of $\Delta v<47$~\ms{} on the net \textit{Hipparcos-Gaia} tangential velocity anomaly, excluding large variations in the stellar motion over the \textit{Hipparcos-Gaia} observing interval.

\subsection{Imaging} \label{subsec:imaging}

\begin{figure*}[t!]
    \centering
    \includegraphics[width=\textwidth]{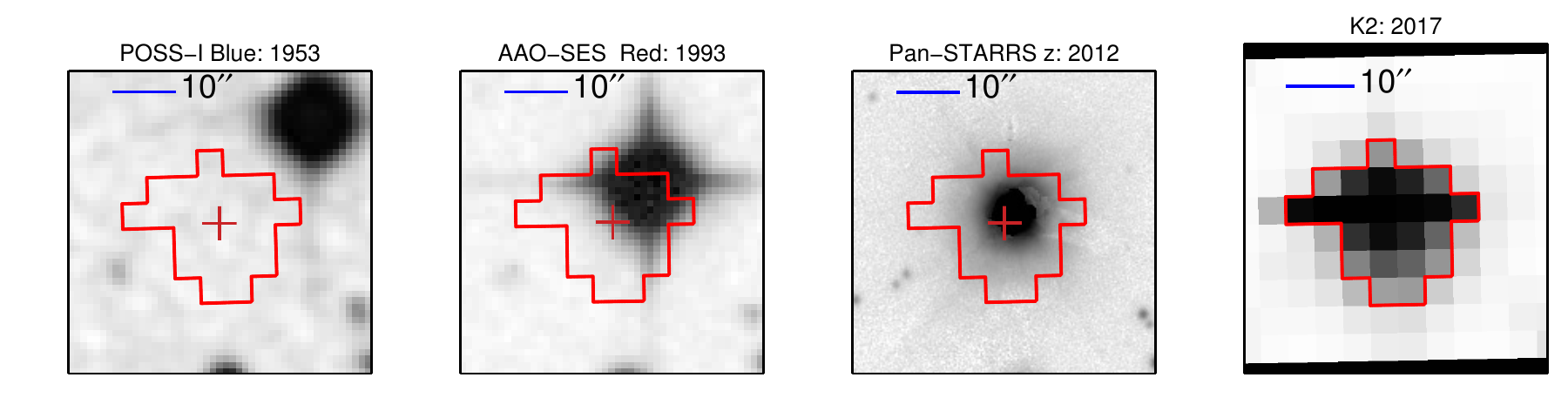}
    \caption{Archival imaging of \thisstar{} over time. The location of the star at the epoch of K2 observations is marked by a red cross, and the adopted K2 photometric aperture is shown by the red outline. (\textit{Far left}) POSS-I blue-wavelength photographic plate image from 1953. (\textit{Centre left}) AAO-SES red-wavelength photographic plate image from 1993. (\textit{Centre right}) Pan-STARRS \textit{z}-band image from 2012. (\textit{Far right}) Summed image from K2 observations. We do not observe any background stars brighter than $\approx$19$^{\text{th}}$ magnitude that fall within the K2 photometric aperture, leading us to conclude that background flux contamination is negligible.}
    \label{fig:archival_imaging}
\end{figure*}

Imaging observations can be used to search for comoving stellar companions as well as background sources which may contaminate observations of \thisstar{}. In particular, the star has a relatively large proper motion of 340~\masyr{} \citep{GaiaDR3}, making it possible to use archival imaging observations to constrain the existence of background stars within the 2017 K2 photometric aperture.

In Figure~\ref{fig:archival_imaging} we show a sequence of imaging observations ranging from 1953 through to the 2017 K2 observations, with the K2 photometric aperture shown to scale. The two nearest visible background stars to \thisstar{}, visible at $\approx$20" separation towards the lower left of the frame in Figure~\ref{fig:archival_imaging}, are detected in \textit{Gaia}~DR3 with \textit{G}-band magnitudes of 19.4 (Gaia~DR3~6254679268787107584) and 20.7 (Gaia~DR3~6254679273084416000). Since these sources are visible in all of the archival imaging, we estimate that these images would be able to detect background stars brighter than $\approx$19$^{\text{th}}$ magnitude within the K2 photometric aperture. However, no such background stars are visible in any of the archival images. We can therefore exclude the existence of any background sources with a flux contrast below $\Delta\lesssim9$~mag that contaminate the K2 photometry.

We obtained new high-resolution speckle imaging observations of \thisstar{} to check for close-separation stellar companions that would be unresolved in the previously discussed imaging data \citep{Schmitt2016}. We used the Zorro speckle instrument at the Gemini South Telescope \citep{Scott2021, Howell2025} to obtain high-contrast speckle imaging data on 2024-07-12. Observations were taken in two medium-band filters with central wavelengths of 562~nm and 832~nm respectively, with a total field-of-view of 2.8$\times$2.8". We reduced the imaging data following methods described in \citet{Howell2011, Howell2016}. We reproduce the reconstructed image and contrast limits in Appendix~\ref{appendix:data}. Our observations achieve a contrast limit of $\Delta=5$~mag at 0.1" separation, through to $\Delta=7-8$~mag around 1". The multiband detection limits exclude contaminants with spectral types later than $\gtrsim$A0 at 0.1" and $\gtrsim$M0 at 1".

We do not detect any additional stars in the speckle imaging field. However, since the speckle imaging epoch postdates the K2 observations by approximately 7 years, \thisstar{} will have moved by over $>$2 arcseconds in the intervening time and our new imaging observations would not capture background sources located in the K2 aperture. Nonetheless, the speckle imaging is informative for excluding comoving companions to \thisstar{}. We explore this further in Section~\ref{subsec:companions_wide}.

In summary, we find that there is no evidence for either background sources or comoving companion stars that could have interfered with or contaminated the K2 photometry.

\section{Analysis} \label{sec:analysis}


\vspace{-8mm}
\begin{deluxetable*}{lrr}[ht!]
\label{tab:star}
\centering
\tablecaption{Properties of \thisstar{}.}
\tablehead{\colhead{Parameter} & \colhead{Value} & \colhead{Reference}}
\startdata
~~Right Ascension $\alpha_{\text{J2000}}$ \dotfill & 15:24:21.25 & \citet{GaiaDR3} \\
~~Declination $\delta_{\text{J2000}}$ \dotfill & -19:44:21.68 & \citet{GaiaDR3} \\
~~$V$ (mag) \dotfill & $10.14\pm0.05$ & \citet{Tycho2} \\
~~$K_p$ (mag) \dotfill & $9.79$ & \citet{Huber2016} \\
~~Parallax $\varpi$ (mas) \dotfill & $22.292\pm0.017$ & \citet{GaiaDR3} \\
~~Distance (pc) \dotfill & $44.86\pm0.03$ & \citet{GaiaDR3} \\
~~R.A. proper motion $\mu_\alpha$ (\masyr{}) \dotfill & $+228.536\pm0.021$ & \citet{GaiaDR3} \\
~~Declination proper motion $\mu_\delta$ (\masyr{}) \dotfill & $-248.158\pm0.014$ & \citet{GaiaDR3} \\
~~Radial velocity (km~s$^{-1}$) \dotfill & $+27.866\pm0.001$ & \citet{Soubiran2018} \\
~~$U$ (km~s$^{-1}$) \dotfill & $+59.18\pm0.03$ & This work \\
~~$V$ (km~s$^{-1}$) \dotfill & $-11.16\pm0.01$ & This work \\
~~$W$ (km~s$^{-1}$) \dotfill & $-47.83\pm0.05$ & This work \\
\hline
~~Spectral type \dotfill & K3.5V & \citet{Gray2006} \\
~~$\log R^{\prime}_{HK}$ \dotfill & $-4.84$ & \citet{GomesSilva2021} \\
~~[Fe/H] (dex) \dotfill & $-0.22\pm0.07$ & \citet{Sousa2011} \\
~~$T_{\text{eff}}$ (K) \dotfill & $4770\pm90$ & This work \\
~~$M_*$ ($M_\odot$) \dotfill & $0.726\pm0.017$ & This work \\
~~$R_*$ ($R_\odot$) \dotfill & $0.707\pm0.023$ & This work \\
~~$\rho_*$ (g~cm$^{-3}$) \dotfill & $2.90^{+0.29}_{-0.26}$ & This work \\
~~log~$g$ (cm~s$^{-2}$) \dotfill & $4.60\pm0.03$ & This work \\
~~$L_*$ ($L_\odot$) \dotfill & $0.232^{+0.023}_{-0.021}$ & This work \\
~~Age (Gyr) \dotfill & $4.8-10$ & This work \\
\enddata
\end{deluxetable*}

\subsection{Stellar Properties} \label{subsec:star_properties}

\thisstar{} is a $V=10.1$~mag star with a spectral type of K3.5V \citep{Gray2006}, located at a distance of $44.86\pm0.03$~pc in the constellation Libra \citep[J2000 R.A. and Declination = 231.0885 -19.7394;][]{GaiaDR3}. In the following we recount the relevant physical properties of the star. We also summarise the most salient of these parameters in Table~\ref{tab:star}.

\subsubsection{Kinematics and Age Estimation} \label{subsec:age}

We first turn to the age of the system. \thisstar{} is appreciably subsolar in mass, so its main sequence lifetime is longer than the age of the Universe and its rate of evolution is sufficiently slow that stellar isochrone models are not strongly age-sensitive. To get a sense for the age of the star, we therefore first consider its kinematics.

Based on the \textit{Hipparcos} astrometric solution and the absolute RV measured by HARPS, \citet{Adibekyan2012} calculate space velocities of $(U_{\text{LSR}}, V_{\text{LSR}}, W_{\text{LSR}})=(+64, +2, -30)$~\kms{}, where $U$ is positive in the direction of the galactic centre, $V$ is positive towards the galactic rotation, and $W$ is positive in the direction of the north galactic pole. Using the methods for assignation of stars to galactic populations from \citet{Bensby2003} and \citet{Robin2003} respectively, \citet{Adibekyan2012} report an 89\% or 95\% probability that \thisstar{} belongs to the thin disk, with the remainder entailing assignment to the thick disk. There is therefore a high probability that this star belongs to the younger component of the galactic disk. There is a consensus that the thin disk began to form $\approx$8-10~Gyr ago \citep{Fuhrmann2011, Xiang2017}, so membership of \thisstar{} in this population implies an age below $<$10 Gyr.

We recalculate the relative space velocities of \thisstar{} following \citet{Johnson1987}, using newer positions and proper motions from \textit{Gaia}~DR3 and the radial velocity from the \citet{Soubiran2018} catalogue of \textit{Gaia} radial velocity standards (data reproduced in Table~\ref{tab:star}). For the conversion between equatorial and galactic coordinates, we use the \textit{Hipparcos} transformation matrix, also utilised in \textit{Gaia}~DR3.\footnote{See Section~4.1.7 (Equation 4.62) of the \textit{Gaia}~DR3 documentation \\(\url{https://www.cosmos.esa.int/web/gaia-users/archive/gdr3-documentation}).} We find $(U, V, W)=(+59.18\pm0.03, -11.16\pm0.01, -47.83\pm0.05)$~\kms{}. Adopting the solar space velocities from \citet{Schonrich2010}, we find absolute motions relative to the local standard of rest $(U_{\text{LSR}}, V_{\text{LSR}}, W_{\text{LSR}})=(+70.3\pm0.8, +1.1\pm0.5, -40.6\pm0.4)$~\kms{}. These values stand in reasonable agreement with those of \citet{Adibekyan2012}, though $V_{\text{LSR}}$ and $W_{\text{LSR}}$ differ fairly substantially; the difference can mainly be attributed to the \textit{Hipparcos} parallax, which is larger than the \textit{Gaia} value but relatively imprecise \citep[$\varpi_{\text{HIP}}=26.65\pm2.65$~mas;][]{HipparcosNew}.

Comparison with the stellar sample of \citet{Adibekyan2012} is sufficient to reconfirm that the star most likely belongs to the thin disk. However, the kinematics of \thisstar{} are comparatively hot for a thin disk member. Our value for the total velocity, $v_\text{tot}\equiv\sqrt{U_{\text{LSR}}^2+V_{\text{LSR}}^2+W_{\text{LSR}}^2}\approx 81$~km~s$^{-1}$, is notably high for a thin disk star. To better quantify the age implied by its kinematics, we use the $UVW$-age relationships of \citet{AlmeidaFernandes2018} and \citet{Veyette2018} to estimate the kinematic age of \thisstar{}. These are calibrated to Sun-like thin disk stars in the solar neighbourhood with known isochrone ages.

Using these velocity-age relationships, we find nominal ages of $T=9.3^{+3.0}_{-3.2}$~Gyr and $T=10.4^{+2.2}_{-2.4}$~Gyr respectively. These values abut with the upper limit on the age of the thin disk, clearly supporting an old age for the star. Given that these posteriors also lie at the maximal extreme of the calibration of the velocity-age relationship \citep[i.e. being limited to thin-disk stars with ages below $<$10~Gyr;][]{AlmeidaFernandes2018}, we opt to use the lower limits provided by the relations. The corresponding 3$\sigma$ lower limits are $T>2.6$~Gyr and $T>4.8$~Gyr respectively; we prefer to adopt the latter of these values, as this relation is less skewed towards young ages \citep{Veyette2018}. We therefore adopt a broad age constraint for \thisstar{} of $[4.8, 10]$~Gyr from its kinematics.

An old age for the star is consistent with the low level of observed stellar activity. \citet{Gray2006} report $\log R^{\prime}_{HK}=-4.96$ from their spectral observations of \thisstar{}, whereas \citet{GomesSilva2021} have more recently reported a mean $\log R^{\prime}_{HK}=-4.84$ from the HARPS spectra. This implies a low level of magnetic activity, as expected for a Sun-like star of supersolar age.

\subsubsection{Stellar Fundamental Parameters} \label{subsec:stellar_parameters}

Having established constraints on the age of \thisstar{}, we next use stellar models to determine its remaining properties. We use the MIST isochrones \citep{Dotter2016, Choi2016} to model the physical parameters of \thisstar{}, using a model applied in our previous work \citep{Venner2024}. We provide details of the data and priors used for our isochrone model in Appendix~\ref{appendix:isochrone}. The photometry we have used in our model includes space-based photometry from Tycho-2, \textit{Gaia}, 2MASS, and WISE which spans optical and infrared wavelengths. We adopt spectroscopic priors of $T_{\text{eff}}=4797\pm121$~K and $\text{[Fe/H]}=-0.22\pm0.07$~dex as determined from the HARPS data by \citet{Sousa2011}. We also use constraints from the preceding section as age priors.

Our isochrone model returns a stellar mass of $0.726\pm0.017~M_\odot$ and radius $0.707\pm0.023~M_\odot$, in line with expectations for a K3.5V star. This scales to a density of $2.90^{+0.29}_{-0.26}~$g~cm$^{-3}$ and $\log g$ of $4.60\pm0.03$ (surface gravity of $10^{4.60\pm0.03}$~cm~s$^{-2}$). We find a posterior effective temperature of $4770\pm90$~K, which coupled with the stellar radius results in a bolometric luminosity of $0.232^{+0.023}_{-0.021}~L_\odot$ using the Stefan-Boltzmann law. We summarise the key stellar parameters in Table~\ref{tab:star}.

\subsection{Limits on Additional Companions} \label{subsec:companions}

We next attempt to place constraints on the existence of additional bodies in the system. This includes placing limits on additional transiting planets in the K2 light curve, and attempting to constrain the existence of additional companions on external orbits using other data.

\subsubsection{Constraints on additional transiting planets} \label{subsec:companions_transit}

We first search for additional transit signals in the K2 photometry. We use the Box Least Square algorithm \citep[BLS;][]{Kovacs2002} implemented in the \texttt{Vartools} package \citep{vartools} to search the light curve time series, with the single transit masked out. We search for planets with period between 0.3 and 60 days, with 150,000 bins in frequency space. We do not detect any signals with a BLS signal-to-pink-noise ratio above $\geq$9, indicating there are no periodic transit signals from additional planets detectable in the light curve.

\begin{figure}
    \centering
    \includegraphics[width=\linewidth, trim=0cm 1.5cm 0cm 0.5cm]{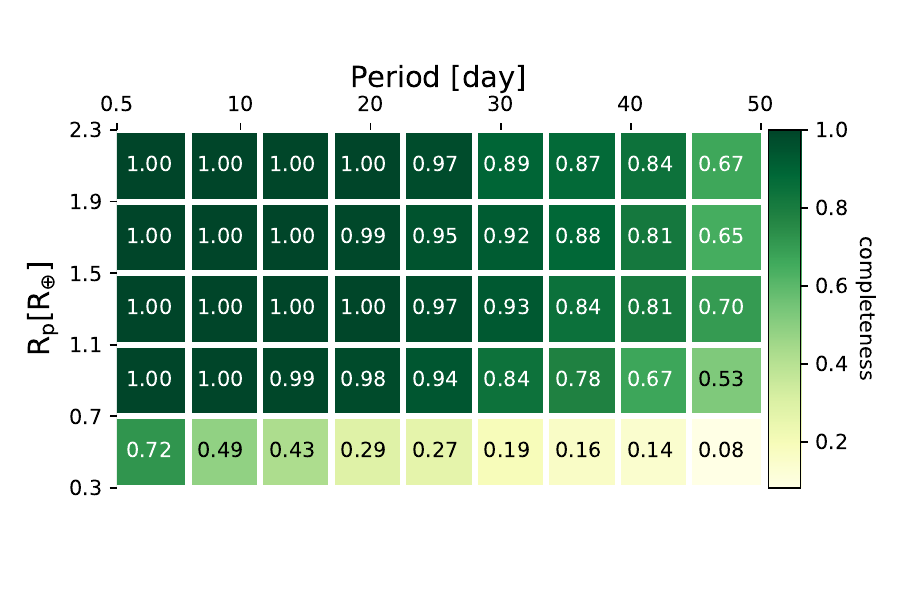}
    \caption{Detectability fractions for transiting planets injected into the K2 light curve across various orbital periods and radii. Owing to the high photometric precision, we are able to almost entirely exclude transiting planets larger than $>$1~$R_\oplus$ with orbital periods shorter than $<$30~days.}
    \label{fig:transit_detectability}
\end{figure}

We then perform an injection and recovery analysis to estimate the parameter space of transiting planets that could be detected in the K2 photometry. For each iteration of the injection and recovery, we simulate the transits of a planet with radii drawn from a uniform distribution between 0.3 to 4 $R_\oplus$, orbital periods drawn from a uniform distribution between 0.5 to 50 days, impact parameter uniformly distributed between 0 and 0.9, and a random orbital phase. The transit signal of the simulated planet is then injected to the systematics-corrected K2 light curve (Figure~\ref{fig:photometry}, top panel) with the known transit signal masked out. We then use a Basis Spline \citep{Vanderburg2016} to detrend the light curve before performing the same BLS search described previously. The injected planet is defined as recovered if it is detected with a BLS signal-to-pink-noise above $>$9, and with a period and epoch matching the originally injected signal following the algorithms described in \citet{Coughlin2014}. We repeat the injection and recovery for 20,000 iterations to sample across the period and radius parameter space. We present the distribution of detectability fractions in Figure~\ref{fig:transit_detectability}. Given the extremely high precision of the K2 light curve, we achieve an exceptionally high detectability completeness for planets larger than $>$1~$R_\oplus$, and can effectively rule out the existence of any additional transiting planets larger than Earth orbiting \thisstar{} with orbital periods below $<$30~days. Towards smaller planetary radii our sensitivity declines, though we may nonetheless largely exclude planets larger than Mars ($>$0.5~$R_\oplus$) with orbital periods shorter than $<$15~days. For orbital periods beyond $>$30~days our detectability fractions begin decline across the board, with the detection efficiency being limited by the 88-day duration of the K2 light curve. We conclude that there is no evidence for additional transiting planets as large as \thisstarb{} in the K2 photometry.

\subsubsection{Constraints on companions on wider orbits} \label{subsec:companions_wide}

We next attempt to place constraints on tertiary companions orbiting exterior to the detected transiting planet candidate. In Section~\ref{subsec:imaging}, we used our speckle imaging observations to put limits on the existence of stellar companions between projected separations of $0.1-1$" ($=5-45$~AU). Based on the 832~nm contrast limits, we estimate that within this range we can rule out stellar companions more massive than $\gtrsim$0.15~$M_\odot$ beyond a projected separation of $>$0.1" ($>$5~AU), and $\gtrsim$0.1~$M_\odot$ beyond $>$0.6" ($>$25~AU).

To extend our companion limits beyond $>$1", we use data from \textit{Gaia}~DR3 \citep{GaiaDR3}. Empirically, \textit{Gaia} is sensitive to binaries with $\Delta G\approx4$~mag contrast at 1" separation and $\Delta G\approx7$~mag at 2", with sensitivity continuing to improve thereafter \citep[][figure 9]{ElBadry2021}. \textit{Gaia} has detected no other sources within 10" of \thisstar{}. This effectively rules out stellar companions more massive than $\gtrsim$0.25~$M_\odot$ at 1" ($45$~AU projected) and companions above $\gtrsim$0.15~$M_\odot$ beyond $>$2" ($>$90~AU). We do not find any stars in \textit{Gaia}~DR3 which share distance and proper motion within 10~arcminutes of \thisstar{}. 

To further expand our sensitivity to wide companions, we next consider constraints from the HARPS RVs and the \textit{Hipparcos-Gaia} astrometry. We do not observe any significant evidence for an acceleration in the astrometry, but in the HARPS data we do see a $\approx$7~\ms{} offset between the 2006 and 2010 observations. Given the limited scope of the HARPS dataset, it is difficult to evaluate the origin of this drift. However, we note that the amplitude and timescale are too large to be explained by the transiting planet candidate, and we do not observe a correlation between the RVs and the width of the line bisectors that might indicate an origin from stellar activity. We therefore consider the possibility that the drift in the HARPS RVs could originate from an additional companion orbiting \thisstar{}.

Given that we have constraints on both the radial and tangential velocity change for the star, it is possible to compute companion masses as a function of velocity change and projected separation. The relevant expression for companion mass $M$ (in $M_J$) is

\begin{equation}
\label{equation:acceleration}
\begin{split}
M=5.599\times 10^{-3}\times~& D^2 \times \rho^2 \times \left(\frac{d TV}{dt}\right)^{-2} \times \\
&\left[\left(\frac{d RV}{dt}\right)^2 + \left(\frac{d TV}{dt}\right)^2\right]^{\frac{3}{2}}
\:,
\end{split}
\end{equation}
where $D$ is the stellar distance in parsecs, $\rho$ is the projected separation in arcseconds, $RV$ represents the radial velocity, and $TV$ the tangential velocity $=\sqrt{\mu_{\alpha}^2 + \mu_{\delta}^2}$ (\citealt{Bowler2021}, equation 2; and see further \citealt{Brandt2019}, section 5). Since $D$ is known from \textit{Gaia} parallax, we can convert from the observed acceleration terms into companion masses as a function of projected separation \citep[for this see also][]{Kunimoto2025}.

\begin{figure*}
    \centering
    \includegraphics[width=\columnwidth]{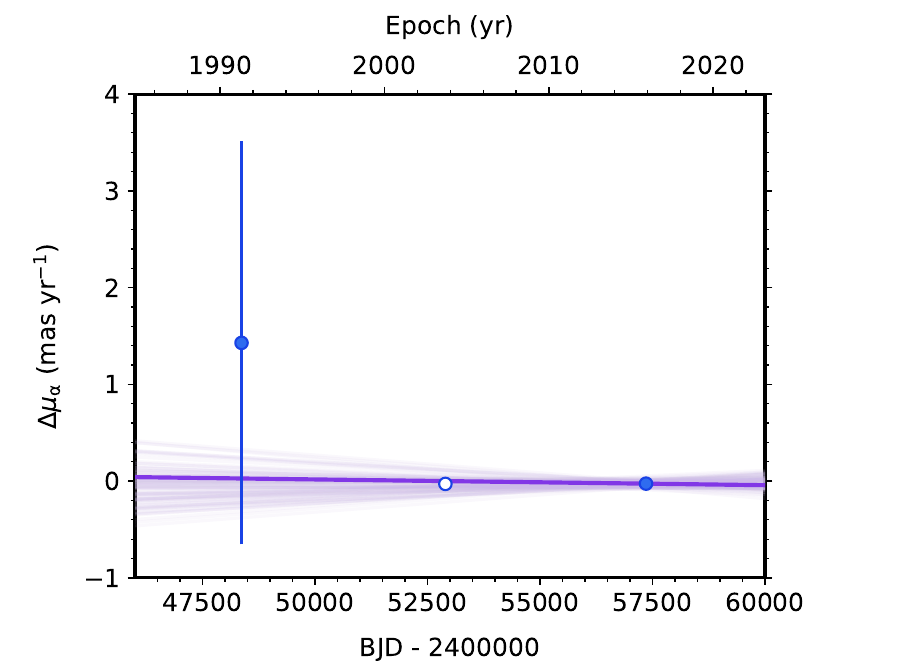}
    \includegraphics[width=\columnwidth]{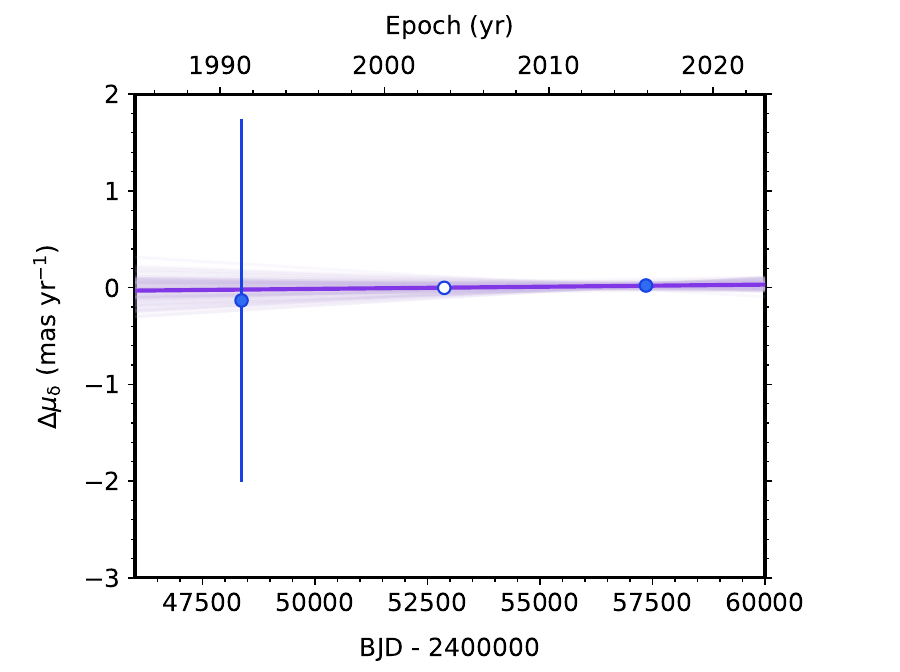}
    \includegraphics[width=\columnwidth]{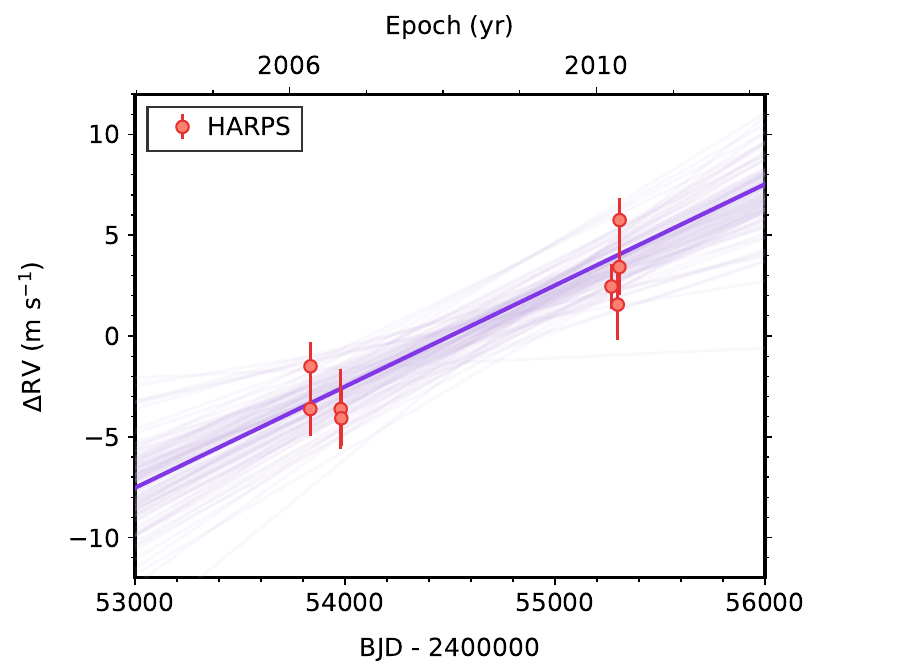}
    \includegraphics[width=\columnwidth]{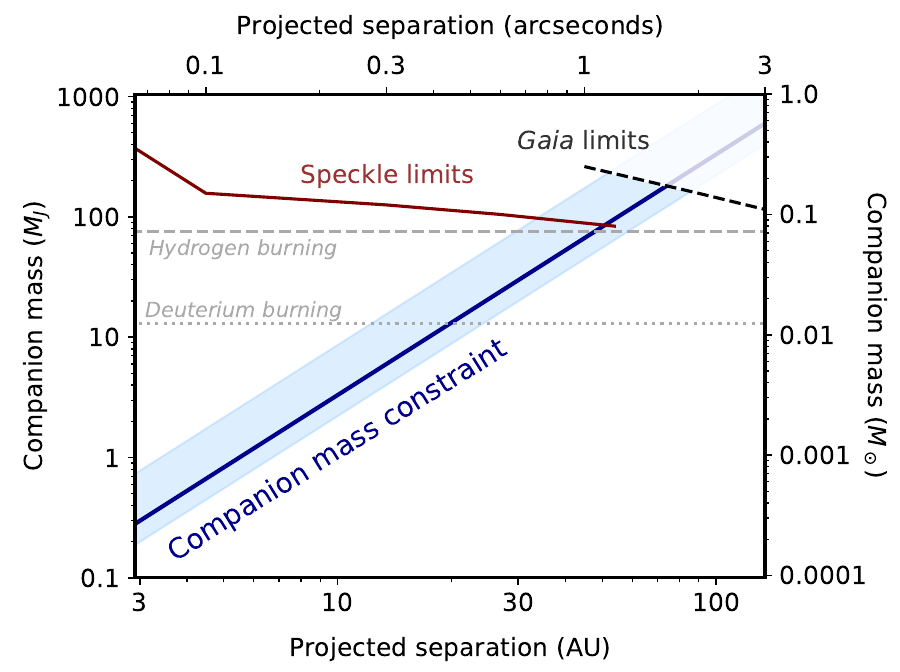}
    \caption{Linear acceleration model for \thisstar{} in \textit{Hipparcos-Gaia} astrometry (right ascension, top left; declination, top right), and radial velocity (bottom left; note different axes). The radial acceleration is nominally over $>$4$\sigma$ from zero ($+1.7\pm0.4$~m~s$^{-1}$~yr$^{-1}$), which may suggest the existence of a third body in the system, while the astrometric accelerations are not significant. We estimate the implied companion mass as a function of projected separation (bottom right). The RV acceleration may be explained by a planetary-mass companion within $\lesssim$20~AU, or a brown dwarf within $\lesssim$50~AU; most stellar companions are excluded by imaging non-detections. However, the limited HARPS dataset prevents us from determining whether the acceleration is truly linear, so further observations are necessary to understand the RV variation.
    }
    \label{fig:acceleration}
\end{figure*}

We perform a simple fit to the RVs and astrometry using linear acceleration terms. Using this model, we find acceleration terms of $\frac{d RV}{dt}=+1.7\pm0.4$ m~s$^{-1}$~yr$^{-1}$, $\frac{d\mu_{\alpha}}{dt}=0.0\pm1.3$ m~s$^{-1}$~yr$^{-1}$, and $\frac{d\mu_{\delta}}{dt}=+0.4\pm1.0$ m~s$^{-1}$~yr$^{-1}$; we plot the best-fit linear acceleration solutions in Figure~\ref{fig:acceleration}. The radial acceleration is nominally significant at above $>$4$\sigma$ confidence; however, the time baseline of the HARPS RVs is substantially less than the \textit{Hipparcos-Gaia} astrometry (4~yr versus 25~yr), so it is not currently possible to demonstrate that the acceleration is linear over long timescales. The method we are using here relies on the assumption that the accelerations are effectively instantaneous \citep{Brandt2019}, which is uncertain in this case. We therefore advise caution in interpreting these results.

Assuming that a linear acceleration model is valid, we find that the companion mass lies under the canonical deuterium-burning limit ($<$13~$M_J$) for projected separations below $\lesssim$20~AU, and under the hydrogen-burning limit ($\lesssim$75~$M_J$) within $\lesssim$50~AU. The absence of any larger acceleration in the RVs and astrometry thoroughly rules out any stellar-mass companions interior to $<$30~AU. Given that our speckle imaging rules out companions above $\gtrsim$0.1~$M_\odot$ between $25-45$~AU, and \textit{Gaia} can exclude companions more massive than $\gtrsim$0.15~$M_\odot$ beyond $>$90~AU, we can reject the existence of all but the least massive stellar companions to \thisstar{} at all projected separations. This strongly suggests that \thisstar{} is a single star, and severely constrains the possible parameter space for stellar companions that could contaminate the K2 photometry.

\subsection{Signal Interpretation} \label{subsec:interpretation}

For exoplanet candidates that have only been detected through transit observations, care must be taken to rule out various false positive scenarios involving architectures without planets that could equally reproduce the photometry. There are now well-established guidelines for the statistical validation of transiting planets \citep[e.g.][]{Vanderburg2019}. We consider the following false positive hypotheses which could potentially reproduce the observed transit of \thisstar{}:

\begin{enumerate}
    \item{\textit{On-target eclipsing binary (EB):} If the observed transit signal were caused by the eclipse of \thisstar{} by a bound stellar companion, we would expect a large (several \kms{}) reflex velocity signal corresponding to a binary orbit of a few years. No such variability is detected, with both the RVs and astrometry being stable at the level of a few \ms{}. We can therefore exclude this hypothesis.
    }
    \item{\textit{Nearby or background eclipsing binary:} An unassociated eclipsing binary could be the source of the transit if it lies sufficiently close to \thisstar{}. Though the K2 photometric aperture is extended due to saturation of the target star ($\approx20-30$"), our analysis of the pixel images suggests that it is unlikely that the transit centroid offset is much larger than a \textit{Kepler} pixel ($\lesssim$4"; Section~\ref{subsec:photometry}). Using archival imaging, we have ruled out background sources with a contrast below $\Delta\lesssim9$~mag at the K2 epoch (Section~\ref{subsec:imaging}).
    If the transit did originate from a blended EB, the observed depth would follow
    \begin{equation}
    \label{eq:blend}
        \delta_{\text{obs}}=\delta_{\text{EB}}\frac{f_{\text{EB}}}{f_{\text{total}}}
        \:,
    \end{equation}
    where $\delta$ is the transit depth as a proportion of total flux, $f_{\text{EB}}$ is the total flux of the blended EB and $f_{\text{total}}$ is the combined flux of the target and blend (\citealt{Vanderburg2019}, equation~1; \citealt{Kunimoto2025}, equation~2). We can convert from magnitude contrast $\Delta m$ to flux ratio following $(f_{\text{EB}}/f_{\text{total}})=10^{-0.4\Delta m}$. Given $\delta_{\text{obs}}=225$~ppm and $\Delta m>9$~mag, blending with an undetected EB would entail $\delta_{\text{EB}}>90\%$ at minimum, i.e. quasi-total eclipse of the source. However, such a deep eclipse would require a large occulter which would produce a ``V-shaped" transit profile, inconsistent with the ``U-shaped" transit we observe. We can therefore reject the background EB hypothesis.
    }
    \item{\textit{Hierarchical triple:} This scenario requires a blended eclipsing binary which is gravitationally bound to the system. Due to the absence of a significant centroid shift, we would still require that the EB lies within $\lesssim$4" ($\lesssim$200~AU projected separation). The absence of any large acceleration in the HARPS RVs and \textit{Hipparcos-Gaia} astrometry rule out any stellar-mass companions within $\lesssim$0.7" (Section~\ref{subsec:companions_wide}). Between $0.6-1$" our speckle imaging rules out companions above $\gtrsim$0.1~$M_\odot$; \textit{Gaia} excludes companions above $\gtrsim$0.25~$M_\odot$ between $1-2$", and above $\gtrsim$0.15~$M_\odot$ beyond $>$2". This severely restricts the potential parameter space for undetected stellar companions, but does not rule them out entirely.
    However, we may still rule out the hierarchical triple hypothesis using information from the transit shape:
    \begin{equation}
    \label{eq:shape}
        \delta_{\text{EB}}\leq\frac{(1-t_\text{F}/t_\text{T})^2}{(1+t_\text{F}/t_\text{T})^2}
        \:,
    \end{equation}
    where $t_\text{T}$ is the total transit duration (first to fourth contact) whereas $t_\text{F}$ is the transit duration excluding ingress and egress (second to third contact; \citealt{Kunimoto2025}, equation~3, cf. \citealt{Vanderburg2019}, equation~2). We may then insert this into Equation~\ref{eq:blend} to derive a lower limit on the flux contrast for hierarchical triples that could reproduce the observed transit. In our case, all we can empirically prove is that transit ingress/egress is not clearly resolved in the 29.4-minute cadence of the K2 photometry (Figure~\ref{fig:photometry}); we set $t_\text{T}=10$~hr and $t_\text{F}=9$~hr to allow for a $\sim$30-minute ingress/egress, which result in $\Delta m<2.7$~mag. Given the \textit{Kepler}-band $K_p=9.79$ of \thisstar{}, this would require the blended companion to be brighter than $K_p>12.5$, which at the system distance corresponds to a $\gtrsim$M0V companion more massive than $\gtrsim$0.5~$M_\odot$. Given that we have already ruled out such massive stellar companions at all separations, we can reject the hierarchical triple hypothesis.
    }
\end{enumerate}

In conclusion, we find that none of the classical false positive scenarios for transit signals can be invoked to explain the transit observed on \thisstar{}. It is customary at this point to quantify a false positive probability for the transit signal using tools for statistical validation, but this is not possible in this case since we have only a single transit event to work with. Nonetheless, the only further evidence required by the transiting planet hypothesis is periodic repetition of transits. Though we cannot prove that these occur for \thisstar{} due to our limited data, there is equally no  hypothesis other than a planetary transit that can reasonably explain the event. We therefore consider the transit signal observed on \thisstar{} to originate from a planet candidate \citep[\textit{sensu}][]{Vanderburg2018}, which we henceforth refer to as \thisstarb{}.

\subsection{Transit Model} \label{subsec:transit_model}

\vspace{-8mm}
\begin{deluxetable*}{lr}[t!]
\label{tab:planet}
\centering
\tablecaption{Parameters of \thisstarb{} from our transit model.}
\tablehead{\colhead{Parameter} & \colhead{Value}}
\startdata
~~Transit midtime $T_0$ (BJD) \dotfill & $2458055.0969^{+0.0035}_{-0.0032}$ \\
~~Transit duration $T_D$ (hr) \dotfill & $9.76^{+0.21}_{-0.18}$ \\
~~Impact parameter $b$ \dotfill & $<0.85$ (3$\sigma$) \\
~~Relative transit velocity $R_*~\text{d}^{-1}$ \dotfill & $4.76^{+0.23}_{-0.65}$ \\
~~Transit depth $\delta$ (ppm) \dotfill & $225\pm10$ \\
~~Radius ratio $R_p/R_*$ \dotfill & $0.01368^{+0.0006}_{-0.0004}$ \\
~~Linear limb darkening $u_1$ \dotfill & $0.56\pm0.07$ \\
~~Quadratic limb darkening $u_2$ \dotfill & $0.10\pm0.07$ \\
\hline
~~Orbital period $P$ (days) \dotfill & $355^{+200}_{-59}$ \\
~~Semi-major axis $a$ (AU) \dotfill & $0.88^{+0.32}_{-0.10}$ \\
~~Semi-major axis ratio $a/R_*$ \dotfill & $270^{+93}_{-37}$ \\
~~Orbital inclination $i$ ($\degree$) \dotfill & $>89.82^{+0.05}_{-0.03}$ (3$\sigma$) \\
~~Transit velocity (km~s$^{-1}$) \dotfill & $27.0^{+1.8}_{-3.6}$ \\
~~Radius $R_p$ ($R_\oplus$) \dotfill & $1.06^{+0.06}_{-0.05}$ \\
~~Incident flux $I$ ($I_\oplus$) \dotfill & $0.29^{+0.11}_{-0.13}$ \\
~~Equilibrium temperature ($T_\text{eq}$) for $\alpha=0.0$ \dotfill & $205^{+17}_{-28}$ \\
~~Equilibrium temperature ($T_\text{eq}$) for $\alpha=0.3$ \dotfill & $188^{+16}_{-25}$ \\
~~Equilibrium temperature ($T_\text{eq}$) for $\alpha=0.5$ \dotfill & $173^{+14}_{-23}$ \\
\enddata
\end{deluxetable*}

While confirmation of the exoplanet candidate \thisstarb{} is not currently possible within the limitations of the available data, we may still provisionally investigate its properties through a more involved analysis of the observed transit. Modelling of planets detected from single transits presents a unique challenge in that we do not possess an input constraint on the orbital period, upon which several of the other transit parameters depend. However, the orbital period can be estimated from the transit duration when assuming a prior on the stellar density and orbital eccentricity \citep[e.g.][]{Sandford2019}. Whereas one approach is to fit for the transit duration and use this to estimate the period, we opt to directly fit for the period while applying the \citet{Kipping2018} single transit probability prior. This allows us to simultaneously weight against solutions with orbital periods too short to reproduce the long transit duration and solutions with long period orbital configurations less likely to be observed in transit.

For the purposes of our transit model we make an assumption of zero orbital eccentricity. Since we have only one transit of \thisstarb{}, we must effectively constrain two physical parameters (period, eccentricity) from the transit duration, which is also substantially dependent on the impact parameter; hence some amount of prior information is necessary. Evidence from the \textit{Kepler} sample suggests that small planets preferentially have low orbital eccentricities \citep{Kane2012}, however this group is skewed towards short orbital periods less analogous to \thisstarb{}. Though eccentricity information for longer-period terrestrial planets is limited, low eccentricities do appear to be common. In the solar system, the orbital eccentricities of the terrestrial planets range between 0.007 (Venus) to 0.206 (Mercury). Recently, \citet{Kipping2025} have investigated the eccentricity distribution of a small sample of transiting exoplanets and candidates with Earth-like radii and incident fluxes orbiting KM-type stars, and found that the mean eccentricity of the sample is likely to be small ($\leq$0.15). To first order, the effect of increasing orbital eccentricity on the transit duration is an increase in the uncertainty on the orbital period as the transit velocity diverges from the mean orbital velocity. However, in our case the uncertainty on the orbital period is already large due to our poor constraint on the transit impact parameter; for small orbital eccentricities ($\lesssim$0.2), the impact on the orbital period uncertainty is not larger than the uncertainty arising from the impact parameter. We therefore make the simplifying assumption that \thisstarb{} has zero orbital eccentricity. If the orbital period of \thisstarb{} can be further constrained in the future, it may become possible to validate this assumption.

For our transit model we use the \texttt{batman} package \citep{batman}, which implements the \citet{MandelAgol2002} analytic transit model. To account for the 29.4-min \textit{Kepler} long-cadence integration time, we supersample our model by a factor of 10. As variables in the model, we fit for the stellar mass $M_*$ and density $\rho_*$, transit time $T_0$, orbital period $P$, impact parameter $b$, planet-to-star radius ratio $R_p/R_*$, and quadratic limb darkening coefficients $u_1$ and $u_2$. We fit $T_0$, $b$, $u_1$, and $u_2$ uniformly, whereas all other variables are fitted log-uniformly. Furthermore, we apply Gaussian priors on the stellar mass and density using the posteriors from our isochrone model ($\log M_*=-0.139\pm0.010$, $\log \rho_*=0.462\pm0.041$; see Table~\ref{tab:star}) and priors on the limb darkening coefficients derived from the \citet{Claret2018} models in the \textit{Kepler} bandpass ($u_1=0.60\pm0.07$, $u_2=0.12\pm0.07$).

We use the Markov Chain Monte Carlo ensemble sampler \texttt{emcee} \citep{emcee} to explore the parameter space. A total of 50 walkers were used to sample our 9-parameter model. The posterior space is substantially non-Gaussian, especially in orbital period, and we found that large chain lengths were required to reach a satisfactory degree of convergence. We opted to set a fixed upper limit on the orbital period of $P<2000$~d, which we may justify \textit{a posteriori} as such long periods are improbable when allowed, but significantly hinder convergence of the model. Our final model was run for a total of $1\times10^8$ steps, requiring about 1~day of computing time. To derive our posterior samples, we discarded the first half of the chain as burn-in and saved every hundredth step across each of the walkers.

We present our posterior values for the planetary parameters of \thisstarb{} in Table~\ref{tab:planet}. We report the medians and 68.3\% confidence intervals, except for the poorly constrained impact parameter $b$ for which we report our 3$\sigma$ upper limit ($<$0.85, where the mode of the posterior is $b=0$), and a corresponding lower limit on the orbital inclination. We determine that the transit mid-time occurred at $T_0=2458055.0969^{+0.0035}_{-0.0032}$~BJD, had a total duration of $T_D=9.76^{+0.21}_{-0.18}$~hr, and entailed a radius ratio of $R_p/R_*=0.01368^{+0.0006}_{-0.0004}$. Combining these parameters, we may derive the transverse velocity of \thisstarb{} during transit using the equation

\begin{equation}
    v_\text{tr}=\frac{2\sqrt{(1+R_p/R_*)^2-b^2}}{T_D}
\end{equation}

in relative units of $R_*/\text{d}$ \citep[][equation~1]{Osborn2016}. We therefore derive a transverse velocity during transit of $4.76^{+0.23}_{-0.65}$~$R_*/\text{d}$, which converts to $27.0^{+1.8}_{-3.6}$~\kms{} in absolute units when accounting for the stellar radius. The posterior uncertainties on this value are largely generated by the uncertainties on the impact parameter, which also drives the asymmetric confidence interval. Within the uncertainties, the transit velocity of \thisstarb{} is intermediate to the mean orbital velocities of Earth and Mars in the solar system (30~\kms{} and 24~\kms{} respectively), suggesting a comparable orbital separation.

Informed by the stellar mass and density, we ultimately estimate an orbital period of $355^{+200}_{-59}$~d for \thisstarb{}, with a mode of $\approx$323~d. Emphasising our assumption of zero orbital eccentricity, shorter orbital periods are rendered improbable since they cannot reproduce the 10-hour transit duration even as $b\rightarrow0$, whereas longer orbital periods are mainly penalised by the $\propto P^{-5/3}$ scaling of the \citet{Kipping2018} single transit probability prior. As for the transit velocity, much of the remaining posterior uncertainty on the orbital period originates from the uncertain impact parameter. From the orbital period of \thisstarb{} we derive a semi-major axis $a=0.88^{+0.32}_{-0.10}$~AU, and a relative semi-major axis $a/R_*=270^{+93}_{-37}$. Compared to the solar system planets, these orbital properties are most similar to Earth.

The observed transit depth of \thisstarb{} is only $225\pm10$~ppm, resulting in the aforementioned planet-to-star radius ratio of $0.01368^{+0.0006}_{-0.0004}$. Given the stellar radius of $0.707\pm0.023~R_\odot$, we therefore derive a planetary radius of $1.06^{+0.06}_{-0.05}~R_\oplus$. \thisstarb{} is therefore only marginally larger than Earth. To our knowledge, this is the smallest high-confidence planet candidate reported from a single transit around a Sun-like star; among all transiting exoplanets it may be exceeded only by TRAPPIST-1~h, originally discovered from a single transit by \textit{Spitzer} and with orbital period resolved by K2 \citep{Gillon2017, Luger2017}.

We may estimate the amount of stellar flux received by \thisstarb{} relative to Earth using the following equation:

\begin{equation}
    I=\frac{L_*}{a}
    \;,
\end{equation}

\begin{figure*}[ht]
    \centering
    \includegraphics[width=\textwidth]{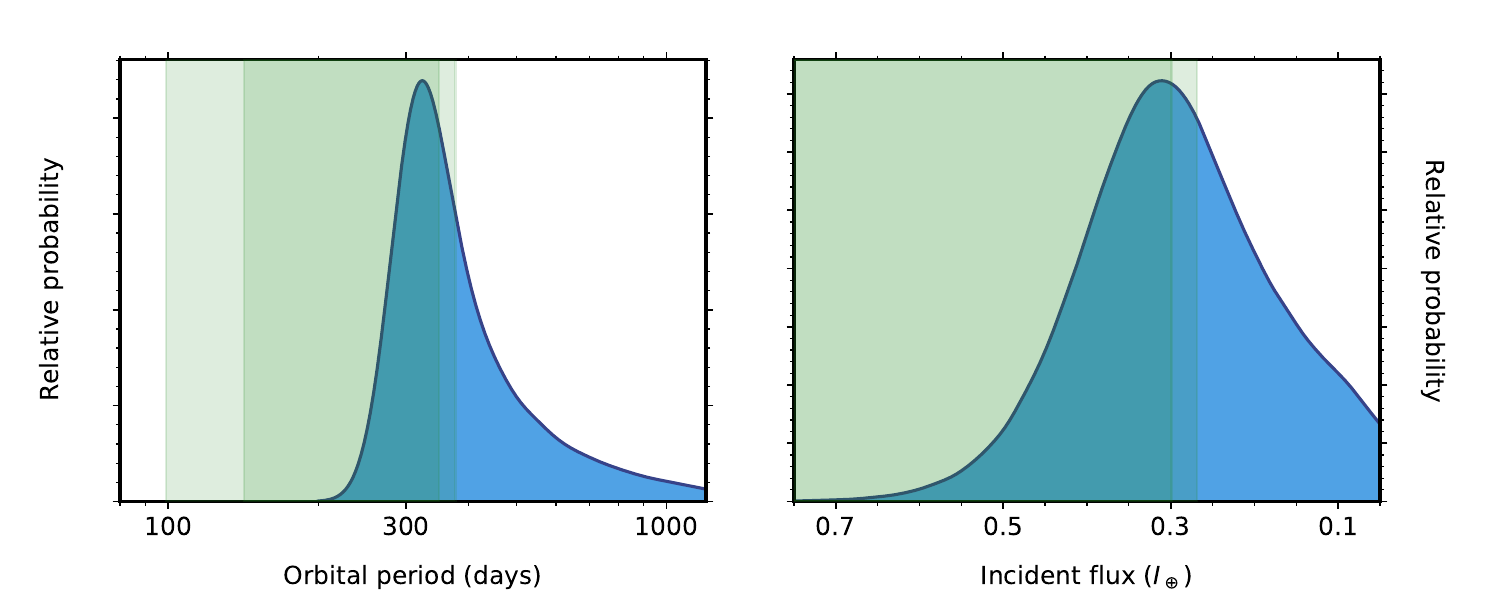}
    \caption{Selected confidence intervals derived from our model for the single transit of \thisstarb{}. (\textit{Left}) Orbital period ($355^{+200}_{-59}$~days). (\textit{Right}) Stellar flux incident on the planet relative to the Earth equivalent ($0.29^{+0.11}_{-0.13}~I_\oplus$). We overplot the conservative  (dark green) and optimistic (pale green) habitable zone limits calculated following \citet{Kopparapu2013}. We find that 40\% and 51\% of our models include \thisstarb{} in the conservative and optimistic habitable zones respectively.
    }
    \label{fig:posteriors}
\end{figure*}

Where $L_*$ is the stellar luminosity relative to the Sun and $a$ is the planetary semi-major axis in AU. Given $L_*=0.232^{+0.023}_{-0.021}~L_\odot$ (Table~\ref{tab:star}) and $a=0.88^{+0.32}_{-0.10}$~AU, we estimate that \thisstarb{} receives an incident flux of $I=0.29^{+0.11}_{-0.13}$ times that which Earth receives from the Sun. Among the major solar system bodies this is most comparable to Mars, which receives $I=0.43~I_\oplus$. We plot the posterior distributions in orbital period and incident flux in Figure~\ref{fig:posteriors}. While \thisstarb{} appears to have a similar orbital period and radius to Earth, it is likely to be significantly cooler on account of its less luminous K-type host star. We discuss the implications of this in the following sections.

\section{Discussion} \label{sec:discussion}

\subsection{System Properties}

In this work, we have reported the discovery of \thisstarb{}, an Earth-sized candidate planet detected in K2 transit photometry around a $V=10$ K-dwarf star. Through detailed analysis of available observations of \thisstar{}, we have excluded all conventional hypotheses for the transit event other than a transiting exoplanet; however, confirmation of the planetary hypothesis will require observation of a second transit or another form of additional detection. While we therefore present \thisstarb{} as a planet \textit{candidate}, we anticipate there is a high likelihood that it is a genuine planet and we analyse it here as if it were confirmed as such.

We have performed a fit to the single transit of \thisstarb{} assuming a zero-eccentricity planetary transit model. We estimate a radius of $1.06^{+0.06}_{-0.05}~R_\oplus$ and an orbital period of $355^{+200}_{-59}$~days. To our knowledge this is the smallest planet candidate detected from a single transit around a Sun-like star, and is particularly remarkable for its Earth-like radius and orbital properties. We discuss the temperature and potential habitability of \thisstarb{} further in the following section (Section~\ref{subsec:HZ}).

The discovery of \thisstarb{} demonstrates the detectability of temperate and cool Earth-sized exoplanets orbiting Sun-like stars through single transits. However this has only been made possible thanks to an especially favourable set of circumstances, namely the exceptionally high precision of the K2 photometry (CDPP~= $\sim$8.5~ppm, Section~\ref{subsec:photometry}), and further facilitated by the sub-solar radius of \thisstar{}. Whereas \thisstarb{} has a transit depth of $225\pm10$~ppm, the transit of Earth across the Sun would have a depth of only $\approx$100~ppm; hence, while the transit of \thisstarb{} has been detected with a white-noise SNR~= 30, an equivalent Earth transit would have less than half the SNR ($\approx$14). Given that only a small fraction of solar-type stars observed by \textit{Kepler} and K2 have such precise photometry, it is perhaps unsurprising that no exoplanet similar to \thisstarb{} has previously been reported from a single transit.

The \thisstar{} system is notable in that it contains a long-period transiting terrestrial planet candidate without showing any evidence for additional terrestrial planets. We can largely exclude transiting planets larger than $\geq$1~$R_\oplus$ with orbital periods below $P<30$~days (Section~\ref{subsec:companions_transit}), which contrasts with known transiting exoplanet systems that contain multiple small planets extending from short periods out to the habitable zone \citep[e.g. Kepler-62, Kepler-186;][]{Borucki2013, Quintana2014}. However, this does not necessarily mean that \thisstarb{} is isolated. Due to the finite duration of K2 observations, the detectability of transiting planets declines towards longer periods, and for beyond the 88 day baseline of the K2 photometry it would be entirely possible for additionally planets to simply not be observed in transit.\footnote{Given an orbital period of $355^{+200}_{-59}$~days, there is only a $25^{+5}_{-8}$\% probability that a transit of \thisstarb{} occurred during the 88 days of K2 Campaign 15 observations.} There is therefore considerable space for additional planets of similar size interior or exterior to the orbit of \thisstarb{} that could have gone undetected by K2, and we cannot assume that \thisstarb{} is dynamically isolated. For example, it is entirely possible that a solar system-like architecture, with multiple terrestrial planets in the region between $0.2-2$~AU, may simply have failed to produce more than one detectable transit during K2 observations. Alternatively, inter-planet mutual inclinations larger than a few tenths of a degree could result in hypothetical other planets in the system to fail to transit altogether. Thus, while \thisstarb{} currently appears to be isolated, it cannot be ruled out that there are additional terrestrial planets in the system which may be detected through further observations.

In Section~\ref{subsec:companions_wide}, we analysed archival HARPS RVs and \textit{Hipparcos-Gaia} astrometry to constrain the existence of additional companions in the \thisstar{} system. The most salient evidence for a third body is the $\sim$7~\ms{} drift in the HARPS RVs, which is nominally $>$4$\sigma$ significant and does not appear to be explained by stellar activity. Due to the limitations of the data it is not possible to confidently determine the origin of this variation, but assuming it does originate from a third body, we may estimate a provisional companion mass as a function of projected separation, which we have shown in Figure~\ref{fig:acceleration}. Though we can largely rule out the role of a stellar companion (at least above $\gtrsim$0.2~$M_\odot$) from our imaging and \textit{Gaia} constraints, the acceleration could equally be reproduced by anything between a Sub-Jovian planet within a few astronomical units to a brown dwarf at $\approx$50~AU. However, given that brown dwarfs are rarer than giant planets around Sun-like stars \citep[e.g.][]{Grether2006}, it appears most plausible \textit{a posteriori} that the RV variation could originate from a giant planet with an orbit exterior to \thisstarb{}. We note that at a projected separation of $\approx$5~AU the trend would be consistent with an $\approx$1~$M_J$ planet, properties comparable to Jupiter. While the available data is insufficient to draw definite conclusions on the existence and parameters of a third body, we reason that the existence of an additional planet orbiting \thisstar{} is a plausible explanation for the RV variability and suggest that further RV observations may confirm this hypothesis.

If \thisstar{} does additionally host an additional long-period companion, it would be significant for understanding the system context of \thisstarb{}. Any third body that could reproduce the observed RV variability is likely to be more massive than the transiting planet candidate by multiple orders of magnitude, meaning that it would dominate the angular momentum budget of the planetary system. Dependent on its hypothetical orbital properties, a third body may have significant implications on the formation history of \thisstarb{}; previous studies have suggested that there is a correlation between the incidence of short-period Super-Earths and long-period giant planets \citep[][]{Zhu2018, Bryan2019}, suggesting that the existence of giant planets significantly influences the formation of planets on interior orbits. Constraining the properties of the potential long-period companion of \thisstar{} may therefore be an important avenue for further study of the system.

Finally, we note that transits of Earth would not be observed from the perspective of \thisstar{}, as its ecliptic latitude of $-1.06\degree$, places it outside of the Earth Transit Zone which is only 0.5$\degree$ wide \citep{Heller2016}.

\subsection{Potential Habitable Zone Orbit} \label{subsec:HZ}

At the conclusion of Section~\ref{subsec:transit_model}, we used our constraints on the planetary orbit and the stellar luminosity to estimate the relative amount of stellar flux incident on \thisstarb{}. We found that this was equal to $I=0.29^{+0.11}_{-0.13}~I_\oplus$, which is significantly lower than the solar flux experienced by Earth and likely inferior to that of Mars ($0.43~I_\oplus$). We may take this further by estimating the planetary equilibrium temperature through the following equation:

\begin{equation}
    T_{\text{eq}}=T_{\text{eff},*}\sqrt{\frac{R_*}{2a}}(1-\alpha)^{1/4}
    \;,
\end{equation}

where $T_{\text{eff}}$ is the stellar effective temperature, $R_*$ is the stellar radius, $a$ is again the planetary semi-major axis in AU, and $\alpha$ is the bond albedo of the planet. Though the bond albedo is not known, we may vary through physically plausible values for this parameter in order to get a sense for the equilibrium temperature of \thisstarb{}. Assuming a minimal bond albedo of $\alpha=0.0$, the equilibrium temperature is $T_{\text{eq}}=205^{+17}_{-28}$~K ($-68$~$\degree$C). Alternatively, assuming an ``Earth-like" $\alpha=0.3$, we find $T_{\text{eq}}=188^{+16}_{-25}$~K ($-85$~$\degree$C).

It therefore appears likely that \thisstarb{} is among the coolest Earth-sized transiting planets yet discovered orbiting a Sun-like star, and we are motivated to consider whether \thisstarb{} lies in the habitable zone. Though the equilibrium temperature estimates above lie substantially below the 273~K freezing point of water, liquid water may still be facilitated given ideal atmospheric conditions. We follow the \citet{Kopparapu2013} model to define the circumstellar habitable zone. Adopting $T_{\text{eff}}=4770$~K and $L_*=0.232^{+0.023}_{-0.021}~L_\odot$ for \thisstar{}, we find that the the conservative bounds of the HZ are $[0.48, 0.88]$~AU ($[1.00,0.30]~I_\oplus$), whereas the optimistic limits are $[0.38, 0.93]$~AU ($[1.60,0.27]~I_\oplus$). Given the mass of \thisstar{}, this converts to conservative and optimistic limits in orbital period of $[140, 350]$~days and $[100, 380]$~days respectively. We plot the \citet{Kopparapu2013} habitable zone limits over the orbital period and incident flux posterior distributions for \thisstarb{} in Figure~\ref{fig:posteriors}. \thisstarb{} lies within the the conservative and optimistic HZ limits in 40\% and 51\% of our transit models respectively. All of the remaining posteriors lie \textit{beyond} the habitable zone limits, so there is a $\approx$50\% probability that \thisstarb{} overshoots the \citet{Kopparapu2013} HZ.

The properties of \thisstarb{} invite comparison to Kepler-186~f, the first-known Earth-sized exoplanet orbiting in the habitable zone \citep{Quintana2014}. Kepler-186~f has a radius of $1.11\pm0.14~R_\oplus$ and receives a low incident flux of $0.320^{+0.059}_{-0.039}~I_\oplus$ \citep{Quintana2014}. Though Kepler-186 is less massive than \thisstar{} ($0.48~M_\odot$ vs. $0.73~M_\odot$), Kepler-186~f nonetheless constitutes one of the closest analogues to \thisstarb{} among known exoplanets. The potential habitability of Kepler-186~f was studied by \citet{Bolmont2014}, who explored the surface temperatures of model planets with atmospheres composed of CO$_2$, N$_2$, and H$_2$O. Varying the CO$_2$ partial pressure under different planetary properties and N$_2$ abundances, they found that the surface temperature consistently rises above the 273~K freezing point of H$_2$O given $200-500$~mbar of CO$_2$. Assuming that broadly similar outcomes would apply for \thisstarb{}, it appears eminently plausible that a moderately CO$_2$-rich atmosphere would be conducive to liquid surface water.

However, a plausible alternative scenario is that the surface of \thisstarb{} is frozen (a ``snowball" climate). For a model Earth-like planet with a fixed modern CO$_2$ abundance, the results of \citet{Wolf2017} imply that for an early K-type host star snowball climate states develop where $I\lesssim0.8~I_\oplus$, which includes our entire posterior distribution. A fully glaciated planet would be highly reflective, significantly reducing the surface temperature. The later spectral type of \thisstar{} has only a moderate effect on the high reflectivity of surface ice; according to \citet{Wilhelm2022}, ice on an Earth-like planet orbiting an early K-dwarf has an albedo of 0.55, compared to 0.6 for a G-dwarf. Assuming a high bond albedo of $\alpha=0.5$ from the Huronian snowball Earth model of \citet{DelGenio2019} to represent a fully glaciated planet, we derive $T_{\text{eq}}=173^{+14}_{-23}$~K ($-100$~$\degree$C). The prospects for surface water and habitability are therefore highly dependent on atmospheric composition, particularly the relative abundance of CO$_2$.

\subsection{HD 137010 b in Context}

\begin{figure*}[ht]
    \centering
    \includegraphics[width=\textwidth]{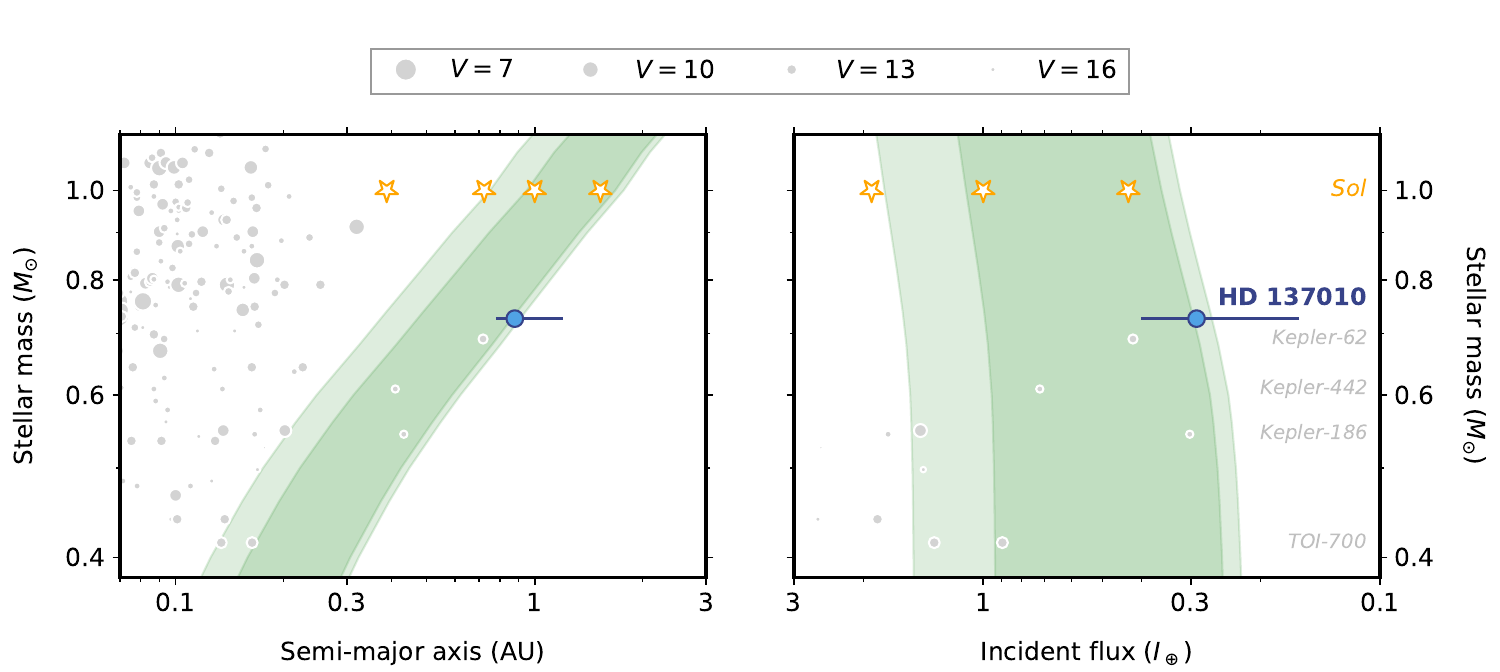}
    \caption{\thisstarb{} in context. We plot known transiting planets smaller than $<$1.6~$R_\oplus$ orbiting $0.4-1.2~M_\odot$ stars with grey points in semi-major axis (left) and incident flux (right) space, with host star masses on the y-axis and point sizes scaled to $V$-band magnitude. We additionally highlight the solar system terrestrial planets using orange stars and mark the \citet{Kopparapu2013} habitable zone limits in green. Systems mentioned in the text are labelled on right. \thisstarb{} is an exceptional example of an Earth-sized planet orbiting around the outer edge of the habitable zone of a bright Sun-like star.}
    \label{fig:population}
\end{figure*}

We place \thisstarb{} in the context of known terrestrial planets in Figure~\ref{fig:population}. Here we compare \thisstarb{} to the solar system terrestrial planets and to known transiting exoplanets with $R_p<1.6~R_\oplus$. The radius cut is intended to limit our sample to terrestrial exoplanets, as it is now conventional wisdom that planets larger than $>1.6$~$R_\oplus$ are unlikely to be rocky \citep{Rogers2015}.\footnote{We note that this radius cut leads to the marginal exclusion of the well-known HZ planets Kepler~62~e \citep[$1.61\pm0.05~R_\oplus$;][]{Borucki2013} and Kepler-452~b \citep[$1.63^{+0.23}_{-0.20}~R_\oplus$;][]{Jenkins2015}.} We plot known planets in both semi-major axis and incident flux space as a function of stellar mass between $0.4-1.2~M_\odot$. The aim of this stellar mass restriction is to approximately limit the sample to FGK-type stars, though we have extended the selection below $\leq$0.6~$M_\odot$ into the early M-dwarf range to enable comparison with the notable HZ planets TOI-700~d and e \citep[M2V, $M_*=0.42~M_\odot$;][]{Gilbert2020, Gilbert2023}. We also plot the \citet{Kopparapu2013} HZ limits over both axes.

Figure~\ref{fig:population} demonstrates that the region of parameter space occupied by \thisstarb{} is otherwise largely barren. The most similar known exoplanet is Kepler-62~f which has $P=267.3$~d, $a=0.72$~AU, $I=0.41\pm0.05~I_\oplus$ and $R_p=1.41\pm0.07~R_\oplus$ and orbits a K2V star \citep{Borucki2013}. Other transiting planets with comparable parameters include Kepler-442~b \citep{Torres2015} and Kepler-186~f \citep{Quintana2014}, both orbiting lower mass stars. \thisstarb{} is therefore a significant addition to the small sample of cool terrestrial planet candidates orbiting K-dwarf stars.

Compared to the other terrestrial habitable zone transiting planet candidates orbiting stars more massive than $\geq$0.6~$M_\odot$ in Figure~\ref{fig:population}, \thisstarb{} can be distinguished from those previously known by its bright host star ($V=10$). The aforementioned \textit{Kepler} planets all orbit stars fainter than $V\geq14$, sufficiently faint to preclude substantial characterisation efforts with current instruments. Previous discoveries of terrestrial planets in the HZ of bright stars have been restricted to M-dwarfs; this has undoubtedly been facilitated by detectability biases, as HZ planets around M-dwarfs have shorter orbital periods and deeper transit signals. However, there are well-known concerns on the ability of terrestrial planets orbiting M-dwarfs to retain their atmospheres in the face of long-term exposure to high-energy radiation from their host stars \citep[e.g.][]{Zahnle2017, Pass2025}. Due to their shorter active lifetimes and the larger orbital separations of their habitable zones,  there is a case to be made for the favourability of K-dwarfs to planetary habitability over M-dwarfs. Indeed, the habitability potential of planets orbiting early-to-mid K-dwarfs was highlighted as early as the foundational work of \citet{Kasting1993}, and has been promoted both in the theoretical aspect \citep[e.g.][]{Cuntz2016} and also from the perspective of biosignature detectability \citep[e.g.][]{Arney2019}. \thisstarb{} is the first example of a transiting terrestrial planet candidate orbiting in or near to the HZ of a K-type dwarf bright enough to allow for substantial follow-up observations, and may be of interest for future work on the habitability of K-dwarf planets.

\subsection{Prospects for Confirmation}

Though the single transit observed by K2 provides evidence for the existence of \thisstarb{}, confirmation and further study both critically depend on redetection of the planetary signal. Here we explore the prospects for confirmation of \thisstarb{} through various methods.

\subsubsection{Transit}

As \thisstarb{} has already been detected in transit, this provides the most obvious avenue for redetection. However, the shallow transit depth means that such an observation would be beyond the capability of many instruments.

At present, the TESS mission \citep{TESS} presents the best prospects for re-detection of the transit of \thisstarb{}. However, TESS employs an observing strategy which generally avoids the ecliptic plane \citep{TESS}. This minimises the impact of scattered light from solar system bodies, but naturally also limits overlap with K2 targets. Nonetheless, in past regions of overlap, TESS reobservation of K2 targets have proved highly informative \citep[e.g.][]{Thygesen2023, Thygesen2024}. Forays into the ecliptic have been made during the TESS extended mission, but until recently these have been restricted to the northern ecliptic.

However TESS has recently executed its first sweep of the southern ecliptic, and observed \thisstar{} for the first time in Sector~91 (April-May 2025). Extrapolating forwards from our orbital period posteriors, we estimated a $\sim$7\% chance that a transit of \thisstarb{} would occur during Sector~91. We proposed for TESS 20-second cadence observations for \thisstar{} to maximise the photometric SNR. The TESS photometry has a CDPP of $\sim$42~ppm (compared to $\sim$8.5~ppm for K2), which makes detection of \thisstarb{} challenging the expected transit SNR would be $\approx$7. We do not observe any visually obvious transits in the TESS light curve that could be attributed to the K2 planet candidate, but it is not impossible that such a transit could evade detection given the photometric precision.

In the event that a second transit of \thisstarb{} is detected, this would only constrain a \textit{maximum} value of the orbital period and would leave the true orbital period as an undetermined alias of this value \citep[see][]{Becker2019, Dholakia2020}. In this scenario, a possible avenue for resolving the true period would be to use CHEOPS \citep{CHEOPS}. CHEOPS observations have been productively used to resolve period aliases of several planets originally detected by TESS \citep[e.g.][]{Osborn2022, Osborn2023}. Though the comparatively low transit incidence means that such an effort would be time-intensive, this may be the most viable path to confirm \thisstarb{} and determine its orbital period.

A near-future mission of relevance is PLATO \citep{PLATO2014, PLATO2024}. PLATO will search for Earth-sized transiting planets orbiting in the HZ using a long-duration ``staring" pointing strategy reminiscent of \textit{Kepler}. The properties of \thisstarb{} align closely with planets expected to be detected \citep{Heller2022}. However, the planned locations of the PLATO long-pointing fields lie near to the ecliptic poles, and do not overlap with the ecliptic locations of the K2 fields \citep{Nascimbeni2022}. It therefore appears unlikely that PLATO could be used to specifically redetect the transit of \thisstarb{}.

\vspace{-2mm}
\subsubsection{Radial Velocity} \label{subsec:RV}

The radial velocity method is one of the main methods used for exoplanet discovery and mass measurement, and RV detection of exoplanets with Earth-like masses and periods has been a long-time goal in the field. We therefore consider the prospects for RV detection of \thisstarb{}. If we assume that a $1.06^{+0.06}_{-0.05}~R_\oplus$ planet has the same density as Earth, we would expect it to have a mass of $1.20^{+0.21}_{-0.15}~M_\oplus$. Using our adopted stellar mass and orbital period and assuming zero orbital eccentricity, we calculate that the expected radial velocity semi-amplitude for \thisstarb{} given this mass would be $0.13\pm0.02$~\ms{}, which is comparable to the reflex signal of Earth. There are considerable challenges in RV planet detection for semi-amplitudes much below the $<$1~\ms{} level due to instrument limitations and stellar effects \citep[ex.][]{Hara2023}, and the expected 13~c\ms{} semi-amplitude of \thisstarb{} lies considerably below the smallest planetary RV signals detected so far.

We may establish a sense for the challenge of detecting \thisstarb{} through comparison to past RV targets with similar observational properties. Here we compare to previous RV observations of the star GJ~9827, which like \thisstar{} is a 10$^{\text{th}}$ magnitude K-dwarf. GJ~9827 has a system of three short-period planets discovered by K2 \citep{Niraula2017, Rodriguez2018} and has gained a substantial body of RV observations used to measure the masses of these planets, each of which have RV semi-amplitudes of a few \ms{}. The most recent RV results from \citet{Passegger2024}, which included 54 RVs from the ESPRESSO spectrograph and a total of 167 RVs from preceding publications, achieved posterior \textit{uncertainties} on the RV semi-amplitudes of $\pm\sim$0.25~\ms{} for the three planets. Detection of the RV signal expected from \thisstarb{} is therefore likely to entail a larger number of observations with higher precision, which would be challenging for current instruments.

Nonetheless, there is substantial interest in working towards the detection of the small RV signals of Earth analogues \citep[see in particular][]{EPRVWG}. We therefore highlight \thisstar{} as a target of interest for current and future extreme-precision RV surveys. \thisstar{} is the first Sun-like star with an Earth-sized habitable zone planet candidate that is also sufficiently bright for precise RV observations. It presents an advantage over blind selection of RV targets in that there is already evidence for a cool Earth-sized planet orbiting the star. Given the location of the star in the ecliptic (declination = ~-20), it is accessible from most observatories with high-precision RV spectrographs. We also note that there is evidence for an additional companion in existing HARPS RV observations of \thisstar{} with a longer orbital period (Section~\ref{subsec:companions_wide}), which may be more easily detected with current instruments.


\subsubsection{Astrometry}

Astrometry has been raised as a potential method for the detection of Earth analogues. However, this largely lies beyond the capabilities of existing instruments. In the case of \thisstarb{}, assuming a mass of $1.20^{+0.21}_{-0.15}~M_\oplus$ as in the previous section we would anticipate an astrometric reflex signal with a semi-amplitude of $\sim$0.1~$\mu$as. This is approximately two orders of magnitude below the astrometric precision of \textit{Gaia}, which is around 10~$\mu$as \citep{Gaia}. The Nancy Grace Roman Space Telescope may achieve an astrometric precision of 1-10~$\mu$as \citep{WFIRSTastrometry}, which is still significantly larger than the expected signal.

Various space mission concepts for the astrometric detection of exoplanets have been proposed over the years \citep[see][]{Janson2018}. Recent mission proposals aiming for the detection of habitable zone exoplanets using sub-$\mu$as precision astrometry include \textit{Theia} \citep{Theia} and CHES \citep{CHES}. \thisstarb{} may be a target of interest for astrometric exoplanet detection efforts.

\subsubsection{Imaging}

In recent years, there has been significant new exploration of the near-future potential for the detection of Earth analogues with direct imaging. A highlight of this is the NASA Habitable Worlds Observatory (HWO) mission recommended in the Astro2020 decadal survey \citep{Astro2020}, which aims for imaging of Earth-like planets orbiting nearby stars using a space-based coronagraphic instrument.

\thisstarb{} lies closer to the parameter space intended to be explored by HWO than almost any known transiting habitable zone exoplanet. With an estimated physical semi-major axis of $0.88^{+0.32}_{-0.10}$~AU at a distance of $44.86\pm0.03$~pc, we estimate that \thisstarb{} has a projected semi-major axis of $20^{+7}_{-2}$~mas.\footnote{We note that since it is transiting and the orbital inclination is near to 90$\degree$, the semi-major axis represents the maximum value for the projected separation, and \thisstarb{} would necessarily spend a significant fraction of its orbit at smaller separations.} However, this may still be too small of a separation to facilitate direct detection. The requirements for stellar target selection for HWO has been explored by \citet{Mamajek2024}, and while the inner working angle (IWA) of the instrument has not yet been defined, this projected separation is around a factor of $3-4$ smaller than the ``Tier C" target IWA constraint \citep[approximately 70~mas;][figure~1]{Mamajek2024}. This means that, with due allowance for the yet-to-be finalised instrument design, it appears likely that \thisstarb{} will be too close to its host star to be detected with HWO.

A second mission concept aiming for the direct detection of Earth-like exoplanets is the Large Interferometer For Exoplanets \citep[LIFE;][]{Quanz2022}. While likewise optimised for the detection planets around nearby stars (\citealt{Quanz2022} consider stars within $<$20~pc), the nulling interferometer design means that it would not have a strict IWA in contrast to a coronagraphic instrument. Depending on the planetary properties, LIFE may therefore be capable of imaging \thisstarb{}; this is a potential avenue for exploration in future work.

\section{Conclusions} \label{sec:conclusions}

In this work, we have presented the discovery of \thisstarb{}, an Earth-sized transiting planet candidate detected from a single transit observed in K2 Campaign~15 on the $V=10.1$ K3.5V star \thisstar{}. A comprehensive analysis of the K2 observations, historical low-resolution imaging and new high-resolution speckle imaging data, archival HARPS RVs, and \textit{Hipparcos-Gaia} astrometry allow us to exclude the conventional false-positive hypotheses for the transit signal, leaving a transiting exoplanet as the most plausible explanation for the photometric event. However, since we only have the evidence of one transit event, we ultimately classify \thisstarb{} as a candidate planet.

We have implemented a single-transit exoplanet model to fit the transit of \thisstarb{}. We determine that it has a radius of $1.06^{+0.06}_{-0.05}~R_\oplus$, and assuming negligible orbital eccentricity, an orbital period of $355^{+200}_{-59}$~days (semi-major axis of $0.88^{+0.32}_{-0.10}$~AU). While these properties are remarkably similar to Earth, due to the sub-solar luminosity of \thisstar{}, we estimate that the candidate planet receives an incident flux of $I=0.29^{+0.11}_{-0.13}~I_\oplus$ which may be less than the flux received by Mars. Following the \citet{Kopparapu2013} definitions for the conservative and optimistic habitable zone, we find that there is a 40\% probability that \thisstarb{} lies within the conservative limits and a 51\% probability that it falls within the optimistic habitable zone; all remaining solutions place it beyond the outer limits. If it does fall within the habitable zone, a moderately CO$_2$-rich atmospheric composition may allow for the formation of liquid surface water, whereas a CO$_2$ abundance more similar to Earth might instead result in a frozen ``snowball" climate.

\thisstarb{} is essentially unique as a candidate terrestrial HZ planet transiting a bright Sun-like star. We encourage follow-up observations for the confirmation and characterisation of \thisstarb{}, and to better understand the architecture of the system.

\section{Acknowledgments}

We acknowledge and pay respect to Australia’s Aboriginal and Torres Strait Islander peoples, who are the traditional custodians of the lands, waterways, and skies all across Australia. We thank the anonymous referee for their helpful comments which have helped to improve this work.
A.Ve. thanks Mary Anne Limbach for thought-provoking discussion (and earnest enthusiasm) which helped to improve this work.
A.Ve. and C.X.H. are supported by ARC DECRA project DE200101840.

We acknowledge all efforts which went into the Planet Hunters citizen science project throughout the \textit{Kepler} and K2 missions, in the absence of which this work would not have been possible.

This research is based on observations collected at the European Southern Observatory under ESO programmes 072.C-0488(E) and 085.C-0019(A). This research has made use of the SIMBAD database and VizieR catalogue access tool, operated at CDS, Strasbourg, France. This research has made use of NASA's Astrophysics Data System.
This paper includes data collected by the \textit{Kepler} mission. Funding for the \textit{Kepler} mission is provided by the NASA Science Mission directorate. Some of the data presented in this paper were obtained from the Mikulski Archive for Space Telescopes (MAST). K2 data used in this paper can be found at \dataset[10.17909/t9-t4y8-eh81]{http://dx.doi.org/10.17909/t9-t4y8-eh81}. STScI is operated by the Association of Universities for Research in Astronomy, Inc., under NASA contract NAS5–26555. Support for MAST for non-HST data is provided by the NASA Office of Space Science via grant NNX13AC07G and by other grants and contracts.


%

\facilities{\textit{Kepler} (K2), TESS, \textit{Hipparcos}, \textit{Gaia}, Gemini~South (Zorro), ESO La~Silla 3.6m (HARPS).}


\software{\texttt{emcee} \citep{emcee},
          \texttt{batman} \citep{batman},     
          \texttt{Vartools} \citep{vartools},   \texttt{LcTools} \citep{Schmitt2019},
          \texttt{minimint} \citep{Koposov2021},
          \texttt{kiauhoku} \citep{Tayar2022}
          }




\appendix

\vspace{-5mm}
\section{Data} \label{appendix:data}

In this section we present the data used in this work.

In Table~\ref{tab:photometry_short} we present the K2 photometry used in this work. See Section~\ref{subsec:photometry} for further information.

\begin{deluxetable}{ccc}[h!]
\label{tab:photometry_short}
\centering
\tablecaption{K2 photometry of \thisstar{}.}
\tablehead{\colhead{BJD - 2454833} & \colhead{Normalised flux} & \colhead{De-trended flux}}
\startdata
3157.8752 & 0.99945683 & 1.0000828  \\
3157.8956 & 0.99943504 & 1.0000273  \\
3157.9160 & 0.99941655 & 0.99997813 \\
3157.9365 & 0.99945957 & 0.99999344 \\
3157.9569 & 0.99948045 & 0.99998964 \\
... & ... & ... \\
3244.3222 & 0.99972678 & 1.0000044  \\
3244.3426 & 0.99972075 & 0.99999964 \\
3244.3630 & 0.99970085 & 0.99998146 \\
3244.3835 & 0.99971981 & 1.0000025  \\
3244.4243 & 0.99971669 & 1.0000050  \\
\enddata
\end{deluxetable}
\vspace{-15mm}
\tablecomments{Table~\ref{tab:photometry_short} is published in its entirety in the machine-readable format. A portion is shown here for guidance regarding its form and content.}

\vspace{4mm}
In Table~\ref{tab:RV} we tabulate the HARPS observations of \thisstar{} extracted from the ESO archive. We have converted all units from k\ms{} to \ms{}, but have not otherwise altered the data.

\begin{deluxetable}{cccc}[h!]
\label{tab:RV}
\centering
\tablecaption{HARPS radial velocity data for \thisstar{}.}
\tablehead{\colhead{BJD} & \colhead{RV (\ms{})} & \colhead{RV error (\ms{})} & \colhead{Bisector span (\ms{})}}
\startdata
2453834.84478 & 27931.19 & 1.34 & 6.20  \\
2453835.79152 & 27933.31 & 1.22 & 13.21 \\
2453979.53447 & 27931.19 & 1.99 & 19.52 \\
2453982.49629 & 27930.73 & 1.39 & 6.77  \\
2455268.89844 & 27937.27 & 1.13 & 18.35 \\
2455298.76206 & 27936.37 & 1.74 & 13.48 \\
2455306.77737 & 27938.23 & 1.36 & 10.51 \\
2455307.78489 & 27940.56 & 1.10 & 18.90 \\
\enddata
\end{deluxetable}
\vspace{-12mm}

In Table~\ref{tab:astrometry} we present the \textit{Hipparcos-Gaia} astrometry for \thisstar{}, as extracted from \textit{Gaia}~EDR3 version of the \textit{Hipparcos-Gaia} Catalog of Accelerations \citep{Brandt2021}. We refer the reader to \citet{Brandt2018, Brandt2021} and further to \citet{Brandt2019} for a full description of this data. We note for clarity that the astrometric solution for \thisstar{} is unchanged between \textit{Gaia}~EDR3 and \textit{Gaia}~DR3.

\vspace{10mm}

\begin{deluxetable}{lrr}[h!]
\label{tab:astrometry}
\centering
\tablecaption{Proper motion data for \thisstar{} extracted from the \textit{Hipparcos-Gaia} Catalog of Accelerations.}
\tablehead{\colhead{Measurement} & \colhead{$\mu_\alpha$ (\masyr{})} & \colhead{$\mu_\delta$ (\masyr{})}}
\startdata
\textit{Hipparcos}      & $229.991\pm2.082$ & $-248.313\pm1.878$ \\
\textit{Gaia}~EDR3      & $228.536\pm0.028$ & $-248.158\pm0.019$ \\
\textit{Hipparcos-Gaia} & $228.533\pm0.068$ & $-248.183\pm0.050$ \\
\enddata
\end{deluxetable}
\vspace{-10mm}

\begin{figure}[h!]
    \centering
    \includegraphics[width=0.5\linewidth]{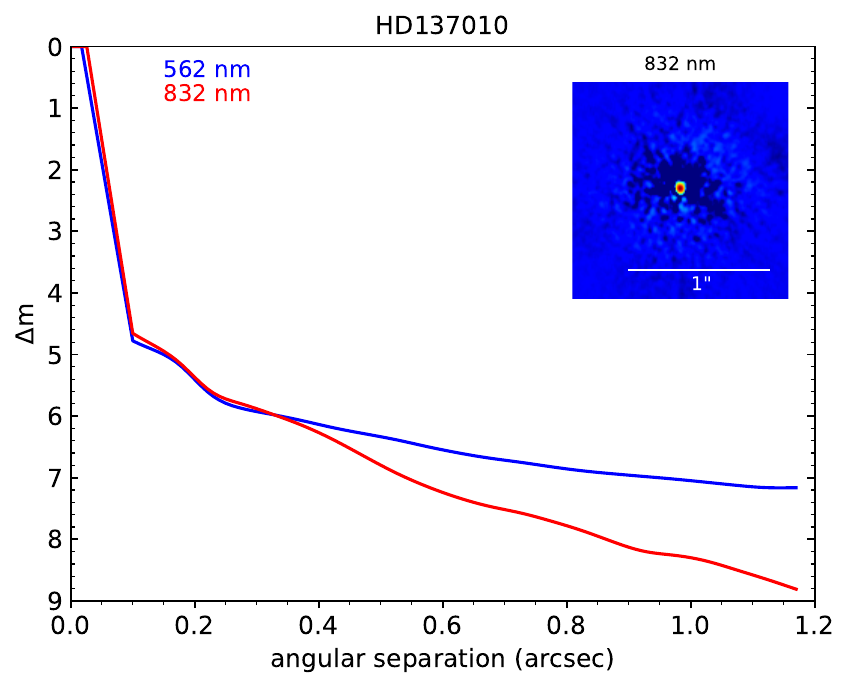}
    \caption{Zorro speckle imaging observations of \thisstar{} obtained on 2024-07-12. The red and blue contrast curves represent the 5$\sigma$ contrast limits as a function of projected separation at 832~nm and 562~nm respectively. The inset figure shows the reconstructed 832~nm image. No additional sources are detected.}
    \label{fig:speckle}
\end{figure}

In Figure~\ref{fig:speckle} we present the Zorro speckle imaging observations of \thisstar{} discussed in Section~\ref{subsec:imaging}.  Observations were taken in the 562~nm and 832~nm filters on 2024-07-12. No additional sources were detected above the 5$\sigma$ contrast limits.

\section{Isochrone Model}  \label{appendix:isochrone}

\begin{deluxetable}{lrr}[ht!]
\label{tab:star_photometry}
\centering
\tablecaption{Photometry of \thisstar{} used by our isochrone model.}
\tablehead{\colhead{Band} & \colhead{Magnitude (mag)} & \colhead{Reference}}
\startdata
$B_\text{T}$    & $11.310\pm0.091$ & \citet{Tycho2} \\
$V_\text{T}$    & $10.245\pm0.053$ & \citet{Tycho2} \\
$G$             & $9.706\pm0.02$ & \citet{GaiaDR3} \\
$G_{\text{BP}}$ & $10.260\pm0.02$ & \citet{GaiaDR3} \\
$G_{\text{RP}}$ & $9.004\pm0.02$ & \citet{GaiaDR3} \\
$J$             & $8.162\pm0.018$ & \citet{2MASS} \\
$H$             & $7.648\pm0.031$ & \citet{2MASS} \\
$K_S$           & $7.526\pm0.020$ & \citet{2MASS} \\
$W1$            & $7.456\pm0.035$ & \citet{WISE} \\
$W2$            & $7.525\pm0.021$ & \citet{WISE} \\
$W3$            & $7.506\pm0.018$ & \citet{WISE} \\
$W4$            & $7.670\pm0.184$ & \citet{WISE} \\
\enddata
\end{deluxetable}
\vspace{-4mm}

Here we detail the data and priors used in our isochrone model. As input data, we have used photometry from Tycho-2 \citep{Tycho2}, \textit{Gaia}~DR3 \citep{GaiaDR3}, 2MASS \citep{2MASS}, and WISE \citep[specifically AllWISE;][]{WISE}. We list the photometry in Table~\ref{tab:star_photometry}. We have applied a floor of $\pm$0.02~mag on the \textit{Gaia} photometric precision.

We use \texttt{minimint} \citep{Koposov2021} for isochrone interpolation and \texttt{emcee} \citep{emcee} to sample the posterior space. Model variables include the stellar parallax, age, mass, $T_{\text{eff}}$, $\log g$, and [Fe/H]; all are fitted log-linearly except for the last two, which are already logarithmic. As priors, we utilise the \textit{Gaia}~DR3 parallax \citep[$\varpi=22.292\pm0.017$~mas;][]{GaiaDR3} and $T_{\text{eff}}=4797\pm121$~K, $\text{[Fe/H]}=-0.22\pm0.07$~dex as derived from the HARPS spectra \citep{Sousa2011}. Following Section~\ref{subsec:age}, we impose an upper bound on the stellar age of $<$10~Gyr, and utilise the age posterior from the \citet{Veyette2018} $UVW$-age relationship as a prior for the model. This helps to penalise implausibly high model ages ($>$10~Gyr), which otherwise have a modest impact on the mass and density estimation. Due to the close proximity of \thisstar{}, we assume zero interstellar extinction.

The posteriors from our isochrone model are recorded in Table~\ref{tab:star}. In comparison to the spectroscopic priors, the addition of photometry affords a slight improvement in the precision on the effective temperature ($T_{\text{eff}}=4770\pm90$~K), but unsurprisingly no change for the [Fe/H]. The age posterior also closely reproduces the \citet{Veyette2018} prior.

Our isochrone model natively returns a stellar mass of $M=0.726\pm0.012~M_\odot$. Reassuringly, this is in $\sim$1$\sigma$ agreement with the mass of $0.709\pm0.013~M_\odot$ previously reported by \citet{Sousa2011}. However, this neglects the contribution to the error budget arising from systematic differences between stellar models, which should not be ignored at this level of precision. We use \texttt{kiauhoku} \citep{Tayar2022} to quasi-empirically estimate the systematic uncertainty on the stellar mass through comparison of the outputs of different stellar models. \texttt{kiauhoku} returns a 1.7\% systematic uncertainty on the stellar mass ($\pm0.012~M_\odot$), which when added in quadrature results in a final value of $0.726\pm0.017~M_\odot$ for \thisstar{}. This is the posterior value recorded in Table~\ref{tab:star} and adopted throughout this work. We have also propagated the increased mass uncertainty through to the stellar density and surface gravity.


\bibliography{bib}

@preamble{"\newcommand{\noop}[1]{}"}

@ARTICLE{K2,
       author = {{Howell}, Steve B. and others},
        title = "{The K2 Mission: Characterization and Early Results}",
      journal = {\pasp},
         year = 2014,
        month = apr,
       volume = {126},
       number = {938},
        pages = {398},
          doi = {10.1086/676406},
archivePrefix = {arXiv},
       eprint = {1402.5163},
 primaryClass = {astro-ph.IM},
       adsurl = {https://ui.adsabs.harvard.edu/abs/2014PASP..126..398H}
}

@ARTICLE{K2SFF,
       author = {{Vanderburg}, Andrew and {Johnson}, John Asher},
        title = "{A Technique for Extracting Highly Precise Photometry for the Two-Wheeled Kepler Mission}",
      journal = {\pasp},
         year = 2014,
        month = oct,
       volume = {126},
       number = {944},
        pages = {948},
          doi = {10.1086/678764},
archivePrefix = {arXiv},
       eprint = {1408.3853},
 primaryClass = {astro-ph.IM},
       adsurl = {https://ui.adsabs.harvard.edu/abs/2014PASP..126..948V}
}

@ARTICLE{K2SFF_joint,
       author = {{Vanderburg}, Andrew and {Latham}, David W. and {Buchhave}, Lars A. and {Bieryla}, Allyson and {Berlind}, Perry and {Calkins}, Michael L. and {Esquerdo}, Gilbert A. and {Welsh}, Sophie and {Johnson}, John Asher},
        title = "{Planetary Candidates from the First Year of the K2 Mission}",
      journal = {\apjs},
         year = 2016,
        month = jan,
       volume = {222},
       number = {1},
          eid = {14},
        pages = {14},
          doi = {10.3847/0067-0049/222/1/14},
archivePrefix = {arXiv},
       eprint = {1511.07820},
 primaryClass = {astro-ph.EP},
       adsurl = {https://ui.adsabs.harvard.edu/abs/2016ApJS..222...14V}
}

@ARTICLE{Kepler,
       author = {{Borucki}, William J. and others},
        title = "{Kepler Planet-Detection Mission: Introduction and First Results}",
      journal = {Science},
         year = 2010,
        month = feb,
       volume = {327},
       number = {5968},
        pages = {977},
          doi = {10.1126/science.1185402},
       adsurl = {https://ui.adsabs.harvard.edu/abs/2010Sci...327..977B}
}

@ARTICLE{Koch2010,
       author = {{Koch}, David G. and others},
        title = "{Kepler Mission Design, Realized Photometric Performance, and Early Science}",
      journal = {\apjl},
         year = 2010,
        month = apr,
       volume = {713},
       number = {2},
        pages = {L79-L86},
          doi = {10.1088/2041-8205/713/2/L79},
archivePrefix = {arXiv},
       eprint = {1001.0268},
 primaryClass = {astro-ph.EP},
       adsurl = {https://ui.adsabs.harvard.edu/abs/2010ApJ...713L..79K}
}

@ARTICLE{TESS,
       author = {{Ricker}, George R. and others},
        title = "{Transiting Exoplanet Survey Satellite (TESS)}",
      journal = "SPIE",
    booktitle = {Space Telescopes and Instrumentation 2014: Optical, Infrared, and Millimeter Wave},
         year = 2014,
       editor = {{Oschmann}, Jacobus M., Jr. and {Clampin}, Mark and {Fazio}, Giovanni G. and {MacEwen}, Howard A.},
       series = {Society of Photo-Optical Instrumentation Engineers (SPIE) Conference Series},
       volume = {9143},
        month = aug,
          eid = {914320},
        pages = {914320},
          doi = {10.1117/12.2063489},
archivePrefix = {arXiv},
       eprint = {1406.0151},
 primaryClass = {astro-ph.EP},
       adsurl = {https://ui.adsabs.harvard.edu/abs/2014SPIE.9143E..20R}
}

@ARTICLE{Schmitt2019,
       author = {{Schmitt}, Allan R. and {Hartman}, Joel D. and {Kipping}, David M.},
        title = "{LcTools: A Windows-Based Software System for Finding and Recording Signals in Lightcurves from NASA Space Missions}",
      journal = {arXiv e-prints},
         year = 2019,
        month = oct,
          eid = {arXiv:1910.08034},
        pages = {arXiv:1910.08034},
          doi = {10.48550/arXiv.1910.08034},
archivePrefix = {arXiv},
       eprint = {1910.08034},
 primaryClass = {astro-ph.IM},
       adsurl = {https://ui.adsabs.harvard.edu/abs/2019arXiv191008034S}
}

@ARTICLE{VSG,
       author = {{Kristiansen}, Martti H.~K. and {Rappaport}, Saul A. and {Vanderburg}, Andrew M. and {Jacobs}, Thomas L. and {Martin Schwengeler}, Hans and {Gagliano}, Robert and {Terentev}, Ivan A. and {LaCourse}, Daryll M. and {Omohundro}, Mark R. and {Schmitt}, Allan R. and {Powell}, Brian P. and {Kostov}, Veselin B.},
        title = "{The Visual Survey Group: A Decade of Hunting Exoplanets and Unusual Stellar Events with Space-based Telescopes}",
      journal = {\pasp},
         year = 2022,
        month = jul,
       volume = {134},
       number = {1037},
          eid = {074401},
        pages = {074401},
          doi = {10.1088/1538-3873/ac6e06},
archivePrefix = {arXiv},
       eprint = {2205.07832},
 primaryClass = {astro-ph.EP},
       adsurl = {https://ui.adsabs.harvard.edu/abs/2022PASP..134g4401K}
}

@ARTICLE{Borucki2012,
       author = {{Borucki}, William J. and {Koch}, David G. and {Batalha}, Natalie and {Bryson}, Stephen T. and {Rowe}, Jason and {Fressin}, Francois and {Torres}, Guillermo and {Caldwell}, Douglas A. and {Christensen-Dalsgaard}, J{\o}rgen and {Cochran}, William D. and {DeVore}, Edna and {Gautier}, Thomas N. and {Geary}, John C. and {Gilliland}, Ronald and {Gould}, Alan and {Howell}, Steve B. and {Jenkins}, Jon M. and {Latham}, David W. and {Lissauer}, Jack J. and {Marcy}, Geoffrey W. and {Sasselov}, Dimitar and {Boss}, Alan and {Charbonneau}, David and {Ciardi}, David and {Kaltenegger}, Lisa and {Doyle}, Laurance and {Dupree}, Andrea K. and {Ford}, Eric B. and {Fortney}, Jonathan and {Holman}, Matthew J. and {Steffen}, Jason H. and {Mullally}, Fergal and {Still}, Martin and {Tarter}, Jill and {Ballard}, Sarah and {Buchhave}, Lars A. and {Carter}, Josh and {Christiansen}, Jessie L. and {Demory}, Brice-Olivier and {D{\'e}sert}, Jean-Michel and {Dressing}, Courtney and {Endl}, Michael and {Fabrycky}, Daniel and {Fischer}, Debra and {Haas}, Michael R. and {Henze}, Christopher and {Horch}, Elliott and {Howard}, Andrew W. and {Isaacson}, Howard and {Kjeldsen}, Hans and {Johnson}, John Asher and {Klaus}, Todd and {Kolodziejczak}, Jeffery and {Barclay}, Thomas and {Li}, Jie and {Meibom}, S{\o}ren and {Prsa}, Andrej and {Quinn}, Samuel N. and {Quintana}, Elisa V. and {Robertson}, Paul and {Sherry}, William and {Shporer}, Avi and {Tenenbaum}, Peter and {Thompson}, Susan E. and {Twicken}, Joseph D. and {Van Cleve}, Jeffrey and {Welsh}, William F. and {Basu}, Sarbani and {Chaplin}, William and {Miglio}, Andrea and {Kawaler}, Steven D. and {Arentoft}, Torben and {Stello}, Dennis and {Metcalfe}, Travis S. and {Verner}, Graham A. and {Karoff}, Christoffer and {Lundkvist}, Mia and {Lund}, Mikkel N. and {Handberg}, Rasmus and {Elsworth}, Yvonne and {Hekker}, Saskia and {Huber}, Daniel and {Bedding}, Timothy R. and {Rapin}, William},
        title = "{Kepler-22b: A 2.4 Earth-radius Planet in the Habitable Zone of a Sun-like Star}",
      journal = {\apj},
         year = 2012,
        month = feb,
       volume = {745},
       number = {2},
          eid = {120},
        pages = {120},
          doi = {10.1088/0004-637X/745/2/120},
archivePrefix = {arXiv},
       eprint = {1112.1640},
 primaryClass = {astro-ph.EP},
       adsurl = {https://ui.adsabs.harvard.edu/abs/2012ApJ...745..120B}
}

@ARTICLE{Borucki2013,
       author = {{Borucki}, William J. and others},
        title = "{Kepler-62: A Five-Planet System with Planets of 1.4 and 1.6 Earth Radii in the Habitable Zone}",
      journal = {Science},
         year = 2013,
        month = may,
       volume = {340},
       number = {6132},
        pages = {587-590},
          doi = {10.1126/science.1234702},
archivePrefix = {arXiv},
       eprint = {1304.7387},
 primaryClass = {astro-ph.EP},
       adsurl = {https://ui.adsabs.harvard.edu/abs/2013Sci...340..587B}
}

@ARTICLE{Quintana2014,
       author = {{Quintana}, Elisa V. and {Barclay}, Thomas and {Raymond}, Sean N. and {Rowe}, Jason F. and {Bolmont}, Emeline and {Caldwell}, Douglas A. and {Howell}, Steve B. and {Kane}, Stephen R. and {Huber}, Daniel and {Crepp}, Justin R. and {Lissauer}, Jack J. and {Ciardi}, David R. and {Coughlin}, Jeffrey L. and {Everett}, Mark E. and {Henze}, Christopher E. and {Horch}, Elliott and {Isaacson}, Howard and {Ford}, Eric B. and {Adams}, Fred C. and {Still}, Martin and {Hunter}, Roger C. and {Quarles}, Billy and {Selsis}, Franck},
        title = "{An Earth-Sized Planet in the Habitable Zone of a Cool Star}",
      journal = {Science},
         year = 2014,
        month = apr,
       volume = {344},
       number = {6181},
        pages = {277-280},
          doi = {10.1126/science.1249403},
archivePrefix = {arXiv},
       eprint = {1404.5667},
 primaryClass = {astro-ph.EP},
       adsurl = {https://ui.adsabs.harvard.edu/abs/2014Sci...344..277Q}
}

@ARTICLE{Torres2015,
       author = {{Torres}, Guillermo and {Kipping}, David M. and {Fressin}, Francois and {Caldwell}, Douglas A. and {Twicken}, Joseph D. and {Ballard}, Sarah and {Batalha}, Natalie M. and {Bryson}, Stephen T. and {Ciardi}, David R. and {Henze}, Christopher E. and {Howell}, Steve B. and {Isaacson}, Howard T. and {Jenkins}, Jon M. and {Muirhead}, Philip S. and {Newton}, Elisabeth R. and {Petigura}, Erik A. and {Barclay}, Thomas and {Borucki}, William J. and {Crepp}, Justin R. and {Everett}, Mark E. and {Horch}, Elliott P. and {Howard}, Andrew W. and {Kolbl}, Rea and {Marcy}, Geoffrey W. and {McCauliff}, Sean and {Quintana}, Elisa V.},
        title = "{Validation of 12 Small Kepler Transiting Planets in the Habitable Zone}",
      journal = {\apj},
         year = 2015,
        month = feb,
       volume = {800},
       number = {2},
          eid = {99},
        pages = {99},
          doi = {10.1088/0004-637X/800/2/99},
archivePrefix = {arXiv},
       eprint = {1501.01101},
 primaryClass = {astro-ph.EP},
       adsurl = {https://ui.adsabs.harvard.edu/abs/2015ApJ...800...99T}
}

@ARTICLE{Jenkins2015,
       author = {{Jenkins}, Jon M. and others},
        title = "{Discovery and Validation of Kepler-452b: A 1.6 R $_{⨁}$ Super Earth Exoplanet in the Habitable Zone of a G2 Star}",
      journal = {\aj},
         year = 2015,
        month = aug,
       volume = {150},
       number = {2},
          eid = {56},
        pages = {56},
          doi = {10.1088/0004-6256/150/2/56},
archivePrefix = {arXiv},
       eprint = {1507.06723},
 primaryClass = {astro-ph.EP},
       adsurl = {https://ui.adsabs.harvard.edu/abs/2015AJ....150...56J}
}

@ARTICLE{Mullally2018,
       author = {{Mullally}, Fergal and {Thompson}, Susan E. and {Coughlin}, Jeffrey L. and {Burke}, Christopher J. and {Rowe}, Jason F.},
        title = "{Kepler{\textquoteright}s Earth-like Planets Should Not Be Confirmed without Independent Detection: The Case of Kepler-452b}",
      journal = {\aj},
         year = 2018,
        month = may,
       volume = {155},
       number = {5},
          eid = {210},
        pages = {210},
          doi = {10.3847/1538-3881/aabae3},
archivePrefix = {arXiv},
       eprint = {1803.11307},
 primaryClass = {astro-ph.EP},
       adsurl = {https://ui.adsabs.harvard.edu/abs/2018AJ....155..210M}
}

@ARTICLE{Burke2019,
       author = {{Burke}, Christopher J. and {Mullally}, F. and {Thompson}, Susan E. and {Coughlin}, Jeffrey L. and {Rowe}, Jason F.},
        title = "{Re-evaluating Small Long-period Confirmed Planets from Kepler}",
      journal = {\aj},
         year = 2019,
        month = apr,
       volume = {157},
       number = {4},
          eid = {143},
        pages = {143},
          doi = {10.3847/1538-3881/aafb79},
archivePrefix = {arXiv},
       eprint = {1901.00506},
 primaryClass = {astro-ph.EP},
       adsurl = {https://ui.adsabs.harvard.edu/abs/2019AJ....157..143B}
}

@ARTICLE{Gilbert2020,
       author = {{Gilbert}, Emily A. and {Barclay}, Thomas and {Schlieder}, Joshua E. and {Quintana}, Elisa V. and {Hord}, Benjamin J. and {Kostov}, Veselin B. and {Lopez}, Eric D. and {Rowe}, Jason F. and {Hoffman}, Kelsey and {Walkowicz}, Lucianne M. and {Silverstein}, Michele L. and {Rodriguez}, Joseph E. and {Vanderburg}, Andrew and {Suissa}, Gabrielle and {Airapetian}, Vladimir S. and {Clement}, Matthew S. and {Raymond}, Sean N. and {Mann}, Andrew W. and {Kruse}, Ethan and {Lissauer}, Jack J. and {Col{\'o}n}, Knicole D. and {Kopparapu}, Ravi kumar and {Kreidberg}, Laura and {Zieba}, Sebastian and {Collins}, Karen A. and {Quinn}, Samuel N. and {Howell}, Steve B. and {Ziegler}, Carl and {Vrijmoet}, Eliot Halley and {Adams}, Fred C. and {Arney}, Giada N. and {Boyd}, Patricia T. and {Brande}, Jonathan and {Burke}, Christopher J. and {Cacciapuoti}, Luca and {Chance}, Quadry and {Christiansen}, Jessie L. and {Covone}, Giovanni and {Daylan}, Tansu and {Dineen}, Danielle and {Dressing}, Courtney D. and {Essack}, Zahra and {Fauchez}, Thomas J. and {Galgano}, Brianna and {Howe}, Alex R. and {Kaltenegger}, Lisa and {Kane}, Stephen R. and {Lam}, Christopher and {Lee}, Eve J. and {Lewis}, Nikole K. and {Logsdon}, Sarah E. and {Mandell}, Avi M. and {Monsue}, Teresa and {Mullally}, Fergal and {Mullally}, Susan E. and {Paudel}, Rishi R. and {Pidhorodetska}, Daria and {Plavchan}, Peter and {Reyes}, Naylynn Ta{\~n}{\'o}n and {Rinehart}, Stephen A. and {Rojas-Ayala}, B{\'a}rbara and {Smith}, Jeffrey C. and {Stassun}, Keivan G. and {Tenenbaum}, Peter and {Vega}, Laura D. and {Villanueva}, Geronimo L. and {Wolf}, Eric T. and {Youngblood}, Allison and {Ricker}, George R. and {Vanderspek}, Roland K. and {Latham}, David W. and {Seager}, Sara and {Winn}, Joshua N. and {Jenkins}, Jon M. and {Bakos}, G{\'a}sp{\'a}r {\r{A}}. and {Brice{\~n}o}, C{\'e}sar and {Ciardi}, David R. and {Cloutier}, Ryan and {Conti}, Dennis M. and {Couperus}, Andrew and {Di Sora}, Mario and {Eisner}, Nora L. and {Everett}, Mark E. and {Gan}, Tianjun and {Hartman}, Joel D. and {Henry}, Todd and {Isopi}, Giovanni and {Jao}, Wei-Chun and {Jensen}, Eric L.~N. and {Law}, Nicholas and {Mallia}, Franco and {Matson}, Rachel A. and {Shappee}, Benjamin J. and {Le Wood}, Mackennae and {Winters}, Jennifer G.},
        title = "{The First Habitable-zone Earth-sized Planet from TESS. I. Validation of the TOI-700 System}",
      journal = {\aj},
         year = 2020,
        month = sep,
       volume = {160},
       number = {3},
          eid = {116},
        pages = {116},
          doi = {10.3847/1538-3881/aba4b2},
archivePrefix = {arXiv},
       eprint = {2001.00952},
 primaryClass = {astro-ph.EP},
       adsurl = {https://ui.adsabs.harvard.edu/abs/2020AJ....160..116G}
}

@ARTICLE{Gilbert2023,
       author = {{Gilbert}, Emily A. and {Vanderburg}, Andrew and {Rodriguez}, Joseph E. and {Hord}, Benjamin J. and {Clement}, Matthew S. and {Barclay}, Thomas and {Quintana}, Elisa V. and {Schlieder}, Joshua E. and {Kane}, Stephen R. and {Jenkins}, Jon M. and {Twicken}, Joseph D. and {Kunimoto}, Michelle and {Vanderspek}, Roland and {Arney}, Giada N. and {Charbonneau}, David and {G{\"u}nther}, Maximilian N. and {Huang}, Chelsea X. and {Isopi}, Giovanni and {Kostov}, Veselin B. and {Kristiansen}, Martti H. and {Latham}, David W. and {Mallia}, Franco and {Mamajek}, Eric E. and {Mireles}, Ismael and {Quinn}, Samuel N. and {Ricker}, George R. and {Schulte}, Jack and {Seager}, S. and {Suissa}, Gabrielle and {Winn}, Joshua N. and {Youngblood}, Allison and {Zapparata}, Aldo},
        title = "{A Second Earth-sized Planet in the Habitable Zone of the M Dwarf, TOI-700}",
      journal = {\apjl},
         year = 2023,
        month = feb,
       volume = {944},
       number = {2},
          eid = {L35},
        pages = {L35},
          doi = {10.3847/2041-8213/acb599},
archivePrefix = {arXiv},
       eprint = {2301.03617},
 primaryClass = {astro-ph.EP},
       adsurl = {https://ui.adsabs.harvard.edu/abs/2023ApJ...944L..35G}
}

@ARTICLE{Rodriguez2020,
       author = {{Rodriguez}, Joseph E. and {Vanderburg}, Andrew and {Zieba}, Sebastian and {Kreidberg}, Laura and {Morley}, Caroline V. and {Eastman}, Jason D. and {Kane}, Stephen R. and {Spencer}, Alton and {Quinn}, Samuel N. and {Cloutier}, Ryan and {Huang}, Chelsea X. and {Collins}, Karen A. and {Mann}, Andrew W. and {Gilbert}, Emily and {Schlieder}, Joshua E. and {Quintana}, Elisa V. and {Barclay}, Thomas and {Suissa}, Gabrielle and {Kopparapu}, Ravi kumar and {Dressing}, Courtney D. and {Ricker}, George R. and {Vanderspek}, Roland K. and {Latham}, David W. and {Seager}, Sara and {Winn}, Joshua N. and {Jenkins}, Jon M. and {Berta-Thompson}, Zachory and {Boyd}, Patricia T. and {Charbonneau}, David and {Caldwell}, Douglas A. and {Chiang}, Eugene and {Christiansen}, Jessie L. and {Ciardi}, David R. and {Col{\'o}n}, Knicole D. and {Doty}, John and {Gan}, Tianjun and {Guerrero}, Natalia and {G{\"u}nther}, Maximilian N. and {Lee}, Eve J. and {Levine}, Alan M. and {Lopez}, Eric and {Muirhead}, Philip S. and {Newton}, Elisabeth and {Rose}, Mark E. and {Twicken}, Joseph D. and {Villase{\~n}or}, Jesus Noel},
        title = "{The First Habitable-zone Earth-sized Planet from TESS. II. Spitzer Confirms TOI-700 d}",
      journal = {\aj},
         year = 2020,
        month = sep,
       volume = {160},
       number = {3},
          eid = {117},
        pages = {117},
          doi = {10.3847/1538-3881/aba4b3},
archivePrefix = {arXiv},
       eprint = {2001.00954},
 primaryClass = {astro-ph.EP},
       adsurl = {https://ui.adsabs.harvard.edu/abs/2020AJ....160..117R}
}

@ARTICLE{Dholakia2024,
       author = {{Dholakia}, Shishir and {Palethorpe}, Larissa and others},
        title = "{Gliese 12 b, a temperate Earth-sized planet at 12 parsecs discovered with TESS and CHEOPS}",
      journal = {\mnras},
         year = 2024,
        month = jun,
       volume = {531},
       number = {1},
        pages = {1276-1293},
          doi = {10.1093/mnras/stae1152},
archivePrefix = {arXiv},
       eprint = {2405.13118},
 primaryClass = {astro-ph.EP},
       adsurl = {https://ui.adsabs.harvard.edu/abs/2024MNRAS.531.1276D}
}

@ARTICLE{Kuzuhara2024,
       author = {{Kuzuhara}, Masayuki and {Fukui}, Akihiko and others},
        title = "{Gliese 12 b: A Temperate Earth-sized Planet at 12 pc Ideal for Atmospheric Transmission Spectroscopy}",
      journal = {\apjl},
         year = 2024,
        month = jun,
       volume = {967},
       number = {2},
          eid = {L21},
        pages = {L21},
          doi = {10.3847/2041-8213/ad3642},
archivePrefix = {arXiv},
       eprint = {2405.14708},
 primaryClass = {astro-ph.EP},
       adsurl = {https://ui.adsabs.harvard.edu/abs/2024ApJ...967L..21K}
}

@ARTICLE{Dransfield2024,
       author = {{Dransfield}, Georgina and {Timmermans}, Mathilde and {Triaud}, Amaury H.~M.~J. and {D{\'e}vora-Pajares}, Mart{\'\i}n and {Aganze}, Christian and {Barkaoui}, Khalid and {Burgasser}, Adam J. and {Collins}, Karen A. and {Cointepas}, Marion and {Ducrot}, Elsa and {G{\"u}nther}, Maximilian N. and {Howell}, Steve B. and {Murray}, Catriona A. and {Niraula}, Prajwal and {Rackham}, Benjamin V. and {Sebastian}, Daniel and {Stassun}, Keivan G. and {Z{\'u}{\~n}iga-Fern{\'a}ndez}, Sebasti{\'a}n and {Almenara}, Jos{\'e} Manuel and {Bonfils}, Xavier and {Bouchy}, Fran{\c{c}}ois and {Burke}, Christopher J. and {Charbonneau}, David and {Christiansen}, Jessie L. and {Delrez}, Laetitia and {Gan}, Tianjun and {Garc{\'\i}a}, Lionel J. and {Gillon}, Micha{\"e}l and {G{\'o}mez Maqueo Chew}, Yilen and {Hesse}, Katharine M. and {Hooton}, Matthew J. and {Isopi}, Giovanni and {Jehin}, Emmanu{\"e}l and {Jenkins}, Jon M. and {Latham}, David W. and {Mallia}, Franco and {Murgas}, Felipe and {Pedersen}, Peter P. and {Pozuelos}, Francisco J. and {Queloz}, Didier and {Rodriguez}, David R. and {Schanche}, Nicole and {Seager}, Sara and {Srdoc}, Gregor and {Stockdale}, Chris and {Twicken}, Joseph D. and {Vanderspek}, Roland and {Wells}, Robert and {Winn}, Joshua N. and {de Wit}, Julien and {Zapparata}, Aldo},
        title = "{A 1.55 R$_{{\ensuremath{\oplus}}}$ habitable-zone planet hosted by TOI-715, an M4 star near the ecliptic South Pole}",
      journal = {\mnras},
         year = 2024,
        month = jan,
       volume = {527},
       number = {1},
        pages = {35-52},
          doi = {10.1093/mnras/stad1439},
archivePrefix = {arXiv},
       eprint = {2305.06206},
 primaryClass = {astro-ph.EP},
       adsurl = {https://ui.adsabs.harvard.edu/abs/2024MNRAS.527...35D}
}

@ARTICLE{Scott2025,
       author = {{Scott}, Madison G. and {Triaud}, Amaury H.~M.~J. and {Barkaoui}, Khalid and {Sebastian}, Daniel and {Burgasser}, Adam J. and {Collins}, Karen A. and {Dransfield}, Georgina and {Hellier}, Coel and {Howell}, Steve B. and {Piette}, Anjali A.~A. and {Rackham}, Benjamin V. and {Stassun}, Keivan G. and {Stokholm}, Amalie and {Timmermans}, Mathilde and {Watkins}, Cristilyn N. and {Fausnaugh}, Michael and {Fukui}, Akihiko and {Jenkins}, Jon M. and {Narita}, Norio and {Ricker}, George and {Softich}, Emma and {Schwarz}, Richard P. and {Seager}, Sara and {Shporer}, Avi and {Theissen}, Christopher and {Twicken}, Joseph D. and {Winn}, Joshua N. and {Watanabe}, David},
        title = "{TOI-6478 b: a cold underdense Neptune transiting a fully convective M dwarf from the thick disc}",
      journal = {\mnras},
         year = 2025,
        month = jun,
       volume = {540},
       number = {2},
        pages = {1909-1927},
          doi = {10.1093/mnras/staf684},
archivePrefix = {arXiv},
       eprint = {2504.06848},
 primaryClass = {astro-ph.EP},
       adsurl = {https://ui.adsabs.harvard.edu/abs/2025MNRAS.540.1909S}
}

@ARTICLE{PLATO2014,
       author = {{Rauer}, H. and {Catala}, C. and {Aerts}, C. and {Appourchaux}, T. and {Benz}, W. and {Brandeker}, A. and {Christensen-Dalsgaard}, J. and {Deleuil}, M. and {Gizon}, L. and {Goupil}, M. -J. and {G{\"u}del}, M. and {Janot-Pacheco}, E. and {Mas-Hesse}, M. and {Pagano}, I. and {Piotto}, G. and {Pollacco}, D. and {Santos}, {\.{C}}. and {Smith}, A. and {Su{\'a}rez}, J. -C. and {Szab{\'o}}, R. and {Udry}, S. and {Adibekyan}, V. and {Alibert}, Y. and {Almenara}, J. -M. and {Amaro-Seoane}, P. and {Eiff}, M. Ammler-von and {Asplund}, M. and {Antonello}, E. and {Barnes}, S. and {Baudin}, F. and {Belkacem}, K. and {Bergemann}, M. and {Bihain}, G. and {Birch}, A.~C. and {Bonfils}, X. and {Boisse}, I. and {Bonomo}, A.~S. and {Borsa}, F. and {Brand{\~a}o}, I.~M. and {Brocato}, E. and {Brun}, S. and {Burleigh}, M. and {Burston}, R. and {Cabrera}, J. and {Cassisi}, S. and {Chaplin}, W. and {Charpinet}, S. and {Chiappini}, C. and {Church}, R.~P. and {Csizmadia}, Sz. and {Cunha}, M. and {Damasso}, M. and {Davies}, M.~B. and {Deeg}, H.~J. and {D{\'\i}az}, R.~F. and {Dreizler}, S. and {Dreyer}, C. and {Eggenberger}, P. and {Ehrenreich}, D. and {Eigm{\"u}ller}, P. and {Erikson}, A. and {Farmer}, R. and {Feltzing}, S. and {de Oliveira Fialho}, F. and {Figueira}, P. and {Forveille}, T. and {Fridlund}, M. and {Garc{\'\i}a}, R.~A. and {Giommi}, P. and {Giuffrida}, G. and {Godolt}, M. and {Gomes da Silva}, J. and {Granzer}, T. and {Grenfell}, J.~L. and {Grotsch-Noels}, A. and {G{\"u}nther}, E. and {Haswell}, C.~A. and {Hatzes}, A.~P. and {H{\'e}brard}, G. and {Hekker}, S. and {Helled}, R. and {Heng}, K. and {Jenkins}, J.~M. and {Johansen}, A. and {Khodachenko}, M.~L. and {Kislyakova}, K.~G. and {Kley}, W. and {Kolb}, U. and {Krivova}, N. and {Kupka}, F. and {Lammer}, H. and {Lanza}, A.~F. and {Lebreton}, Y. and {Magrin}, D. and {Marcos-Arenal}, P. and {Marrese}, P.~M. and {Marques}, J.~P. and {Martins}, J. and {Mathis}, S. and {Mathur}, S. and {Messina}, S. and {Miglio}, A. and {Montalban}, J. and {Montalto}, M. and {Monteiro}, M.~J.~P.~F.~G. and {Moradi}, H. and {Moravveji}, E. and {Mordasini}, C. and {Morel}, T. and {Mortier}, A. and {Nascimbeni}, V. and {Nelson}, R.~P. and {Nielsen}, M.~B. and {Noack}, L. and {Norton}, A.~J. and {Ofir}, A. and {Oshagh}, M. and {Ouazzani}, R. -M. and {P{\'a}pics}, P. and {Parro}, V.~C. and {Petit}, P. and {Plez}, B. and {Poretti}, E. and {Quirrenbach}, A. and {Ragazzoni}, R. and {Raimondo}, G. and {Rainer}, M. and {Reese}, D.~R. and {Redmer}, R. and {Reffert}, S. and {Rojas-Ayala}, B. and {Roxburgh}, I.~W. and {Salmon}, S. and {Santerne}, A. and {Schneider}, J. and {Schou}, J. and {Schuh}, S. and {Schunker}, H. and {Silva-Valio}, A. and {Silvotti}, R. and {Skillen}, I. and {Snellen}, I. and {Sohl}, F. and {Sousa}, S.~G. and {Sozzetti}, A. and {Stello}, D. and {Strassmeier}, K.~G. and {{\v{S}}vanda}, M. and {Szab{\'o}}, Gy. M. and {Tkachenko}, A. and {Valencia}, D. and {Van Grootel}, V. and {Vauclair}, S.~D. and {Ventura}, P. and {Wagner}, F.~W. and {Walton}, N.~A. and {Weingrill}, J. and {Werner}, S.~C. and {Wheatley}, P.~J. and {Zwintz}, K.},
        title = "{The PLATO 2.0 mission}",
      journal = {Experimental Astronomy},
         year = 2014,
        month = nov,
       volume = {38},
       number = {1-2},
        pages = {249-330},
          doi = {10.1007/s10686-014-9383-4},
archivePrefix = {arXiv},
       eprint = {1310.0696},
 primaryClass = {astro-ph.EP},
       adsurl = {https://ui.adsabs.harvard.edu/abs/2014ExA....38..249R}
}

@ARTICLE{PLATO2024,
       author = {{Rauer}, Heike and {Aerts}, Conny and {Cabrera}, Juan and {Deleuil}, Magali and {Erikson}, Anders and {Gizon}, Laurent and {Goupil}, Mariejo and {Heras}, Ana and {Walloschek}, Thomas and {Lorenzo-Alvarez}, Jose and {Marliani}, Filippo and {Martin-Garcia}, C{\'e}sar and {Mas-Hesse}, J. Miguel and {O'Rourke}, Laurence and {Osborn}, Hugh and {Pagano}, Isabella and {Piotto}, Giampaolo and {Pollacco}, Don and {Ragazzoni}, Roberto and {Ramsay}, Gavin and {Udry}, St{\'e}phane and {Appourchaux}, Thierry and {Benz}, Willy and {Brandeker}, Alexis and {G{\"u}del}, Manuel and {Janot-Pacheco}, Eduardo and {Kabath}, Petr and {Kjeldsen}, Hans and {Min}, Michiel and {Santos}, Nuno and {Smith}, Alan and {Suarez}, Juan-Carlos and {Werner}, Stephanie C. and {Aboudan}, Alessio and {Abreu}, Manuel and {Acu{\~n}a}, Lorena and {Adams}, Moritz and {Adibekyan}, Vardan and {Affer}, Laura and {Agneray}, Fran{\c{c}}ois and {Agnor}, Craig and {Aguirre B{\o}rsen-Koch}, Victor and {Ahmed}, Saad and {Aigrain}, Suzanne and {Al-Bahlawan}, Ashraf and {Alcacera Gil}, Ma de los Angeles and {Alei}, Eleonora and {Alencar}, Silvia and {Alexander}, Richard and {Alfonso-Garz{\'o}n}, Julia and {Alibert}, Yann and {Allende Prieto}, Carlos and {Almeida}, Leonardo and {Alonso Sobrino}, Roi and {Altavilla}, Giuseppe and {Althaus}, Christian and {Alvarez Trujillo}, Luis Alonso and {Amarsi}, Anish and {Ammler-von Eiff}, Matthias and {Am{\^o}res}, Eduardo and {Andrade}, Laerte and {Antoniadis-Karnavas}, Alexandros and {Ant{\'o}nio}, Carlos and {Aparicio del Moral}, Beatriz and {Appolloni}, Matteo and {Arena}, Claudio and {Armstrong}, David and {Aroca Aliaga}, Jose and {Asplund}, Martin and {Audenaert}, Jeroen and {Auricchio}, Natalia and {Avelino}, Pedro and {Baeke}, Ann and {Bailli{\'e}}, Kevin and {Balado}, Ana and {Ballber Balaguer{\'o}}, Pau and {Balestra}, Andrea and {Ball}, Warrick and {Ballans}, Herve and {Ballot}, Jerome and {Barban}, Caroline and {Barbary}, Ga{\"e}le and {Barbieri}, Mauro and {Barcel{\'o} Forteza}, Sebasti{\`a} and {Barker}, Adrian and {Barklem}, Paul and {Barnes}, Sydney and {Barrado Navascues}, David and {Barragan}, Oscar and {Baruteau}, Cl{\'e}ment and {Basu}, Sarbani and {Baudin}, Frederic and {Baumeister}, Philipp and {Bayliss}, Daniel and {Bazot}, Michael and {Beck}, Paul G. and {Belkacem}, Kevin and {Bellinger}, Earl and {Benatti}, Serena and {Benomar}, Othman and {B{\'e}rard}, Diane and {Bergemann}, Maria and {Bergomi}, Maria and {Bernardo}, Pierre and {Biazzo}, Katia and {Bignamini}, Andrea and {Bigot}, Lionel and {Billot}, Nicolas and {Binet}, Martin and {Biondi}, David and {Biondi}, Federico and {Birch}, Aaron C. and {Bitsch}, Bertram and {Bluhm Ceballos}, Paz Victoria and {B{\'o}di}, Attila and {Bogn{\'a}r}, Zs{\'o}fia and {Boisse}, Isabelle and {Bolmont}, Emeline and {Bonanno}, Alfio and {Bonavita}, Mariangela and {Bonfanti}, Andrea and {Bonfils}, Xavier and {Bonito}, Rosaria and {Bonomo}, Aldo Stefano and {B{\"o}rner}, Anko and {Boro Saikia}, Sudeshna and {Borreguero Mart{\'\i}n}, Elisa and {Borsa}, Francesco and {Borsato}, Luca and {Bossini}, Diego and {Bouchy}, Francois and {Bou{\'e}}, Gwena{\"e}l and {Boufleur}, Rodrigo and {Boumier}, Patrick and {Bourrier}, Vincent and {Bowman}, Dominic M. and {Bozzo}, Enrico and {Bradley}, Louisa and {Bray}, John and {Bressan}, Alessandro and {Breton}, Sylvain and {Brienza}, Daniele and {Brito}, Ana and {Brogi}, Matteo and {Brown}, Beverly and {Brown}, David J.~A. and {Brun}, Allan Sacha and {Bruno}, Giovanni and {Bruns}, Michael and {Buchhave}, Lars A. and {Bugnet}, Lisa and {Buldgen}, Ga{\"e}l and {Burgess}, Patrick and {Busatta}, Andrea and {Busso}, Giorgia and {Buzasi}, Derek and {Caballero}, Jos{\'e} A. and {Cabral}, Alexandre and {Cabrero Gomez}, Juan-Francisco and {Calderone}, Flavia and {Cameron}, Robert and {Cameron}, Andrew and {Campante}, Tiago and {Campos Gestal}, N{\'e}stor and {Canto Martins}, Bruno Leonardo and {Cara}, Christophe and {Carone}, Ludmila and {Carrasco}, Josep Manel and {Casagrande}, Luca and {Casewell}, Sarah L. and {Cassisi}, Santi and {Castellani}, Marco and {Castro}, Matthieu and {Catala}, Claude and {Catal{\'a}n Fern{\'a}ndez}, Irene and {Catelan}, M{\'a}rcio and {Cegla}, Heather and {Cerruti}, Chiara and {Cessa}, Virginie and {Chadid}, Merieme and {Chaplin}, William and {Charpinet}, Stephane and {Chiappini}, Cristina and {Chiarucci}, Simone and {Chiavassa}, Andrea and {Chinellato}, Simonetta and {Chirulli}, Giovanni and {Christensen-Dalsgaard}, J{\o}rgen and {Church}, Ross and {Claret}, Antonio and {Clarke}, Cathie and {Claudi}, Riccardo and {Clermont}, Lionel and {Coelho}, Hugo and {Coelho}, Joao and {Cogato}, Fabrizio and {Colom{\'e}}, Josep and {Condamin}, Mathieu and {Conde Garc{\'\i}a}, Fernando and {Conseil}, Simon},
        title = "{The PLATO mission}",
      journal = {Experimental Astronomy},
         year = 2025,
        month = jun,
       volume = {59},
       number = {3},
          eid = {26},
        pages = {26},
          doi = {10.1007/s10686-025-09985-9},
archivePrefix = {arXiv},
       eprint = {2406.05447},
 primaryClass = {astro-ph.IM},
       adsurl = {https://ui.adsabs.harvard.edu/abs/2025ExA....59...26R}
}

@ARTICLE{Ge2022,
       author = {{Ge}, Jian and {Zhang}, Hui and {Zang}, Weicheng and {Deng}, Hongping and {Mao}, Shude and {Xie}, Ji-Wei and {Liu}, Hui-Gen and {Zhou}, Ji-Lin and {Willis}, Kevin and {Huang}, Chelsea and {Howell}, Steve B. and {Feng}, Fabo and {Zhu}, Jiapeng and {Yao}, Xinyu and {Liu}, Beibei and {Aizawa}, Masataka and {Zhu}, Wei and {Li}, Ya-Ping and {Ma}, Bo and {Ye}, Quanzhi and {Yu}, Jie and {Xiang}, Maosheng and {Yu}, Cong and {Liu}, Shangfei and {Yang}, Ming and {Wang}, Mu-Tian and {Shi}, Xian and {Fang}, Tong and {Zong}, Weikai and {Liu}, Jinzhong and {Zhang}, Yu and {Zhang}, Liyun and {El-Badry}, Kareem and {Shen}, Rongfeng and {Tam}, Pak-Hin Thomas and {Hu}, Zhecheng and {Yang}, Yanlv and {Zou}, Yuan-Chuan and {Wu}, Jia-Li and {Lei}, Wei-Hua and {Wei}, Jun-Jie and {Wu}, Xue-Feng and {Sun}, Tian-Rui and {Wang}, Fa-Yin and {Zhang}, Bin-Bin and {Xu}, Dong and {Yang}, Yuan-Pei and {Li}, Wen-Xiong and {Xiang}, Dan-Feng and {Wang}, Xiaofeng and {Wang}, Tinggui and {Zhang}, Bing and {Jia}, Peng and {Yuan}, Haibo and {Zhang}, Jinghua and {Xuesong Wang}, Sharon and {Gan}, Tianjun and {Wang}, Wei and {Zhao}, Yinan and {Liu}, Yujuan and {Wei}, Chuanxin and {Kang}, Yanwu and {Yang}, Baoyu and {Qi}, Chao and {Liu}, Xiaohua and {Zhang}, Quan and {Zhu}, Yuji and {Zhou}, Dan and {Zhang}, Congcong and {Yu}, Yong and {Zhang}, Yongshuai and {Li}, Yan and {Tang}, Zhenghong and {Wang}, Chaoyan and {Wang}, Fengtao and {Li}, Wei and {Cheng}, Pengfei and {Shen}, Chao and {Li}, Baopeng and {Pan}, Yue and {Yang}, Sen and {Gao}, Wei and {Song}, Zongxi and {Wang}, Jian and {Zhang}, Hongfei and {Chen}, Cheng and {Wang}, Hui and {Zhang}, Jun and {Wang}, Zhiyue and {Zeng}, Feng and {Zheng}, Zhenhao and {Zhu}, Jie and {Guo}, Yingfan and {Zhang}, Yihao and {Li}, Yudong and {Wen}, Lin and {Feng}, Jie and {Chen}, Wen and {Chen}, Kun and {Han}, Xingbo and {Yang}, Yingquan and {Wang}, Haoyu and {Duan}, Xuliang and {Huang}, Jiangjiang and {Liang}, Hong and {Bi}, Shaolan and {Gai}, Ning and {Ge}, Zhishuai and {Guo}, Zhao and {Huang}, Yang and {Li}, Gang and {Li}, Haining and {Li}, Tanda and {Yuxi} and {Lu} and {Rix}, Hans-Walter and {Shi}, Jianrong and {Song}, Fen and {Tang}, Yanke and {Ting}, Yuan-Sen and {Wu}, Tao and {Wu}, Yaqian and {Yang}, Taozhi and {Yin}, Qing-Zhu and {Gould}, Andrew and {Lee}, Chung-Uk and {Dong}, Subo and {Yee}, Jennifer C. and {Shvartzvald}, Yossi and {Yang}, Hongjing and {Kuang}, Renkun and {Zhang}, Jiyuan and {Liao}, Shilong and {Qi}, Zhaoxiang and {Yang}, Jun and {Zhang}, Ruisheng and {Jiang}, Chen and {Ou}, Jian-Wen and {Li}, Yaguang and {Beck}, Paul and {Bedding}, Timothy R. and {Campante}, Tiago L. and {Chaplin}, William J. and {Christensen-Dalsgaard}, J{\o}rgen and {Garc{\'\i}a}, Rafael A. and {Gaulme}, Patrick and {Gizon}, Laurent and {Hekker}, Saskia and {Huber}, Daniel and {Khanna}, Shourya and {Li}, Yan and {Mathur}, Savita and {Miglio}, Andrea and {Mosser}, Beno{\^\i}t and {Ong}, J.~M. Joel and {Santos}, {\^A}ngela R.~G. and {Stello}, Dennis and {Bowman}, Dominic M. and {Lares-Martiz}, Mariel and {Murphy}, Simon and {Niu}, Jia-Shu and {Ma}, Xiao-Yu and {Moln{\'a}r}, L{\'a}szl{\'o} and {Fu}, Jian-Ning and {De Cat}, Peter and {Su}, Jie and {consortium}, the ET},
        title = "{ET White Paper: To Find the First Earth 2.0}",
      journal = {arXiv e-prints},
         year = 2022,
        month = jun,
          eid = {arXiv:2206.06693},
        pages = {arXiv:2206.06693},
          doi = {10.48550/arXiv.2206.06693},
archivePrefix = {arXiv},
       eprint = {2206.06693},
 primaryClass = {astro-ph.IM},
       adsurl = {https://ui.adsabs.harvard.edu/abs/2022arXiv220606693G}
}

@ARTICLE{Dragomir2019,
       author = {{Dragomir}, Diana and {Teske}, Johanna and {G{\"u}nther}, Maximilian N. and {S{\'e}gransan}, Damien and {Burt}, Jennifer A. and {Huang}, Chelsea X. and {Vanderburg}, Andrew and {Matthews}, Elisabeth and {Dumusque}, Xavier and {Stassun}, Keivan G. and {Pepper}, Joshua and {Ricker}, George R. and {Vanderspek}, Roland and {Latham}, David W. and {Seager}, Sara and {Winn}, Joshua N. and {Jenkins}, Jon M. and {Beatty}, Thomas and {Bouchy}, Fran{\c{c}}ois and {Brown}, Timothy M. and {Butler}, R. Paul and {Ciardi}, David R. and {Crane}, Jeffrey D. and {Eastman}, Jason D. and {Fossati}, Luca and {Francis}, Jim and {Fulton}, Benjamin J. and {Gaudi}, B. Scott and {Goeke}, Robert F. and {James}, David and {Klaus}, Todd C. and {Kuhn}, Rudolf B. and {Lovis}, Christophe and {Lund}, Michael B. and {McDermott}, Scott and {Paegert}, Martin and {Pepe}, Francesco and {Rodriguez}, Joseph E. and {Sha}, Lizhou and {Shectman}, Stephen A. and {Shporer}, Avi and {Siverd}, Robert J. and {Garcia Soto}, Aylin and {Stevens}, Daniel J. and {Twicken}, Joseph D. and {Udry}, St{\'e}phane and {Villanueva}, Jr., Steven and {Wang}, Sharon X. and {Wohler}, Bill and {Yao}, Xinyu and {Zhan}, Zhuchang},
        title = "{TESS Delivers Its First Earth-sized Planet and a Warm Sub-Neptune}",
      journal = {\apjl},
         year = 2019,
        month = apr,
       volume = {875},
       number = {2},
          eid = {L7},
        pages = {L7},
          doi = {10.3847/2041-8213/ab12ed},
archivePrefix = {arXiv},
       eprint = {1901.00051},
 primaryClass = {astro-ph.EP},
       adsurl = {https://ui.adsabs.harvard.edu/abs/2019ApJ...875L...7D}
}

@ARTICLE{Kunimoto2025,
       author = {{Kunimoto}, Michelle and {Lin}, Zifan and {Millholland}, Sarah and {Venner}, Alexander and {Hinkel}, Natalie R. and {Shporer}, Avi and {Vanderburg}, Andrew and {Bailey}, Jeremy and {Brahm}, Rafael and {Burt}, Jennifer A. and {Butler}, R. Paul and {Carter}, Brad and {Ciardi}, David R. and {Collins}, Karen A. and {Collins}, Kevin I. and {Col{\'o}n}, Knicole D. and {Crane}, Jeffrey D. and {Daylan}, Tansu and {D{\'\i}az}, Mat{\'\i}as R. and {Doty}, John P. and {Feng}, Fabo and {Guenther}, Eike W. and {Horner}, Jonathan and {Howell}, Steve B. and {Janik}, Jan and {Jones}, Hugh R.~A. and {Kab{\'a}th}, Petr and {Kanodia}, Shubham and {Littlefield}, Colin and {Osborn}, Hugh P. and {O'Toole}, Simon and {Paegert}, Martin and {Pintr}, Pavel and {Schwarz}, Richard P. and {Shectman}, Steve and {Srdoc}, Gregor and {Stassun}, Keivan G. and {Teske}, Johanna K. and {Twicken}, Joseph D. and {Vanzi}, Leonardo and {Wang}, Sharon X. and {Wittenmyer}, Robert A. and {Jenkins}, Jon M. and {Ricker}, George R. and {Seager}, Sara and {Winn}, Joshua},
        title = "{Two Earth-size Planets and an Earth-size Candidate Transiting the nearby Star HD 101581}",
      journal = {\aj},
         year = 2025,
        month = jan,
       volume = {169},
       number = {1},
          eid = {47},
        pages = {47},
          doi = {10.3847/1538-3881/ad9266},
archivePrefix = {arXiv},
       eprint = {2412.08863},
 primaryClass = {astro-ph.EP},
       adsurl = {https://ui.adsabs.harvard.edu/abs/2025AJ....169...47K}
}

@BOOK{Howell2020,
       editor = {{Howell}, Steve B.},
        title = "{The NASA Kepler Mission}",
       series = {AAS-IOP Astronomy},
    publisher = "{IOP Publishing.}",
         year = 2020,
       adsurl = {https://ui.adsabs.harvard.edu/abs/2020nkm..book.....H}
}

@ARTICLE{Huber2016,
       author = {{Huber}, Daniel and {Bryson}, Stephen T. and {Haas}, Michael R. and {Barclay}, Thomas and {Barentsen}, Geert and {Howell}, Steve B. and {Sharma}, Sanjib and {Stello}, Dennis and {Thompson}, Susan E.},
        title = "{The K2 Ecliptic Plane Input Catalog (EPIC) and Stellar Classifications of 138,600 Targets in Campaigns 1-8}",
      journal = {\apjs},
         year = 2016,
        month = may,
       volume = {224},
       number = {1},
          eid = {2},
        pages = {2},
          doi = {10.3847/0067-0049/224/1/2},
archivePrefix = {arXiv},
       eprint = {1512.02643},
 primaryClass = {astro-ph.SR},
       adsurl = {https://ui.adsabs.harvard.edu/abs/2016ApJS..224....2H}
}

@ARTICLE{Wang2015,
       author = {{Wang}, Ji and {Fischer}, Debra A. and {Barclay}, Thomas and {Picard}, Alyssa and {Ma}, Bo and {Bowler}, Brendan P. and {Schmitt}, Joseph R. and {Boyajian}, Tabetha S. and {Jek}, Kian J. and {LaCourse}, Daryll and {Baranec}, Christoph and {Riddle}, Reed and {Law}, Nicholas M. and {Lintott}, Chris and {Schawinski}, Kevin and {Simister}, Dean Joseph and {Gr{\'e}goire}, Boscher and {Babin}, Sean P. and {Poile}, Trevor and {Jacobs}, Thomas Lee and {Jebson}, Tony and {Omohundro}, Mark R. and {Schwengeler}, Hans Martin and {Sejpka}, Johann and {Terentev}, Ivan A. and {Gagliano}, Robert and {Paakkonen}, Jari-Pekka and {Otnes Berge}, Hans Kristian and {Winarski}, Troy and {Green}, Gerald R. and {Schmitt}, Allan R. and {Kristiansen}, Martti H. and {Hoekstra}, Abe},
        title = "{Planet Hunters. VIII. Characterization of 41 Long-period Exoplanet Candidates from Kepler Archival Data}",
      journal = {\apj},
         year = 2015,
        month = dec,
       volume = {815},
       number = {2},
          eid = {127},
        pages = {127},
          doi = {10.1088/0004-637X/815/2/127},
archivePrefix = {arXiv},
       eprint = {1512.02559},
 primaryClass = {astro-ph.EP},
       adsurl = {https://ui.adsabs.harvard.edu/abs/2015ApJ...815..127W}
}

@ARTICLE{Uehara2016,
       author = {{Uehara}, Sho and {Kawahara}, Hajime and {Masuda}, Kento and {Yamada}, Shin'ya and {Aizawa}, Masataka},
        title = "{Transiting Planet Candidates Beyond the Snow Line Detected by Visual Inspection of 7557 Kepler Objects of Interest}",
      journal = {\apj},
         year = 2016,
        month = may,
       volume = {822},
       number = {1},
          eid = {2},
        pages = {2},
          doi = {10.3847/0004-637X/822/1/2},
archivePrefix = {arXiv},
       eprint = {1602.07848},
 primaryClass = {astro-ph.EP},
       adsurl = {https://ui.adsabs.harvard.edu/abs/2016ApJ...822....2U}
}

@ARTICLE{ForemanMackey2016,
       author = {{Foreman-Mackey}, Daniel and {Morton}, Timothy D. and {Hogg}, David W. and {Agol}, Eric and {Sch{\"o}lkopf}, Bernhard},
        title = "{The Population of Long-period Transiting Exoplanets}",
      journal = {\aj},
         year = 2016,
        month = dec,
       volume = {152},
       number = {6},
          eid = {206},
        pages = {206},
          doi = {10.3847/0004-6256/152/6/206},
archivePrefix = {arXiv},
       eprint = {1607.08237},
 primaryClass = {astro-ph.EP},
       adsurl = {https://ui.adsabs.harvard.edu/abs/2016AJ....152..206F}
}

@ARTICLE{Eisner2021,
       author = {{Eisner}, N.~L. and {Barrag{\'a}n}, O. and {Lintott}, C. and {Aigrain}, S. and {Nicholson}, B. and {Boyajian}, T.~S. and {Howell}, S. and {Johnston}, C. and {Lakeland}, B. and {Miller}, G. and {McMaster}, A. and {Parviainen}, H. and {Safron}, E.~J. and {Schwamb}, M.~E. and {Trouille}, L. and {Vaughan}, S. and {Zicher}, N. and {Allen}, C. and {Allen}, S. and {Bouslog}, M. and {Johnson}, C. and {Simon}, M.~N. and {Wolfenbarger}, Z. and {Baeten}, E.~M.~L. and {Bundy}, D.~M. and {Hoffman}, T.},
        title = "{Planet Hunters TESS II: findings from the first two years of TESS}",
      journal = {\mnras},
         year = 2021,
        month = mar,
       volume = {501},
       number = {4},
        pages = {4669-4690},
          doi = {10.1093/mnras/staa3739},
archivePrefix = {arXiv},
       eprint = {2011.13944},
 primaryClass = {astro-ph.EP},
       adsurl = {https://ui.adsabs.harvard.edu/abs/2021MNRAS.501.4669E}
}

@ARTICLE{Dalba2022,
       author = {{Dalba}, Paul A. and {Kane}, Stephen R. and {Dragomir}, Diana and {Villanueva}, Steven and {Collins}, Karen A. and {Jacobs}, Thomas Lee and {LaCourse}, Daryll M. and {Gagliano}, Robert and {Kristiansen}, Martti H. and {Omohundro}, Mark and {Schwengeler}, Hans M. and {Terentev}, Ivan A. and {Vanderburg}, Andrew and {Fulton}, Benjamin and {Isaacson}, Howard and {Van Zandt}, Judah and {Howard}, Andrew W. and {Thorngren}, Daniel P. and {Howell}, Steve B. and {Batalha}, Natalie M. and {Chontos}, Ashley and {Crossfield}, Ian J.~M. and {Dressing}, Courtney D. and {Huber}, Daniel and {Petigura}, Erik A. and {Robertson}, Paul and {Roy}, Arpita and {Weiss}, Lauren M. and {Behmard}, Aida and {Beard}, Corey and {Brinkman}, Casey L. and {Giacalone}, Steven and {Hill}, Michelle L. and {Lubin}, Jack and {Mayo}, Andrew W. and {Mo{\v{c}}nik}, Teo and {Akana Murphy}, Joseph M. and {Polanski}, Alex S. and {Rice}, Malena and {Rosenthal}, Lee J. and {Rubenzahl}, Ryan A. and {Scarsdale}, Nicholas and {Turtelboom}, Emma V. and {Tyler}, Dakotah and {Benni}, Paul and {Boyce}, Pat and {Esposito}, Thomas M. and {Girardin}, E. and {Laloum}, Didier and {Lewin}, Pablo and {Mann}, Christopher R. and {Marchis}, Franck and {Schwarz}, Richard P. and {Srdoc}, Gregor and {Steuer}, Jana and {Sivarani}, Thirupathi and {Unni}, Athira and {Eisner}, Nora L. and {Fetherolf}, Tara and {Li}, Zhexing and {Yao}, Xinyu and {Pepper}, Joshua and {Ricker}, George R. and {Vanderspek}, Roland and {Latham}, David W. and {Seager}, S. and {Winn}, Joshua N. and {Jenkins}, Jon M. and {Burke}, Christopher J. and {Eastman}, Jason D. and {Lund}, Michael B. and {Rodriguez}, David R. and {Rowden}, Pamela and {Ting}, Eric B. and {Villase{\~n}or}, Jesus Noel},
        title = "{The TESS-Keck Survey. VIII. Confirmation of a Transiting Giant Planet on an Eccentric 261 Day Orbit with the Automated Planet Finder Telescope}",
      journal = {\aj},
         year = 2022,
        month = feb,
       volume = {163},
       number = {2},
          eid = {61},
        pages = {61},
          doi = {10.3847/1538-3881/ac415b},
archivePrefix = {arXiv},
       eprint = {2201.04146},
 primaryClass = {astro-ph.EP},
       adsurl = {https://ui.adsabs.harvard.edu/abs/2022AJ....163...61D}
}

@ARTICLE{Mann2023,
       author = {{Mann}, Christopher R. and {Dalba}, Paul A. and {Lafreni{\`e}re}, David and {Fulton}, Benjamin J. and {H{\'e}brard}, Guillaume and {Boisse}, Isabelle and {Dalal}, Shweta and {Deleuil}, Magali and {Delfosse}, Xavier and {Demangeon}, Olivier and {Forveille}, Thierry and {Heidari}, Neda and {Kiefer}, Flavien and {Martioli}, Eder and {Moutou}, Claire and {Endl}, Michael and {Cochran}, William D. and {MacQueen}, Phillip and {Marchis}, Franck and {Dragomir}, Diana and {Gupta}, Arvind F. and {Feliz}, Dax L. and {Nicholson}, Belinda A. and {Ziegler}, Carl and {Villanueva}, Steven and {Rowe}, Jason and {Talens}, Geert Jan and {Thorngren}, Daniel and {LaCourse}, Daryll and {Jacobs}, Tom and {Howard}, Andrew W. and {Bieryla}, Allyson and {Latham}, David W. and {Rabus}, Markus and {Fetherolf}, Tara and {Hellier}, Coel and {Howell}, Steve B. and {Plavchan}, Peter and {Reefe}, Michael and {Combs}, Deven and {Bowen}, Michael and {Wittrock}, Justin and {Ricker}, George R. and {Seager}, S. and {Winn}, Joshua N. and {Jenkins}, Jon M. and {Barclay}, Thomas and {Watanabe}, David and {Collins}, Karen A. and {Eastman}, Jason D. and {Ting}, Eric B.},
        title = "{Giant Outer Transiting Exoplanet Mass (GOT 'EM) Survey. III. Recovery and Confirmation of a Temperate, Mildly Eccentric, Single-transit Jupiter Orbiting TOI-2010}",
      journal = {\aj},
         year = 2023,
        month = dec,
       volume = {166},
       number = {6},
          eid = {239},
        pages = {239},
          doi = {10.3847/1538-3881/ad00bc},
archivePrefix = {arXiv},
       eprint = {2311.10232},
 primaryClass = {astro-ph.EP},
       adsurl = {https://ui.adsabs.harvard.edu/abs/2023AJ....166..239M}
}

@ARTICLE{Sgro2024,
       author = {{Sgro}, Lauren A. and {Dalba}, Paul A. and {Esposito}, Thomas M. and {Marchis}, Franck and {Dragomir}, Diana and {Villanueva}, Steven and {Fulton}, Benjamin and {Billiani}, Mario and {Loose}, Margaret and {Meneghelli}, Nicola and {Rivett}, Darren and {Saibi}, Fadi and {Saibi}, Sophie and {Martin}, Bryan and {Lekkas}, Georgios and {Zaharevitz}, Daniel and {Zellem}, Robert T. and {Terentev}, Ivan A. and {Gagliano}, Robert and {Jacobs}, Thomas Lee and {Kristiansen}, Martti H. and {LaCourse}, Daryll M. and {Omohundro}, Mark and {Schwengeler}, Hans M.},
        title = "{Confirmation and Characterization of the Eccentric, Warm Jupiter TIC 393818343 b with a Network of Citizen Scientists}",
      journal = {\aj},
         year = 2024,
        month = jul,
       volume = {168},
       number = {1},
          eid = {26},
        pages = {26},
          doi = {10.3847/1538-3881/ad5096},
archivePrefix = {arXiv},
       eprint = {2405.15021},
 primaryClass = {astro-ph.EP},
       adsurl = {https://ui.adsabs.harvard.edu/abs/2024AJ....168...26S}
}

@ARTICLE{Vanderburg2015,
       author = {{Vanderburg}, Andrew and {Montet}, Benjamin T. and {Johnson}, John Asher and {Buchhave}, Lars A. and {Zeng}, Li and {Pepe}, Francesco and {Collier Cameron}, Andrew and {Latham}, David W. and {Molinari}, Emilio and {Udry}, St{\'e}phane and {Lovis}, Christophe and {Matthews}, Jaymie M. and {Cameron}, Chris and {Law}, Nicholas and {Bowler}, Brendan P. and {Angus}, Ruth and {Baranec}, Christoph and {Bieryla}, Allyson and {Boschin}, Walter and {Charbonneau}, David and {Cosentino}, Rosario and {Dumusque}, Xavier and {Figueira}, Pedro and {Guenther}, David B. and {Harutyunyan}, Avet and {Hellier}, Coel and {Kuschnig}, Rainer and {Lopez-Morales}, Mercedes and {Mayor}, Michel and {Micela}, Giusi and {Moffat}, Anthony F.~J. and {Pedani}, Marco and {Phillips}, David F. and {Piotto}, Giampaolo and {Pollacco}, Don and {Queloz}, Didier and {Rice}, Ken and {Riddle}, Reed and {Rowe}, Jason F. and {Rucinski}, Slavek M. and {Sasselov}, Dimitar and {S{\'e}gransan}, Damien and {Sozzetti}, Alessandro and {Szentgyorgyi}, Andrew and {Watson}, Chris and {Weiss}, Werner W.},
        title = "{Characterizing K2 Planet Discoveries: A Super-Earth Transiting the Bright K Dwarf HIP 116454}",
      journal = {\apj},
         year = 2015,
        month = feb,
       volume = {800},
       number = {1},
          eid = {59},
        pages = {59},
          doi = {10.1088/0004-637X/800/1/59},
archivePrefix = {arXiv},
       eprint = {1412.5674},
 primaryClass = {astro-ph.EP},
       adsurl = {https://ui.adsabs.harvard.edu/abs/2015ApJ...800...59V}
}

@ARTICLE{Vanderburg2016,
       author = {{Vanderburg}, Andrew and {Becker}, Juliette C. and {Kristiansen}, Martti H. and {Bieryla}, Allyson and {Duev}, Dmitry A. and {Jensen-Clem}, Rebecca and {Morton}, Timothy D. and {Latham}, David W. and {Adams}, Fred C. and {Baranec}, Christoph and {Berlind}, Perry and {Calkins}, Michael L. and {Esquerdo}, Gilbert A. and {Kulkarni}, Shrinivas and {Law}, Nicholas M. and {Riddle}, Reed and {Salama}, Ma{\"\i}ssa and {Schmitt}, Allan R.},
        title = "{Five Planets Transiting a Ninth Magnitude Star}",
      journal = {\apjl},
         year = 2016,
        month = aug,
       volume = {827},
       number = {1},
          eid = {L10},
        pages = {L10},
          doi = {10.3847/2041-8205/827/1/L10},
archivePrefix = {arXiv},
       eprint = {1606.08441},
 primaryClass = {astro-ph.EP},
       adsurl = {https://ui.adsabs.harvard.edu/abs/2016ApJ...827L..10V}
}

@ARTICLE{Vanderburg2018,
       author = {{Vanderburg}, Andrew and {Mann}, Andrew W. and {Rizzuto}, Aaron and {Bieryla}, Allyson and {Kraus}, Adam L. and {Berlind}, Perry and {Calkins}, Michael L. and {Curtis}, Jason L. and {Douglas}, Stephanie T. and {Esquerdo}, Gilbert A. and {Everett}, Mark E. and {Horch}, Elliott P. and {Howell}, Steve B. and {Latham}, David W. and {Mayo}, Andrew W. and {Quinn}, Samuel N. and {Scott}, Nicholas J. and {Stefanik}, Robert P.},
        title = "{Zodiacal Exoplanets in Time (ZEIT). VII. A Temperate Candidate Super-Earth in the Hyades Cluster}",
      journal = {\aj},
         year = 2018,
        month = aug,
       volume = {156},
       number = {2},
          eid = {46},
        pages = {46},
          doi = {10.3847/1538-3881/aac894},
archivePrefix = {arXiv},
       eprint = {1805.11117},
 primaryClass = {astro-ph.EP},
       adsurl = {https://ui.adsabs.harvard.edu/abs/2018AJ....156...46V}
}

@ARTICLE{Giles2018,
       author = {{Giles}, H.~A.~C. and {Osborn}, H.~P. and {Blanco-Cuaresma}, S. and {Lovis}, C. and {Bayliss}, D. and {Eggenberger}, P. and {Collier Cameron}, A. and {Kristiansen}, M.~H. and {Turner}, O. and {Bouchy}, F. and {Udry}, S.},
        title = "{Transiting planet candidate from K2 with the longest period}",
      journal = {\aap},
         year = 2018,
        month = jul,
       volume = {615},
          eid = {L13},
        pages = {L13},
          doi = {10.1051/0004-6361/201833569},
archivePrefix = {arXiv},
       eprint = {1806.08757},
 primaryClass = {astro-ph.EP},
       adsurl = {https://ui.adsabs.harvard.edu/abs/2018A&A...615L..13G}
}

@ARTICLE{Incha2023,
       author = {{Incha}, Elyse and {Vanderburg}, Andrew and {Jacobs}, Tom and {LaCourse}, Daryll and {Bieryla}, Allyson and {Pass}, Emily and {Howell}, Steve B. and {Berlind}, Perry and {Calkins}, Michael and {Esquerdo}, Gilbert and {Latham}, David W. and {Mann}, Andrew W.},
        title = "{Kepler's last planet discoveries: two new planets and one single-transit candidate from K2 campaign 19}",
      journal = {\mnras},
         year = 2023,
        month = jul,
       volume = {523},
       number = {1},
        pages = {474-487},
          doi = {10.1093/mnras/stad1049},
archivePrefix = {arXiv},
       eprint = {2305.18516},
 primaryClass = {astro-ph.EP},
       adsurl = {https://ui.adsabs.harvard.edu/abs/2023MNRAS.523..474I}
}

@ARTICLE{Gilliland2011,
       author = {{Gilliland}, Ronald L. and {Chaplin}, William J. and {Dunham}, Edward W. and {Argabright}, Vic S. and {Borucki}, William J. and {Basri}, Gibor and {Bryson}, Stephen T. and {Buzasi}, Derek L. and {Caldwell}, Douglas A. and {Elsworth}, Yvonne P. and {Jenkins}, Jon M. and {Koch}, David G. and {Kolodziejczak}, Jeffrey and {Miglio}, Andrea and {van Cleve}, Jeffrey and {Walkowicz}, Lucianne M. and {Welsh}, William F.},
        title = "{Kepler Mission Stellar and Instrument Noise Properties}",
      journal = {\apjs},
         year = 2011,
        month = nov,
       volume = {197},
       number = {1},
          eid = {6},
        pages = {6},
          doi = {10.1088/0067-0049/197/1/6},
archivePrefix = {arXiv},
       eprint = {1107.5207},
 primaryClass = {astro-ph.SR},
       adsurl = {https://ui.adsabs.harvard.edu/abs/2011ApJS..197....6G}
}

@ARTICLE{Luger2016,
       author = {{Luger}, Rodrigo and {Agol}, Eric and {Kruse}, Ethan and {Barnes}, Rory and {Becker}, Andrew and {Foreman-Mackey}, Daniel and {Deming}, Drake},
        title = "{EVEREST: Pixel Level Decorrelation of K2 Light Curves}",
      journal = {\aj},
         year = 2016,
        month = oct,
       volume = {152},
       number = {4},
          eid = {100},
        pages = {100},
          doi = {10.3847/0004-6256/152/4/100},
archivePrefix = {arXiv},
       eprint = {1607.00524},
 primaryClass = {astro-ph.EP},
       adsurl = {https://ui.adsabs.harvard.edu/abs/2016AJ....152..100L}
}

@ARTICLE{Kovacs2002,
       author = {{Kov{\'a}cs}, G. and {Zucker}, S. and {Mazeh}, T.},
        title = "{A box-fitting algorithm in the search for periodic transits}",
      journal = {\aap},
         year = 2002,
        month = aug,
       volume = {391},
        pages = {369-377},
          doi = {10.1051/0004-6361:20020802},
archivePrefix = {arXiv},
       eprint = {astro-ph/0206099},
 primaryClass = {astro-ph},
       adsurl = {https://ui.adsabs.harvard.edu/abs/2002A&A...391..369K}
}

@ARTICLE{Rowe2014,
       author = {{Rowe}, Jason F. and others},
        title = "{Validation of Kepler's Multiple Planet Candidates. III. Light Curve Analysis and Announcement of Hundreds of New Multi-planet Systems}",
      journal = {\apj},
         year = 2014,
        month = mar,
       volume = {784},
       number = {1},
          eid = {45},
        pages = {45},
          doi = {10.1088/0004-637X/784/1/45},
archivePrefix = {arXiv},
       eprint = {1402.6534},
 primaryClass = {astro-ph.EP},
       adsurl = {https://ui.adsabs.harvard.edu/abs/2014ApJ...784...45R}
}

@ARTICLE{Kunimoto2020,
       author = {{Kunimoto}, Michelle and {Matthews}, Jaymie M.},
        title = "{Searching the Entirety of Kepler Data. II. Occurrence Rate Estimates for FGK Stars}",
      journal = {\aj},
         year = 2020,
        month = jun,
       volume = {159},
       number = {6},
          eid = {248},
        pages = {248},
          doi = {10.3847/1538-3881/ab88b0},
archivePrefix = {arXiv},
       eprint = {2004.05296},
 primaryClass = {astro-ph.EP},
       adsurl = {https://ui.adsabs.harvard.edu/abs/2020AJ....159..248K}
}

@ARTICLE{HARPS,
	author = {{Mayor}, M. and {Pepe}, F. and {Queloz}, D. and {Bouchy}, F. and {Rupprecht}, G. and {Lo Curto}, G. and {Avila}, G. and {Benz}, W. and {Bertaux}, J. -L. and {Bonfils}, X. and {Dall}, Th. and {Dekker}, H. and {Delabre}, B. and {Eckert}, W. and {Fleury}, M. and {Gilliotte}, A. and {Gojak}, D. and {Guzman}, J.~C. and {Kohler}, D. and {Lizon}, J. -L. and {Longinotti}, A. and {Lovis}, C. and {Megevand}, D. and {Pasquini}, L. and {Reyes}, J. and {Sivan}, J. -P. and {Sosnowska}, D. and {Soto}, R. and {Udry}, S. and {van Kesteren}, A. and {Weber}, L. and {Weilenmann}, U.},
	title = "{Setting New Standards with HARPS}",
	journal = {The Messenger},
	year = 2003,
	month = dec,
	volume = {114},
	pages = {20-24},
	adsurl = {https://ui.adsabs.harvard.edu/abs/2003Msngr.114...20M}
}

@ARTICLE{LoCurto2010,
       author = {{Lo Curto}, G. and {Mayor}, M. and {Benz}, W. and {Bouchy}, F. and {Lovis}, C. and {Moutou}, C. and {Naef}, D. and {Pepe}, F. and {Queloz}, D. and {Santos}, N.~C. and {Segransan}, D. and {Udry}, S.},
        title = "{The HARPS search for southern extra-solar planets . XXII. Multiple planet systems from the HARPS volume limited sample}",
      journal = {\aap},
         year = 2010,
        month = mar,
       volume = {512},
          eid = {A48},
        pages = {A48},
          doi = {10.1051/0004-6361/200913523},
       adsurl = {https://ui.adsabs.harvard.edu/abs/2010A&A...512A..48L}
}

@ARTICLE{Sousa2011,
       author = {{Sousa}, S.~G. and {Santos}, N.~C. and {Israelian}, G. and {Mayor}, M. and {Udry}, S.},
        title = "{Spectroscopic stellar parameters for 582 FGK stars in the HARPS volume-limited sample. Revising the metallicity-planet correlation}",
      journal = {\aap},
         year = 2011,
        month = sep,
       volume = {533},
          eid = {A141},
        pages = {A141},
          doi = {10.1051/0004-6361/201117699},
archivePrefix = {arXiv},
       eprint = {1108.5279},
 primaryClass = {astro-ph.EP},
       adsurl = {https://ui.adsabs.harvard.edu/abs/2011A&A...533A.141S}
}

@ARTICLE{Coughlin2014,
       author = {{Coughlin}, Jeffrey L. and {Thompson}, Susan E. and {Bryson}, Stephen T. and {Burke}, Christopher J. and {Caldwell}, Douglas A. and {Christiansen}, Jessie L. and {Haas}, Michael R. and {Howell}, Steve B. and {Jenkins}, Jon M. and {Kolodziejczak}, Jeffery J. and {Mullally}, Fergal R. and {Rowe}, Jason F.},
        title = "{Contamination in the Kepler Field. Identification of 685 KOIs as False Positives via Ephemeris Matching Based on Q1-Q12 Data}",
      journal = {\aj},
         year = 2014,
        month = may,
       volume = {147},
       number = {5},
          eid = {119},
        pages = {119},
          doi = {10.1088/0004-6256/147/5/119},
archivePrefix = {arXiv},
       eprint = {1401.1240},
 primaryClass = {astro-ph.IM},
       adsurl = {https://ui.adsabs.harvard.edu/abs/2014AJ....147..119C}
}

@ARTICLE{vartools,
       author = {{Hartman}, J.~D. and {Bakos}, G. {\'A}.},
        title = "{VARTOOLS: A program for analyzing astronomical time-series data}",
      journal = {Astronomy and Computing},
         year = 2016,
        month = oct,
       volume = {17},
        pages = {1-72},
          doi = {10.1016/j.ascom.2016.05.006},
archivePrefix = {arXiv},
       eprint = {1605.06811},
 primaryClass = {astro-ph.IM},
       adsurl = {https://ui.adsabs.harvard.edu/abs/2016A&C....17....1H}
}

@ARTICLE{Brandt2018,
       author = {{Brandt}, Timothy D.},
        title = "{The Hipparcos-Gaia Catalog of Accelerations}",
      journal = {\apjs},
         year = 2018,
        month = dec,
       volume = {239},
       number = {2},
          eid = {31},
        pages = {31},
          doi = {10.3847/1538-4365/aaec06},
archivePrefix = {arXiv},
       eprint = {1811.07283},
 primaryClass = {astro-ph.SR},
       adsurl = {https://ui.adsabs.harvard.edu/abs/2018ApJS..239...31B}
}

@ARTICLE{Brandt2021,
       author = {{Brandt}, Timothy D.},
        title = "{The Hipparcos-Gaia Catalog of Accelerations: Gaia EDR3 Edition}",
      journal = {\apjs},
         year = 2021,
        month = jun,
       volume = {254},
       number = {2},
          eid = {42},
        pages = {42},
          doi = {10.3847/1538-4365/abf93c},
archivePrefix = {arXiv},
       eprint = {2105.11662},
 primaryClass = {astro-ph.GA},
       adsurl = {https://ui.adsabs.harvard.edu/abs/2021ApJS..254...42B}
}

@ARTICLE{Hipparcos,
       author = {{Perryman}, M.~A.~C. and {Lindegren}, L. and {Kovalevsky}, J. and {Hoeg}, E. and {Bastian}, U. and {Bernacca}, P.~L. and {Cr{\'e}z{\'e}}, M. and {Donati}, F. and {Grenon}, M. and {Grewing}, M. and {van Leeuwen}, F. and {van der Marel}, H. and {Mignard}, F. and {Murray}, C.~A. and {Le Poole}, R.~S. and {Schrijver}, H. and {Turon}, C. and {Arenou}, F. and {Froeschl{\'e}}, M. and {Petersen}, C.~S.},
        title = "{The HIPPARCOS Catalogue}",
      journal = {\aap},
         year = 1997,
        month = jul,
       volume = {323},
        pages = {L49-L52},
       adsurl = {https://ui.adsabs.harvard.edu/abs/1997A&A...323L..49P}
}

@ARTICLE{HipparcosNew,
	author = {{\noop{Leeuwen}}{van Leeuwen}, F.},
	title = "{Validation of the new Hipparcos reduction}",
	journal = {\aap},
	year = 2007,
	month = nov,
	volume = {474},
	number = {2},
	pages = {653-664},
	doi = {10.1051/0004-6361:20078357},
	archivePrefix = {arXiv},
	eprint = {0708.1752},
	primaryClass = {astro-ph},
	adsurl = {https://ui.adsabs.harvard.edu/abs/2007A&A...474..653V}
}

@ARTICLE{Gaia,
	author = {{Gaia Collaboration} and others},
	title = "{The Gaia mission}",
	journal = {\aap},
	year = 2016,
	month = nov,
	volume = {595},
	eid = {A1},
	pages = {A1},
	doi = {10.1051/0004-6361/201629272},
	archivePrefix = {arXiv},
	eprint = {1609.04153},
	primaryClass = {astro-ph.IM},
	adsurl = {https://ui.adsabs.harvard.edu/abs/2016A&A...595A...1G}
}

@ARTICLE{Venner2021,
       author = {{Venner}, Alexander and {Vanderburg}, Andrew and {Pearce}, Logan A.},
        title = "{True Masses of the Long-period Companions to HD 92987 and HD 221420 from Hipparcos-Gaia Astrometry}",
      journal = {\aj},
         year = 2021,
        month = jul,
       volume = {162},
       number = {1},
          eid = {12},
        pages = {12},
          doi = {10.3847/1538-3881/abf932},
archivePrefix = {arXiv},
       eprint = {2104.13941},
 primaryClass = {astro-ph.EP},
       adsurl = {https://ui.adsabs.harvard.edu/abs/2021AJ....162...12V}
}

@ARTICLE{ElBadry2021,
       author = {{El-Badry}, Kareem and {Rix}, Hans-Walter and {Heintz}, Tyler M.},
        title = "{A million binaries from Gaia eDR3: sample selection and validation of Gaia parallax uncertainties}",
      journal = {\mnras},
         year = 2021,
        month = sep,
       volume = {506},
       number = {2},
        pages = {2269-2295},
          doi = {10.1093/mnras/stab323},
archivePrefix = {arXiv},
       eprint = {2101.05282},
 primaryClass = {astro-ph.SR},
       adsurl = {https://ui.adsabs.harvard.edu/abs/2021MNRAS.506.2269E}
}

@ARTICLE{Schmitt2016,
       author = {{Schmitt}, Joseph R. and {Tokovinin}, Andrei and {Wang}, Ji and {Fischer}, Debra A. and {Kristiansen}, Martti H. and {LaCourse}, Daryll M. and {Gagliano}, Robert and {Tan}, Arvin Joseff V. and {Schwengeler}, Hans Martin and {Omohundro}, Mark R. and {Venner}, Alexander and {Terentev}, Ivan and {Schmitt}, Allan R. and {Jacobs}, Thomas L. and {Winarski}, Troy and {Sejpka}, Johann and {Jek}, Kian J. and {Boyajian}, Tabetha S. and {Brewer}, John M. and {Ishikawa}, Sascha T. and {Lintott}, Chris and {Lynn}, Stuart and {Schawinski}, Kevin and {Schwamb}, Megan E. and {Weiksnar}, Alex},
        title = "{Planet Hunters. X. Searching for Nearby Neighbors of 75 Planet and Eclipsing Binary Candidates from the K2 Kepler extended mission}",
      journal = {\aj},
         year = 2016,
        month = jun,
       volume = {151},
       number = {6},
          eid = {159},
        pages = {159},
          doi = {10.3847/0004-6256/151/6/159},
archivePrefix = {arXiv},
       eprint = {1603.06945},
 primaryClass = {astro-ph.EP},
       adsurl = {https://ui.adsabs.harvard.edu/abs/2016AJ....151..159S}
}

@ARTICLE{Howell2011,
       author = {{Howell}, Steve B. and {Everett}, Mark E. and {Sherry}, William and {Horch}, Elliott and {Ciardi}, David R.},
        title = "{Speckle Camera Observations for the NASA Kepler Mission Follow-up Program}",
      journal = {\aj},
         year = 2011,
        month = jul,
       volume = {142},
       number = {1},
          eid = {19},
        pages = {19},
          doi = {10.1088/0004-6256/142/1/19},
       adsurl = {https://ui.adsabs.harvard.edu/abs/2011AJ....142...19H}
}

@ARTICLE{Howell2016,
       author = {{Howell}, Steve B. and {Everett}, Mark E. and {Horch}, Elliott P. and {Winters}, Jennifer G. and {Hirsch}, Lea and {Nusdeo}, Dan and {Scott}, Nicholas J.},
        title = "{Speckle Imaging Excludes Low-mass Companions Orbiting the Exoplanet Host Star TRAPPIST-1}",
      journal = {\apjl},
         year = 2016,
        month = sep,
       volume = {829},
       number = {1},
          eid = {L2},
        pages = {L2},
          doi = {10.3847/2041-8205/829/1/L2},
archivePrefix = {arXiv},
       eprint = {1610.05269},
 primaryClass = {astro-ph.EP},
       adsurl = {https://ui.adsabs.harvard.edu/abs/2016ApJ...829L...2H}
}

@ARTICLE{Howell2025,
       author = {{Howell}, Steve B. and {Mart{\'\i}nez-V{\'a}zquez}, Clara E. and {Furlan}, Elise and {Scott}, Nicholas J. and {Matson}, Rachel A. and {Littlefield}, Colin and {Clark}, Catherine A. and {Lester}, Kathryn V. and {Hartman}, Zachary D. and {Ciardi}, David R. and {Deveny}, Sarah J.},
        title = "{Nearly a decade of groundbreaking speckle interferometry at the international Gemini observatory}",
      journal = {Frontiers in Astronomy and Space Sciences},
         year = 2025,
        month = jun,
       volume = {12},
          eid = {1608411},
        pages = {1608411},
          doi = {10.3389/fspas.2025.1608411},
archivePrefix = {arXiv},
       eprint = {2503.10765},
 primaryClass = {astro-ph.IM},
       adsurl = {https://ui.adsabs.harvard.edu/abs/2025FrASS..1208411H}
}

@ARTICLE{Scott2021,
       author = {{Scott}, Nicholas J. and {Howell}, Steve B. and {Gnilka}, Crystal L. and {Stephens}, Andrew W. and {Salinas}, Ricardo and {Matson}, Rachel A. and {Furlan}, Elise and {Horch}, Elliott P. and {Everett}, Mark E. and {Ciardi}, David R. and {Mills}, Dave and {Quigley}, Emmett A.},
        title = "{Twin High-resolution, High-speed Imagers for the Gemini Telescopes: Instrument description and science verification results}",
      journal = {Frontiers in Astronomy and Space Sciences},
         year = 2021,
        month = sep,
       volume = {8},
          eid = {138},
        pages = {138},
          doi = {10.3389/fspas.2021.716560},
       adsurl = {https://ui.adsabs.harvard.edu/abs/2021FrASS...8..138S}
}

@ARTICLE{GaiaDR3,
       author = {{Gaia Collaboration} and others},
        title = "{Gaia Data Release 3. Summary of the content and survey properties}",
      journal = {\aap},
         year = 2023,
        month = jun,
       volume = {674},
          eid = {A1},
        pages = {A1},
          doi = {10.1051/0004-6361/202243940},
archivePrefix = {arXiv},
       eprint = {2208.00211},
 primaryClass = {astro-ph.GA},
       adsurl = {https://ui.adsabs.harvard.edu/abs/2023A&A...674A...1G}
}

@ARTICLE{Soubiran2018,
       author = {{Soubiran}, C. and {Jasniewicz}, G. and {Chemin}, L. and {Zurbach}, C. and {Brouillet}, N. and {Panuzzo}, P. and {Sartoretti}, P. and {Katz}, D. and {Le Campion}, J. -F. and {Marchal}, O. and {Hestroffer}, D. and {Th{\'e}venin}, F. and {Crifo}, F. and {Udry}, S. and {Cropper}, M. and {Seabroke}, G. and {Viala}, Y. and {Benson}, K. and {Blomme}, R. and {Jean-Antoine}, A. and {Huckle}, H. and {Smith}, M. and {Baker}, S.~G. and {Damerdji}, Y. and {Dolding}, C. and {Fr{\'e}mat}, Y. and {Gosset}, E. and {Guerrier}, A. and {Guy}, L.~P. and {Haigron}, R. and {Jan{\ss}en}, K. and {Plum}, G. and {Fabre}, C. and {Lasne}, Y. and {Pailler}, F. and {Panem}, C. and {Riclet}, F. and {Royer}, F. and {Tauran}, G. and {Zwitter}, T. and {Gueguen}, A. and {Turon}, C.},
        title = "{Gaia Data Release 2. The catalogue of radial velocity standard stars}",
      journal = {\aap},
         year = 2018,
        month = aug,
       volume = {616},
          eid = {A7},
        pages = {A7},
          doi = {10.1051/0004-6361/201832795},
archivePrefix = {arXiv},
       eprint = {1804.09370},
 primaryClass = {astro-ph.GA},
       adsurl = {https://ui.adsabs.harvard.edu/abs/2018A&A...616A...7S}
}

@ARTICLE{Gray2006,
       author = {{Gray}, R.~O. and {Corbally}, C.~J. and {Garrison}, R.~F. and {McFadden}, M.~T. and {Bubar}, E.~J. and {McGahee}, C.~E. and {O'Donoghue}, A.~A. and {Knox}, E.~R.},
        title = "{Contributions to the Nearby Stars (NStars) Project: Spectroscopy of Stars Earlier than M0 within 40 pc-The Southern Sample}",
      journal = {\aj},
         year = 2006,
        month = jul,
       volume = {132},
       number = {1},
        pages = {161-170},
          doi = {10.1086/504637},
archivePrefix = {arXiv},
       eprint = {astro-ph/0603770},
 primaryClass = {astro-ph},
       adsurl = {https://ui.adsabs.harvard.edu/abs/2006AJ....132..161G}
}

@ARTICLE{GomesSilva2021,
       author = {{Gomes da Silva}, J. and {Santos}, N.~C. and {Adibekyan}, V. and {Sousa}, S.~G. and {Campante}, T.~L. and {Figueira}, P. and {Bossini}, D. and {Delgado-Mena}, E. and {Monteiro}, M.~J.~P.~F.~G. and {de Laverny}, P. and {Recio-Blanco}, A. and {Lovis}, C.},
        title = "{Stellar chromospheric activity of 1674 FGK stars from the AMBRE-HARPS sample. I. A catalogue of homogeneous chromospheric activity}",
      journal = {\aap},
         year = 2021,
        month = feb,
       volume = {646},
          eid = {A77},
        pages = {A77},
          doi = {10.1051/0004-6361/202039765},
archivePrefix = {arXiv},
       eprint = {2012.10199},
 primaryClass = {astro-ph.SR},
       adsurl = {https://ui.adsabs.harvard.edu/abs/2021A&A...646A..77G}
}

@ARTICLE{Johnson1987,
       author = {{Johnson}, Dean R.~H. and {Soderblom}, David R.},
        title = "{Calculating Galactic Space Velocities and Their Uncertainties, with an Application to the Ursa Major Group}",
      journal = {\aj},
         year = 1987,
        month = apr,
       volume = {93},
        pages = {864},
          doi = {10.1086/114370},
       adsurl = {https://ui.adsabs.harvard.edu/abs/1987AJ.....93..864J}
}

@ARTICLE{Schonrich2010,
       author = {{Sch{\"o}nrich}, Ralph and {Binney}, James and {Dehnen}, Walter},
        title = "{Local kinematics and the local standard of rest}",
      journal = {\mnras},
         year = 2010,
        month = apr,
       volume = {403},
       number = {4},
        pages = {1829-1833},
          doi = {10.1111/j.1365-2966.2010.16253.x},
archivePrefix = {arXiv},
       eprint = {0912.3693},
 primaryClass = {astro-ph.GA},
       adsurl = {https://ui.adsabs.harvard.edu/abs/2010MNRAS.403.1829S}
}

@ARTICLE{Adibekyan2012,
       author = {{Adibekyan}, V. Zh. and {Sousa}, S.~G. and {Santos}, N.~C. and {Delgado Mena}, E. and {Gonz{\'a}lez Hern{\'a}ndez}, J.~I. and {Israelian}, G. and {Mayor}, M. and {Khachatryan}, G.},
        title = "{Chemical abundances of 1111 FGK stars from the HARPS GTO planet search program. Galactic stellar populations and planets}",
      journal = {\aap},
         year = 2012,
        month = sep,
       volume = {545},
          eid = {A32},
        pages = {A32},
          doi = {10.1051/0004-6361/201219401},
archivePrefix = {arXiv},
       eprint = {1207.2388},
 primaryClass = {astro-ph.EP},
       adsurl = {https://ui.adsabs.harvard.edu/abs/2012A&A...545A..32A}
}

@ARTICLE{Bensby2003,
       author = {{Bensby}, T. and {Feltzing}, S. and {Lundstr{\"o}m}, I.},
        title = "{Elemental abundance trends in the Galactic thin and thick disks as traced by nearby F and G dwarf stars}",
      journal = {\aap},
         year = 2003,
        month = nov,
       volume = {410},
        pages = {527-551},
          doi = {10.1051/0004-6361:20031213},
       adsurl = {https://ui.adsabs.harvard.edu/abs/2003A&A...410..527B}
}

@ARTICLE{Robin2003,
       author = {{Robin}, A.~C. and {Reyl{\'e}}, C. and {Derri{\`e}re}, S. and {Picaud}, S.},
        title = "{A synthetic view on structure and evolution of the Milky Way}",
      journal = {\aap},
         year = 2003,
        month = oct,
       volume = {409},
        pages = {523-540},
          doi = {10.1051/0004-6361:20031117},
       adsurl = {https://ui.adsabs.harvard.edu/abs/2003A&A...409..523R}
}

@ARTICLE{Fuhrmann2011,
       author = {{Fuhrmann}, Klaus},
        title = "{Nearby stars of the Galactic disc and halo - V}",
      journal = {\mnras},
         year = 2011,
        month = jul,
       volume = {414},
       number = {4},
        pages = {2893-2922},
          doi = {10.1111/j.1365-2966.2011.18476.x},
       adsurl = {https://ui.adsabs.harvard.edu/abs/2011MNRAS.414.2893F}
}

@ARTICLE{Xiang2017,
       author = {{Xiang}, Maosheng and {Liu}, Xiaowei and {Shi}, Jianrong and {Yuan}, Haibo and {Huang}, Yang and {Chen}, Bingqiu and {Wang}, Chun and {Tian}, Zhijia and {Wu}, Yaqian and {Yang}, Yong and {Zhang}, Huawei and {Huo}, Zhiying and {Ren}, Juanjuan},
        title = "{The Ages and Masses of a Million Galactic-disk Main-sequence Turnoff and Subgiant Stars from the LAMOST Galactic Spectroscopic Surveys}",
      journal = {\apjs},
         year = 2017,
        month = sep,
       volume = {232},
       number = {1},
          eid = {2},
        pages = {2},
          doi = {10.3847/1538-4365/aa80e4},
archivePrefix = {arXiv},
       eprint = {1707.06236},
 primaryClass = {astro-ph.SR},
       adsurl = {https://ui.adsabs.harvard.edu/abs/2017ApJS..232....2X}
}

@ARTICLE{AlmeidaFernandes2018,
       author = {{Almeida-Fernandes}, F. and {Rocha-Pinto}, H.~J.},
        title = "{A method to estimate stellar ages from kinematical data}",
      journal = {\mnras},
         year = 2018,
        month = may,
       volume = {476},
       number = {1},
        pages = {184-197},
          doi = {10.1093/mnras/sty119},
archivePrefix = {arXiv},
       eprint = {1801.04046},
 primaryClass = {astro-ph.SR},
       adsurl = {https://ui.adsabs.harvard.edu/abs/2018MNRAS.476..184A}
}

@ARTICLE{Veyette2018,
       author = {{Veyette}, Mark J. and {Muirhead}, Philip S.},
        title = "{Chemo-kinematic Ages of Eccentric-planet-hosting M Dwarf Stars}",
      journal = {\apj},
         year = 2018,
        month = aug,
       volume = {863},
       number = {2},
          eid = {166},
        pages = {166},
          doi = {10.3847/1538-4357/aad40e},
archivePrefix = {arXiv},
       eprint = {1807.06017},
 primaryClass = {astro-ph.SR},
       adsurl = {https://ui.adsabs.harvard.edu/abs/2018ApJ...863..166V}
}

@ARTICLE{Venner2024,
       author = {{Venner}, Alexander and {An}, Qier and {Huang}, Chelsea X. and {Brandt}, Timothy D. and {Wittenmyer}, Robert A. and {Vanderburg}, Andrew},
        title = "{HD 28185 revisited: an outer planet, instead of a brown dwarf, on a Saturn-like orbit}",
      journal = {\mnras},
         year = 2024,
        month = nov,
       volume = {535},
       number = {1},
        pages = {90-106},
          doi = {10.1093/mnras/stae2336},
archivePrefix = {arXiv},
       eprint = {2410.14218},
 primaryClass = {astro-ph.EP},
       adsurl = {https://ui.adsabs.harvard.edu/abs/2024MNRAS.535...90V}
}

@ARTICLE{Dotter2016,
       author = {{Dotter}, Aaron},
        title = "{MESA Isochrones and Stellar Tracks (MIST) 0: Methods for the Construction of Stellar Isochrones}",
      journal = {\apjs},
         year = 2016,
        month = jan,
       volume = {222},
       number = {1},
          eid = {8},
        pages = {8},
          doi = {10.3847/0067-0049/222/1/8},
archivePrefix = {arXiv},
       eprint = {1601.05144},
 primaryClass = {astro-ph.SR},
       adsurl = {https://ui.adsabs.harvard.edu/abs/2016ApJS..222....8D}
}

@ARTICLE{Choi2016,
       author = {{Choi}, Jieun and {Dotter}, Aaron and {Conroy}, Charlie and {Cantiello}, Matteo and {Paxton}, Bill and {Johnson}, Benjamin D.},
        title = "{Mesa Isochrones and Stellar Tracks (MIST). I. Solar-scaled Models}",
      journal = {\apj},
         year = 2016,
        month = jun,
       volume = {823},
       number = {2},
          eid = {102},
        pages = {102},
          doi = {10.3847/0004-637X/823/2/102},
archivePrefix = {arXiv},
       eprint = {1604.08592},
 primaryClass = {astro-ph.SR},
       adsurl = {https://ui.adsabs.harvard.edu/abs/2016ApJ...823..102C}
}

@ARTICLE{Tycho2,
	author = {{H{\o}g}, E. and {Fabricius}, C. and {Makarov}, V.~V. and {Urban}, S. and
	{Corbin}, T. and {Wycoff}, G. and {Bastian}, U. and {Schwekendiek}, P. and
	{Wicenec}, A.},
	title = "{The Tycho-2 catalogue of the 2.5 million brightest stars}",
	journal = {\aap},
	year = 2000,
	month = mar,
	volume = {355},
	pages = {L27-L30},
	adsurl = {https://ui.adsabs.harvard.edu/abs/2000A&A...355L..27H}
}

@ARTICLE{2MASS,
	author = {{Skrutskie}, M.~F. and {Cutri}, R.~M. and {Stiening}, R. and {Weinberg}, M.~D. and {Schneider}, S. and {Carpenter}, J.~M. and {Beichman}, C. and {Capps}, R. and {Chester}, T. and {Elias}, J. and {Huchra}, J. and {Liebert}, J. and {Lonsdale}, C. and {Monet}, D.~G. and {Price}, S. and {Seitzer}, P. and {Jarrett}, T. and {Kirkpatrick}, J.~D. and {Gizis}, J.~E. and {Howard}, E. and {Evans}, T. and {Fowler}, J. and {Fullmer}, L. and {Hurt}, R. and {Light}, R. and {Kopan}, E.~L. and {Marsh}, K.~A. and {McCallon}, H.~L. and {Tam}, R. and {Van Dyk}, S. and {Wheelock}, S.},
	title = "{The Two Micron All Sky Survey (2MASS)}",
	journal = {\aj},
	year = 2006,
	month = feb,
	volume = {131},
	number = {2},
	pages = {1163-1183},
	doi = {10.1086/498708},
	adsurl = {https://ui.adsabs.harvard.edu/abs/2006AJ....131.1163S}
}

@ARTICLE{WISE,
	author = {{Cutri}, R.~M. and {et al.}},
	title = "{VizieR Online Data Catalog: WISE All-Sky Data Release (Cutri+ 2012)}",
	journal = {VizieR Online Data Catalog},
	year = 2012,
	month = apr,
	eid = {II/311},
	pages = {II/311},
	adsurl = {https://ui.adsabs.harvard.edu/abs/2012yCat.2311....0C}
}

@MISC{Koposov2021,
       author = {{Koposov}, Sergey},
        title = "{segasai/minimint: Minimint 0.3.0}",
 howpublished = {Zenodo},
         year = 2021,
        month = oct,
          eid = {10.5281/zenodo.5610692},
          doi = {10.5281/zenodo.5610692},
      version = {v0.3.0},
    publisher = {Zenodo},
       adsurl = {https://ui.adsabs.harvard.edu/abs/2021zndo...5610692K}
}

@ARTICLE{Tayar2022,
       author = {{Tayar}, Jamie and {Claytor}, Zachary R. and {Huber}, Daniel and {van Saders}, Jennifer},
        title = "{A Guide to Realistic Uncertainties on the Fundamental Properties of Solar-type Exoplanet Host Stars}",
      journal = {\apj},
         year = 2022,
        month = mar,
       volume = {927},
       number = {1},
          eid = {31},
        pages = {31},
          doi = {10.3847/1538-4357/ac4bbc},
archivePrefix = {arXiv},
       eprint = {2012.07957},
 primaryClass = {astro-ph.EP},
       adsurl = {https://ui.adsabs.harvard.edu/abs/2022ApJ...927...31T}
}

@ARTICLE{Brandt2019,
       author = {{Brandt}, Timothy D. and {Dupuy}, Trent J. and {Bowler}, Brendan P.},
        title = "{Precise Dynamical Masses of Directly Imaged Companions from Relative Astrometry, Radial Velocities, and Hipparcos-Gaia DR2 Accelerations}",
      journal = {\aj},
         year = 2019,
        month = oct,
       volume = {158},
       number = {4},
          eid = {140},
        pages = {140},
          doi = {10.3847/1538-3881/ab04a8},
archivePrefix = {arXiv},
       eprint = {1811.07285},
 primaryClass = {astro-ph.SR},
       adsurl = {https://ui.adsabs.harvard.edu/abs/2019AJ....158..140B}
}

@ARTICLE{Bowler2021,
       author = {{Bowler}, Brendan P. and {Cochran}, William D. and {Endl}, Michael and {Franson}, Kyle and {Brandt}, Timothy D. and {Dupuy}, Trent J. and {MacQueen}, Phillip J. and {Kratter}, Kaitlin M. and {Mawet}, Dimitri and {Ruane}, Garreth},
        title = "{The McDonald Accelerating Stars Survey (MASS): White Dwarf Companions Accelerating the Sun-like Stars 12 Psc and HD 159062}",
      journal = {\aj},
         year = 2021,
        month = mar,
       volume = {161},
       number = {3},
          eid = {106},
        pages = {106},
          doi = {10.3847/1538-3881/abd243},
archivePrefix = {arXiv},
       eprint = {2012.04847},
 primaryClass = {astro-ph.SR},
       adsurl = {https://ui.adsabs.harvard.edu/abs/2021AJ....161..106B}
}

@ARTICLE{Vanderburg2019,
       author = {{Vanderburg}, Andrew and {Huang}, Chelsea X. and {Rodriguez}, Joseph E. and {Becker}, Juliette C. and {Ricker}, George R. and {Vanderspek}, Roland K. and {Latham}, David W. and {Seager}, Sara and {Winn}, Joshua N. and {Jenkins}, Jon M. and {Addison}, Brett and {Bieryla}, Allyson and {Brice{\~n}o}, Cesar and {Bowler}, Brendan P. and {Brown}, Timothy M. and {Burke}, Christopher J. and {Burt}, Jennifer A. and {Caldwell}, Douglas A. and {Clark}, Jake T. and {Crossfield}, Ian and {Dittmann}, Jason A. and {Dynes}, Scott and {Fulton}, Benjamin J. and {Guerrero}, Natalia and {Harbeck}, Daniel and {Horner}, Jonathan and {Kane}, Stephen R. and {Kielkopf}, John and {Kraus}, Adam L. and {Kreidberg}, Laura and {Law}, Nicolas and {Mann}, Andrew W. and {Mengel}, Matthew W. and {Morton}, Timothy D. and {Okumura}, Jack and {Pearce}, Logan A. and {Plavchan}, Peter and {Quinn}, Samuel N. and {Rabus}, Markus and {Rose}, Mark E. and {Rowden}, Pam and {Shporer}, Avi and {Siverd}, Robert J. and {Smith}, Jeffrey C. and {Stassun}, Keivan and {Tinney}, C.~G. and {Wittenmyer}, Rob and {Wright}, Duncan J. and {Zhang}, Hui and {Zhou}, George and {Ziegler}, Carl A.},
        title = "{TESS Spots a Compact System of Super-Earths around the Naked-eye Star HR 858}",
      journal = {\apjl},
         year = 2019,
        month = aug,
       volume = {881},
       number = {1},
          eid = {L19},
        pages = {L19},
          doi = {10.3847/2041-8213/ab322d},
archivePrefix = {arXiv},
       eprint = {1905.05193},
 primaryClass = {astro-ph.EP},
       adsurl = {https://ui.adsabs.harvard.edu/abs/2019ApJ...881L..19V}
}

@ARTICLE{MandelAgol2002,
       author = {{Mandel}, Kaisey and {Agol}, Eric},
        title = "{Analytic Light Curves for Planetary Transit Searches}",
      journal = {\apjl},
         year = 2002,
        month = dec,
       volume = {580},
       number = {2},
        pages = {L171-L175},
          doi = {10.1086/345520},
archivePrefix = {arXiv},
       eprint = {astro-ph/0210099},
 primaryClass = {astro-ph},
       adsurl = {https://ui.adsabs.harvard.edu/abs/2002ApJ...580L.171M}
}

@ARTICLE{batman,
       author = {{Kreidberg}, Laura},
        title = "{batman: BAsic Transit Model cAlculatioN in Python}",
      journal = {\pasp},
         year = 2015,
        month = nov,
       volume = {127},
       number = {957},
        pages = {1161},
          doi = {10.1086/683602},
archivePrefix = {arXiv},
       eprint = {1507.08285},
 primaryClass = {astro-ph.EP},
       adsurl = {https://ui.adsabs.harvard.edu/abs/2015PASP..127.1161K}
}

@ARTICLE{emcee,
       author = {{Foreman-Mackey}, Daniel and {Hogg}, David W. and {Lang}, Dustin and {Goodman}, Jonathan},
        title = "{emcee: The MCMC Hammer}",
      journal = {\pasp},
         year = 2013,
        month = mar,
       volume = {125},
       number = {925},
        pages = {306},
          doi = {10.1086/670067},
archivePrefix = {arXiv},
       eprint = {1202.3665},
 primaryClass = {astro-ph.IM},
       adsurl = {https://ui.adsabs.harvard.edu/abs/2013PASP..125..306F}
}

@ARTICLE{Claret2018,
       author = {{Claret}, Antonio},
        title = "{A new method to compute limb-darkening coefficients for stellar atmosphere models with spherical symmetry: the space missions TESS, Kepler, CoRoT, and MOST}",
      journal = {\aap},
         year = 2018,
        month = oct,
       volume = {618},
          eid = {A20},
        pages = {A20},
          doi = {10.1051/0004-6361/201833060},
archivePrefix = {arXiv},
       eprint = {1804.10135},
 primaryClass = {astro-ph.SR},
       adsurl = {https://ui.adsabs.harvard.edu/abs/2018A&A...618A..20C}
}

@ARTICLE{Kipping2018,
       author = {{Kipping}, David},
        title = "{The Orbital Period Prior for Single Transits}",
      journal = {Research Notes of the American Astronomical Society},
         year = 2018,
        month = dec,
       volume = {2},
       number = {4},
          eid = {223},
        pages = {223},
          doi = {10.3847/2515-5172/aaf50c},
       adsurl = {https://ui.adsabs.harvard.edu/abs/2018RNAAS...2..223K}
}

@ARTICLE{Becker2019,
       author = {{Becker}, Juliette C. and {Vanderburg}, Andrew and {Rodriguez}, Joseph E. and {Omohundro}, Mark and {Adams}, Fred C. and {Stassun}, Keivan G. and {Yao}, Xinyu and {Hartman}, Joel and {Pepper}, Joshua and {Bakos}, Gaspar and {Barentsen}, Geert and {Beatty}, Thomas G. and {Bhatti}, Waqas and {Chontos}, Ashley and {Collier Cameron}, Andrew and {Hellier}, Coel and {Huber}, Daniel and {James}, David and {Kuhn}, Rudolf B. and {Lund}, Michael B. and {Pollacco}, Don and {Siverd}, Robert J. and {Stevens}, Daniel J. and {Cardoso}, Jos{\'e} Vin{\'\i}cius de Miranda and {West}, Richard},
        title = "{A Discrete Set of Possible Transit Ephemerides for Two Long-period Gas Giants Orbiting HIP 41378}",
      journal = {\aj},
         year = 2019,
        month = jan,
       volume = {157},
       number = {1},
          eid = {19},
        pages = {19},
          doi = {10.3847/1538-3881/aaf0a2},
archivePrefix = {arXiv},
       eprint = {1809.10688},
 primaryClass = {astro-ph.EP},
       adsurl = {https://ui.adsabs.harvard.edu/abs/2019AJ....157...19B}
}

@ARTICLE{Dholakia2020,
       author = {{Dholakia}, S. and {Dholakia}, S. and {Mayo}, Andrew W. and {Dressing}, Courtney D.},
        title = "{Constraining Orbital Periods from Nonconsecutive Observations: Period Estimates for Long-period Planets in Six Systems Observed by K2 During Multiple Campaigns}",
      journal = {\aj},
         year = 2020,
        month = mar,
       volume = {159},
       number = {3},
          eid = {93},
        pages = {93},
          doi = {10.3847/1538-3881/ab594c},
archivePrefix = {arXiv},
       eprint = {1912.04287},
 primaryClass = {astro-ph.EP},
       adsurl = {https://ui.adsabs.harvard.edu/abs/2020AJ....159...93D}
}

@ARTICLE{Kane2012,
       author = {{Kane}, Stephen R. and {Ciardi}, David R. and {Gelino}, Dawn M. and {von Braun}, Kaspar},
        title = "{The exoplanet eccentricity distribution from Kepler planet candidates}",
      journal = {\mnras},
         year = 2012,
        month = sep,
       volume = {425},
       number = {1},
        pages = {757-762},
          doi = {10.1111/j.1365-2966.2012.21627.x},
archivePrefix = {arXiv},
       eprint = {1203.1631},
 primaryClass = {astro-ph.EP},
       adsurl = {https://ui.adsabs.harvard.edu/abs/2012MNRAS.425..757K}
}

@ARTICLE{Kipping2025,
       author = {{Kipping}, David and {Solano-Oropeza}, Diana and {Yahalomi}, Daniel A. and {Li}, Madison and {Poddar}, Avishi and {Zhang}, Xunhe},
        title = "{Near-circular orbits for planets around M/K-type stars with Earth-like sizes and instellations}",
      journal = {arXiv e-prints},
         year = 2025,
        month = jan,
          eid = {arXiv:2501.10571},
        pages = {arXiv:2501.10571},
          doi = {10.48550/arXiv.2501.10571},
archivePrefix = {arXiv},
       eprint = {2501.10571},
 primaryClass = {astro-ph.EP},
       adsurl = {https://ui.adsabs.harvard.edu/abs/2025arXiv250110571K}
}

@ARTICLE{Osborn2016,
       author = {{Osborn}, H.~P. and {Armstrong}, D.~J. and {Brown}, D.~J.~A. and {McCormac}, J. and {Doyle}, A.~P. and {Louden}, T.~M. and {Kirk}, J. and {Spake}, J.~J. and {Lam}, K.~W.~F. and {Walker}, S.~R. and {Faedi}, F. and {Pollacco}, D.~L.},
        title = "{Single transit candidates from K2: detection and period estimation}",
      journal = {\mnras},
         year = 2016,
        month = apr,
       volume = {457},
       number = {3},
        pages = {2273-2286},
          doi = {10.1093/mnras/stw137},
archivePrefix = {arXiv},
       eprint = {1512.03722},
 primaryClass = {astro-ph.EP},
       adsurl = {https://ui.adsabs.harvard.edu/abs/2016MNRAS.457.2273O}
}

@ARTICLE{Sandford2019,
       author = {{Sandford}, Emily and {Espinoza}, N{\'e}stor and {Brahm}, Rafael and {Jord{\'a}n}, Andr{\'e}s},
        title = "{Estimation of singly transiting K2 planet periods with Gaia parallaxes}",
      journal = {\mnras},
         year = 2019,
        month = nov,
       volume = {489},
       number = {3},
        pages = {3149-3161},
          doi = {10.1093/mnras/stz2348},
archivePrefix = {arXiv},
       eprint = {1908.08548},
 primaryClass = {astro-ph.EP},
       adsurl = {https://ui.adsabs.harvard.edu/abs/2019MNRAS.489.3149S}
}

@ARTICLE{Gillon2017,
       author = {{Gillon}, Micha{\"e}l and {Triaud}, Amaury H.~M.~J. and {Demory}, Brice-Olivier and {Jehin}, Emmanu{\"e}l and {Agol}, Eric and {Deck}, Katherine M. and {Lederer}, Susan M. and {de Wit}, Julien and {Burdanov}, Artem and {Ingalls}, James G. and {Bolmont}, Emeline and {Leconte}, Jeremy and {Raymond}, Sean N. and {Selsis}, Franck and {Turbet}, Martin and {Barkaoui}, Khalid and {Burgasser}, Adam and {Burleigh}, Matthew R. and {Carey}, Sean J. and {Chaushev}, Aleksander and {Copperwheat}, Chris M. and {Delrez}, Laetitia and {Fernandes}, Catarina S. and {Holdsworth}, Daniel L. and {Kotze}, Enrico J. and {Van Grootel}, Val{\'e}rie and {Almleaky}, Yaseen and {Benkhaldoun}, Zouhair and {Magain}, Pierre and {Queloz}, Didier},
        title = "{Seven temperate terrestrial planets around the nearby ultracool dwarf star TRAPPIST-1}",
      journal = {\nat},
         year = 2017,
        month = feb,
       volume = {542},
       number = {7642},
        pages = {456-460},
          doi = {10.1038/nature21360},
archivePrefix = {arXiv},
       eprint = {1703.01424},
 primaryClass = {astro-ph.EP},
       adsurl = {https://ui.adsabs.harvard.edu/abs/2017Natur.542..456G}
}

@ARTICLE{Luger2017,
       author = {{Luger}, Rodrigo and {Sestovic}, Marko and {Kruse}, Ethan and {Grimm}, Simon L. and {Demory}, Brice-Olivier and {Agol}, Eric and {Bolmont}, Emeline and {Fabrycky}, Daniel and {Fernandes}, Catarina S. and {Van Grootel}, Val{\'e}rie and {Burgasser}, Adam and {Gillon}, Micha{\"e}l and {Ingalls}, James G. and {Jehin}, Emmanu{\"e}l and {Raymond}, Sean N. and {Selsis}, Franck and {Triaud}, Amaury H.~M.~J. and {Barclay}, Thomas and {Barentsen}, Geert and {Howell}, Steve B. and {Delrez}, Laetitia and {de Wit}, Julien and {Foreman-Mackey}, Daniel and {Holdsworth}, Daniel L. and {Leconte}, J{\'e}r{\'e}my and {Lederer}, Susan and {Turbet}, Martin and {Almleaky}, Yaseen and {Benkhaldoun}, Zouhair and {Magain}, Pierre and {Morris}, Brett M. and {Heng}, Kevin and {Queloz}, Didier},
        title = "{A seven-planet resonant chain in TRAPPIST-1}",
      journal = {Nature Astronomy},
         year = 2017,
        month = jun,
       volume = {1},
          eid = {0129},
        pages = {0129},
          doi = {10.1038/s41550-017-0129},
archivePrefix = {arXiv},
       eprint = {1703.04166},
 primaryClass = {astro-ph.EP},
       adsurl = {https://ui.adsabs.harvard.edu/abs/2017NatAs...1E.129L}
}

@ARTICLE{Grether2006,
       author = {{Grether}, Daniel and {Lineweaver}, Charles H.},
        title = "{How Dry is the Brown Dwarf Desert? Quantifying the Relative Number of Planets, Brown Dwarfs, and Stellar Companions around Nearby Sun-like Stars}",
      journal = {\apj},
         year = 2006,
        month = apr,
       volume = {640},
       number = {2},
        pages = {1051-1062},
          doi = {10.1086/500161},
archivePrefix = {arXiv},
       eprint = {astro-ph/0412356},
 primaryClass = {astro-ph},
       adsurl = {https://ui.adsabs.harvard.edu/abs/2006ApJ...640.1051G}
}

@ARTICLE{Zhu2018,
       author = {{Zhu}, Wei and {Wu}, Yanqin},
        title = "{The Super Earth-Cold Jupiter Relations}",
      journal = {\aj},
         year = 2018,
        month = sep,
       volume = {156},
       number = {3},
          eid = {92},
        pages = {92},
          doi = {10.3847/1538-3881/aad22a},
archivePrefix = {arXiv},
       eprint = {1805.02660},
 primaryClass = {astro-ph.EP},
       adsurl = {https://ui.adsabs.harvard.edu/abs/2018AJ....156...92Z}
}

@ARTICLE{Bryan2019,
       author = {{Bryan}, Marta L. and {Knutson}, Heather A. and {Lee}, Eve J. and {Fulton}, B.~J. and {Batygin}, Konstantin and {Ngo}, Henry and {Meshkat}, Tiffany},
        title = "{An Excess of Jupiter Analogs in Super-Earth Systems}",
      journal = {\aj},
         year = 2019,
        month = feb,
       volume = {157},
       number = {2},
          eid = {52},
        pages = {52},
          doi = {10.3847/1538-3881/aaf57f},
archivePrefix = {arXiv},
       eprint = {1806.08799},
 primaryClass = {astro-ph.EP},
       adsurl = {https://ui.adsabs.harvard.edu/abs/2019AJ....157...52B}
}

@ARTICLE{Rogers2015,
       author = {{Rogers}, Leslie A.},
        title = "{Most 1.6 Earth-radius Planets are Not Rocky}",
      journal = {\apj},
         year = 2015,
        month = mar,
       volume = {801},
       number = {1},
          eid = {41},
        pages = {41},
          doi = {10.1088/0004-637X/801/1/41},
archivePrefix = {arXiv},
       eprint = {1407.4457},
 primaryClass = {astro-ph.EP},
       adsurl = {https://ui.adsabs.harvard.edu/abs/2015ApJ...801...41R}
}

@ARTICLE{Kopparapu2013,
       author = {{Kopparapu}, Ravi Kumar and {Ramirez}, Ramses and {Kasting}, James F. and {Eymet}, Vincent and {Robinson}, Tyler D. and {Mahadevan}, Suvrath and {Terrien}, Ryan C. and {Domagal-Goldman}, Shawn and {Meadows}, Victoria and {Deshpande}, Rohit},
        title = "{Habitable Zones around Main-sequence Stars: New Estimates}",
      journal = {\apj},
         year = 2013,
        month = mar,
       volume = {765},
       number = {2},
          eid = {131},
        pages = {131},
          doi = {10.1088/0004-637X/765/2/131},
archivePrefix = {arXiv},
       eprint = {1301.6674},
 primaryClass = {astro-ph.EP},
       adsurl = {https://ui.adsabs.harvard.edu/abs/2013ApJ...765..131K}
}

@ARTICLE{Zahnle2017,
       author = {{Zahnle}, Kevin J. and {Catling}, David C.},
        title = "{The Cosmic Shoreline: The Evidence that Escape Determines which Planets Have Atmospheres, and what this May Mean for Proxima Centauri B}",
      journal = {\apj},
         year = 2017,
        month = jul,
       volume = {843},
       number = {2},
          eid = {122},
        pages = {122},
          doi = {10.3847/1538-4357/aa7846},
archivePrefix = {arXiv},
       eprint = {1702.03386},
 primaryClass = {astro-ph.EP},
       adsurl = {https://ui.adsabs.harvard.edu/abs/2017ApJ...843..122Z}
}

@ARTICLE{Pass2025,
       author = {{Pass}, Emily K. and {Charbonneau}, David and {Vanderburg}, Andrew},
        title = "{The Receding Cosmic Shoreline of Mid-to-late M Dwarfs: Measurements of Active Lifetimes Worsen Challenges for Atmosphere Retention by Rocky Exoplanets}",
      journal = {\apjl},
         year = 2025,
        month = jun,
       volume = {986},
       number = {1},
          eid = {L3},
        pages = {L3},
          doi = {10.3847/2041-8213/adda39},
       adsurl = {https://ui.adsabs.harvard.edu/abs/2025ApJ...986L...3P}
}

@ARTICLE{Cuntz2016,
       author = {{Cuntz}, M. and {Guinan}, E.~F.},
        title = "{About Exobiology: The Case for Dwarf K Stars}",
      journal = {\apj},
         year = 2016,
        month = aug,
       volume = {827},
       number = {1},
          eid = {79},
        pages = {79},
          doi = {10.3847/0004-637X/827/1/79},
archivePrefix = {arXiv},
       eprint = {1606.09580},
 primaryClass = {astro-ph.SR},
       adsurl = {https://ui.adsabs.harvard.edu/abs/2016ApJ...827...79C}
}

@ARTICLE{Arney2019,
       author = {{Arney}, Giada N.},
        title = "{The K Dwarf Advantage for Biosignatures on Directly Imaged Exoplanets}",
      journal = {\apjl},
         year = 2019,
        month = mar,
       volume = {873},
       number = {1},
          eid = {L7},
        pages = {L7},
          doi = {10.3847/2041-8213/ab0651},
archivePrefix = {arXiv},
       eprint = {2001.10458},
 primaryClass = {astro-ph.EP},
       adsurl = {https://ui.adsabs.harvard.edu/abs/2019ApJ...873L...7A}
}

@ARTICLE{Bolmont2014,
       author = {{Bolmont}, Emeline and {Raymond}, Sean N. and {von Paris}, Philip and {Selsis}, Franck and {Hersant}, Franck and {Quintana}, Elisa V. and {Barclay}, Thomas},
        title = "{Formation, Tidal Evolution, and Habitability of the Kepler-186 System}",
      journal = {\apj},
         year = 2014,
        month = sep,
       volume = {793},
       number = {1},
          eid = {3},
        pages = {3},
          doi = {10.1088/0004-637X/793/1/3},
archivePrefix = {arXiv},
       eprint = {1404.4368},
 primaryClass = {astro-ph.EP},
       adsurl = {https://ui.adsabs.harvard.edu/abs/2014ApJ...793....3B}
}

@ARTICLE{Kasting1993,
       author = {{Kasting}, James F. and {Whitmire}, Daniel P. and {Reynolds}, Ray T.},
        title = "{Habitable Zones around Main Sequence Stars}",
      journal = {\icarus},
         year = 1993,
        month = jan,
       volume = {101},
       number = {1},
        pages = {108-128},
          doi = {10.1006/icar.1993.1010},
       adsurl = {https://ui.adsabs.harvard.edu/abs/1993Icar..101..108K}
}

@BOOK{Astro2020,
       author = "{National Academies of Sciences, Engineering, and Medicine}",
        title = "{Pathways to Discovery in Astronomy and Astrophysics for the 2020s}",
    publisher = "{The National Academies Press}",
         year = 2021,
          doi = {10.17226/26141},
       adsurl = {https://ui.adsabs.harvard.edu/abs/2021pdaa.book.....N}
}
\bibliographystyle{mnras}



\end{document}